\documentclass[11pt,leqno]{article}

\usepackage{tikz}
\usetikzlibrary{arrows.meta}
\tikzset{%
	>={Latex[width=2mm,length=2mm]},
	base/.style = {rectangle, rounded corners, draw=black,
		minimum width=2cm, minimum height=1cm,
		text centered, font=\footnotesize, align=left},
	activityStarts/.style = {base, fill=blue!5},
	startstop/.style = {base, fill=blue!1},
	activityRuns/.style = {base, fill=red!5},
	squarednode/.style={base, fill=red!13},
	process/.style = {base, fill=orange!1}
}
\usepackage{xcolor}
\usepackage{ragged2e}
\usepackage{blindtext}
\usepackage{pdfpages}
\usepackage{lmodern}
\usepackage{enumitem}
\usepackage{amsmath}
\usepackage{amssymb}
\usepackage{amsthm}
\usepackage{adjustbox}
\usepackage{blindtext}
\usepackage{import}

\usepackage{titlesec} 
\usepackage{tikz}
\usepackage{amsmath}
\usepackage{lmodern}

\usepackage{titlesec}
\usepackage{eurosym}
\usepackage{geometry}
\usepackage{stackengine}
\usepackage{listings}
\usepackage{float}
\usepackage{graphicx}
\usepackage{stmaryrd}
\usepackage{framed}
\usepackage{tcolorbox}
\usepackage{adjustbox}
\usepackage{placeins}
\usepackage[font=small,labelfont=bf,justification=centering]{caption}
\usepackage{multirow}
\usepackage{epstopdf}
\usepackage{subcaption}

\usepackage{pgf,tikz,pgfplots}
\pgfplotsset{compat=1.15}
\usepackage{mathrsfs}
\usetikzlibrary{arrows}
\usepackage{chronosys}
\usepackage{hyperref}
\usepackage[nameinlink]{cleveref}
\usepackage{pdfpages}
\usepackage{setspace}
\usepackage{hyphenat}
\hypersetup{
	colorlinks,
	linkcolor={red!50!black},
	citecolor={blue!50!black},
	urlcolor={blue!50!black},
}

\usepackage{siunitx}
\renewrobustcmd{\bfseries}{\fontseries{b}\selectfont}
\renewrobustcmd{\boldmath}{}
\newrobustcmd{\B}{\bfseries}

\usepackage{footnotebackref} 
\usepackage[hang,bottom,]{footmisc}
\usetikzlibrary{positioning,shadings}
\usetikzlibrary{arrows}

\usepackage{dcolumn}
\newcolumntype{d}[1]{D..{#1}} 

\usepackage{tabularx}
\newcolumntype{L}[1]{>{\raggedright\arraybackslash}p{#1}} 
\newcolumntype{C}[1]{>{\centering\arraybackslash}p{#1}} 
\newcolumntype{R}[1]{>{\raggedleft\arraybackslash}p{#1}} 

\usepackage{booktabs}
\usepackage{multirow}
\usepackage{pdflscape}
\usepackage{fancyhdr}
\usepackage{afterpage}
\usepackage{longtable}
\usepackage{afterpage}
\usepackage{pifont}
\usepackage{lipsum}

\makeatletter
\def\ifGm@preamble#1{\@firstofone}
\appto\restoregeometry{%
  \pdfpagewidth=\paperwidth
  \pdfpageheight=\paperheight}
\apptocmd\newgeometry{%
  \pdfpagewidth=\paperwidth
  \pdfpageheight=\paperheight}{}{}
\makeatother

\makeatletter
\def\chron@selectmonth#1{\ifcase#1\or January\or February\or March\or April\or%
  May\or June\or July\or August\or September\or October\or November\or December\fi}
\makeatother

\usepackage{etoolbox}

\usepackage[toc,page,header]{appendix} 
\usepackage{minitoc} 
\mtcsetfont{parttoc}{chapter}{\normalsize\bfseries\sffamily}           
\mtcsetfont{parttoc}{section}{\normalsize\sffamily}            
\mtcsetfont{parttoc}{subsection}{\normalsize\sffamily}         
\mtcsettitlefont{parttoc}{\LARGE\bfseries\sffamily}            

\title{Censorship in Democracy}
\date{\today}
\author{Marcel Caesmann\thanks{University of Zurich, e-mail: \textit{marcel.caesmann@econ.uzh.ch}}, $\quad$Janis Goldzycher\thanks{University of Zurich, e-mail: \textit{goldzycher@cl.uzh.ch}}, $\quad$Matteo Grigoletto\thanks{University of Bern, Wyss Academy for Nature at the University of Bern, e-mail: \textit{matteo.grigoletto@unibe.ch}}, $\quad$Lorenz Gschwent\thanks{University of Duisburg-Essen, RTG Regional Disparities and Economic Policy, e-mail: \textit{lorenz.gschwent@uni-due.de}}}

\usepackage[backend=bibtex,style=authoryear,natbib=true]{biblatex}
\DeclareCiteCommand{\cite}
  {\usebibmacro{prenote}}
  {\usebibmacro{citeindex}%
   \printtext[bibhyperref]{\printnames{labelname}}%
   \setunit{\nameyeardelim}%
   \printtext{(\printfield{year})}}
  {\multicitedelim}
  {\usebibmacro{postnote}}
\renewcommand*{\nameyeardelim}{\addspace}

\AtEveryBibitem{\clearfield{doi}}

\AtEveryBibitem{\clearfield{url}}
\AtEveryBibitem{\clearfield{urldate}}

\AtEveryBibitem{\clearfield{isbn}}

\DeclareFieldFormat{urldate}{}
\AtEveryBibitem{\clearfield{urldate}}

\begin{document}
\maketitle
\bigskip
\thispagestyle{empty}
\begin{abstract} 
The spread of propaganda, misinformation, and biased narratives from autocratic regimes, especially on social media, is a growing concern in many democracies. Can censorship be an effective tool to curb the spread of such slanted narratives? In this paper, we study the European Union’s ban on Russian state-led news outlets after the 2022 Russian invasion of Ukraine. We analyze 775,616 tweets from 133,276 users on Twitter/X, employing a difference-in-differences strategy. We show that the ban reduced pro-Russian slant among users who had previously directly interacted with banned outlets. The impact is most pronounced among users with the highest pre-ban slant levels. However, this effect was short-lived, with slant returning to its pre-ban levels within two weeks post-enforcement. Additionally, we find a detectable albeit less pronounced indirect effect on users who had not directly interacted with the outlets before the ban. We provide evidence that other suppliers of propaganda may have actively sought to mitigate the ban's influence by intensifying their activity, effectively counteracting the persistence and reach of the ban.
\end{abstract} 
\smallskip


\vspace{4cm} \noindent \begin{footnotesize} \textbf{Acknowledgments:} \\ We are grateful to Joop Adema, Elliot Ash, Kai Gehring, Carlo Schwarz, Nikita Zakharov and seminar participants at the University of Zurich, RTG Regional Disparities and Economic Policy, University of Duisburg, as well as participants and discussants at the CINCH Research Agenda for Ukraine workshop, $5^{\text{th}}$ Monash-Warwick-Zurich Text-as-Data workshop, Annual Meeting of the Austrian Economic Association (NOeG) 2023, Digital Democracy Workshop DigDemLab Zurich, $17^\text{th}$ CESifo Workshop on Political Economy, $2^\text{nd}$ CESifo/ifo Junior Workshop on Big Data, $7^\text{th}$ Conference on Political Economy of Democracy and Dictatorship, $17^\text{th}$ RGS Doctoral Conference, European Public Choice Society 2024, FRIAS conference Understanding the Rise of Autocrats in the 21$^{st}$ Century, for their helpful comments. \end{footnotesize}
\newpage
\clearpage
\pagenumbering{arabic}
\onehalfspacing
\setlength{\parskip}{0cm}
\setlength\parindent{2em}

\pagestyle{plain}
\fancyhf{}
\fancyhead[L]{\leftmark}
\fancyfoot[C]{\thepage}
\fancypagestyle{plain}{%
	\fancyhf{}%
	\renewcommand{\headrulewidth}{1pt}%
	\renewcommand{\footrulewidth}{1pt}
}
\cfoot{\vspace*{1\baselineskip}\thepage}

\doparttoc 
\faketableofcontents 

\section{Introduction}
\label{sec:intro}

Slanted media can be very effective in swaying opinions and behavior \citep{enikolopov_media_2011, yanagizawa-drott_propaganda_2014, adena_radio_2015}. Recognizing this, autocratic regimes extensively use censorship of media to suppress dissent and manage the flow of information within their borders, while employing slanted narratives to exert influence abroad \citep{guriev_sergei_spin_2022}. Concerns regarding the spread of misinformation, foreign propaganda and biased narratives, especially on social media, have been rising in recent years in democratic societies, in particular since the Russian interference in the 2016 US presidential election.\footnote{For analyses of the event, see for example work by \cite{badawy_characterizing_2019} and \cite{eady_exposure_2023}.} This pivotal event sparked heightened attention on information manipulation threats, such as the weaponization of information as part of Russia's 'asymmetric warfare' strategy in the context of its invasion of Ukraine or Chinese influence over TikTok, leading to the PAFACA act passed by US congress in 2024 to counteract foreign media influence (\href{https://www.congress.gov/bill/118th-congress/house-bill/8038}{US 118th Congress, 2024}). Given the trade-off between using censorship and upholding the conviction that free speech and independent media are the backbones of the democratic order itself, it is crucial to understand the effectiveness and consequences of media regulation in the context of liberal democracies.  While there is some evidence on the effects of censorship in authoritarian regimes \citep{chen_impact_2019, becker_freedom_2021}, systematic evidence on the effectiveness and consequences of government-imposed censorship in a democratic context is very limited. 

 To shed light on the effects of censorship in democracies, we study a ban on Russian state-backed media outlets implemented by the European Union in March 2022 in the context of Russia's invasion of Ukraine. The ban aimed to counteract the spread of Russian narratives provided by seemingly independent news sources to justify the invasion and influence public opinion.\footnote{As highlighted by the European Court of Justice's upholding, the ban was implemented on the grounds of the prohibition of ``propaganda for war'' in international law (\href{https://www.ohchr.org/en/instruments-mechanisms/instruments/international-covenant-civil-and-political-rights\#:\~:text=Article 20}{International Covenant on Civil and Political Rights, UN 1966}).} The unprecedented decision to completely ban all activity of the two most prominent of these outlets -- Russia Today and Sputnik -- was taken virtually overnight. It affected all their channels, including online platforms, in the European Union from the 2$^{nd}$ of March onwards. We focus on the social media platform Twitter/X, a pivotal platform in shaping public opinion and fueling both offline and online political activity \citep{allcott_social_2017, acemoglu_power_2018}. Twitter serves as a platform where narratives -- often led by misinformation or radical influential users -- can become extreme \citep{muller_hashtag_2023}. Its influence extends beyond digital boundaries, with narratives and stories that gain traction on the platform frequently making their way into mainstream and traditional media \citep{cage_social_2020}.

To measure pro-Russian or pro-Ukrainian media slant, we build on \cite{gentzkow_what_2010} and \cite{gennaro_emotion_2023}, establishing two distinct poles of slant: one favoring Russia and the other Ukraine. The slant of a tweet is determined by its proximity ratio to these two poles and captures the nuance of discussion in the context of Twitter. Positive numbers indicate pro-Russian while negative numbers refer to pro-Ukrainian slant.\footnote{Throughout this paper, references to a ``pro-Russia'' or ``pro-Russian'' slant specifically denote content that aligns with the Russian Government's perspectives at the time of the Ukraine invasion. This terminology is not intended to generalize or suggest that the Russian government's actions or policies reflect the opinions of the entire Russian population (analogously for ``pro-Ukrainian'' slant).}

To assess the effects of censorship in democratic contexts, we exploit geographic and temporal variation in the implementation of the ban on Russian state-controlled media, applying a difference-in-differences approach. Our analysis compares social media discourse in EU countries enforcing the ban -- Austria, France, Germany, Ireland, and Italy -- with that in non-EU countries -- Switzerland and the United Kingdom -- where no restrictions on Russia Today or Sputnik were applied during our study period. We analyze 775,616 English tweets from 133,276 users on the war in Ukraine. We conceptualize the ban as a supply shock and study its effect from three analytical angles. First, we focus on the intensive margin, i.e. the degree of slant used in tweets. Second, we study the extensive margin, i.e. the volume of slanted content created in the form of tweets. Third, we investigate the spread of slanted content through retweets. By examining these three aspects -- extremism, production, and spread -- we aim to comprehensively understand the effects of the ban. Based on this conceptual and analytical framework, we proceed in three steps to investigate how the ban affected users directly and indirectly.

First, we examine the impact of the ban on a specific subset of users who, at any time before the ban, had interacted at least once with the banned outlets. We refer to this group as \textit{interaction users}. Despite their relatively small number in our sample, these users are highly active and make up for a significant fraction of tweets on the conflict. They are more likely to support Russian propaganda and serve as a crucial link in the network, potentially bridging extreme views and misinformation from government propaganda outlets to the broader user base. Considering the aim of the ban in decreasing the spread of narratives originating from state-led Russia Today and Sputnik, these users represent a prime sample of interest to assess the effect of the ban.

We find that the ban reduced the average slant of the \textit{interaction users}, 63.1\% compared to the pre-ban mean. In turn, we observe no significant change in the proportion of tweets and retweets that are pro-Russia -- defined as tweets with more than one standard deviation from the neutral point towards the Russian pole. Exploring the heterogeneity of effects, we find that effects are most pronounced among \textit{interaction users} who used the highest pro-Russian slant before the ban on average. Specifically, users in the top 25\% of the pre-ban average slant distribution experienced a reduction of 0.15 standard deviations in their slant compared to their counterparts in non-EU countries while there is no clear effect for the bottom 75\% of the distribution.

Second, we investigate the temporal impact observed among the \textit{interaction users}. Despite detecting a significant effect on our slant measure in the pooled pre-post comparison, results using an event-study design on a daily level suggest that the influence of the ban faded within just a few days. This pattern is confirmed when we detect a meaningful and significant effect on the average slant in the first week after the ban, while effects in the second week are smaller in magnitude and with confidence intervals including zero.\footnote{We limit the analysis to two weeks after the ban, as in the third week post-ban, the UK adopted a ban on the same outlets as well depriving us of a sizeable control group. Further, in the course of this project the access to Twitter/X API for researchers was restricted and limited the option to collect additional data.} Overall, the results suggest an immediate but short-lived effect on the users who interacted with Russia Today and Sputnik before the ban.

Next, we study the indirect effects of the ban on users who did not directly interact with the banned outlets. We label these individuals as \textit{non-interaction users}. We assess the ban's impact on these users across the same metrics of intensive and extensive margins as we did for the \textit{interaction users}. We find that the ban reduced pro-Russian slant among the \textit{non-interaction users}, despite to a lesser degree, resulting in a decrease of approximately 17.3\% from pre-ban levels of slant, in contrast to the 63.1\% observed among \textit{interaction users}. Notably, we do find effects on the extensive margin for this group of users, specifically in a reduction of the share of pro-Russian retweets. This suggests that \textit{non-interaction users} are deprived of slanted content they are able and willing to share.

Our results show that the ban had an effect, particularly on those users who interacted with the banned outlets before the implementation. However, this effect fades quickly. In the third step of our analysis, we investigate the mechanisms that might have compensated for the ban's effect, effectively re-balancing the supply of pro-Russia slanted content. This part of our study specifically examines users identified as \textit{suppliers} of slanted content. We define as \textit{supplier} any user who, during a given time period, produced or shared any content that was more than one standard deviation towards the pro-Russia pole.

The first potential mechanism we examine is the entry of new \textit{suppliers} following the ban. We assess the proportion of users identified as \textit{suppliers} in both EU and non-EU countries, before and after the ban was implemented. Additionally, we explore a potential rise in the number of pro-Russian bots. In both analyses, we find that although there is an increase in the number of \textit{suppliers} within the EU post-ban this increase is actually outpaced by the rise in \textit{suppliers} observed in non-EU regions. Thus, while the augmented number of \textit{suppliers} might contribute to mitigate the impact of the ban, it seems unfit to explain the different trajectories in EU and non-EU countries.

The second mechanism we investigate concerns the behavior of users who were already disseminating pro-Russia content prior to the ban. It is plausible that these individuals were either competing with or aiding the banned outlets in spreading slanted content. With the ban in place, they might have moved to occupy the void created. Our analysis of the ban's impact on these users yields several key findings. Overall, it appears that European \textit{suppliers} affected by the ban reduced both their content's slant and their share of pro-Russian retweets compared to their non-EU counterparts. When examining the heterogeneity of this effect based on pre-ban activity levels, it becomes evident that the primary contributors to this effect were \textit{suppliers} who were moderately active before the ban. In contrast, the most active \textit{suppliers} seemed unaffected by the ban. Moreover, an analysis of the very top \textit{suppliers} in the activity distribution prior to the ban suggests -- with caution due to the small sample size -- that the most active \textit{suppliers} may have even increased the production of new pro-Russian content in response to the ban.

This paper contributes to several strands of the literature. First, we contribute to the literature on censorship. Conceptual works by \cite{shadmehr_state_2015} and \cite{gehlbach_government_2014} highlight a trade-off between censoring and allowing unbiased information that autocratic rulers face. Censoring comes with the cost of signaling to the population the attempt to control the information and readers might decide to take distance from the news outlets if they do not meet their need for informational content. Empirical evidence on the effects of censorship is limited. In autocratic contexts, \cite{chen_impact_2019} study the impact of providing citizens with access to uncensored internet and find effects on beliefs, attitudes, and intended behavior. \cite{becker_freedom_2021} show the impact of censorship imposed by the Catholic Church during the Counter-Reformation preventing the diffusion of Protestant material. In the context of Russia, \cite{simonov_demand_2022} show how outlet-specific characteristics attract readers to government-controlled media and how readers, once there, do not change information sources. Overall, the existing evidence suggests that censorship can be effective in impeding the spread of information and changing citizens' beliefs and behaviors in autocratic contexts. 

In democratic contexts, there is even more limited evidence of the effectiveness of censorship. \cite{bjornskov_is_2021} study the effect of constitutional provisions in preventing media censorship in the aftermath of terrorist attacks. This is one of the few examples investigating the interaction between functioning constitutional systems and censorship of media, in line also with the work by \cite{kellam_silencing_2016}, providing evidence that powerful presidents can be threatening to media freedom even in democratic contexts. There is growing attention to weaker forms of self or state-mandated regulation of media platforms. Information withholding that is only targeting certain outlets is referred to as ``selective censorship" \citep{guriev_sergei_spin_2022}. So far, empirical analyses of this form of censorship have been rare. \cite{corduneanu-huci_selective_2022} find in a cross-country analysis that media outlets that likely reach the median voter have a higher chance of being censored in both autocracies and democracies. They further argue that censors undertake a cost-benefit analysis weighing an outlet's audience's perceived political danger against legal and reputation costs. In the case of the EU's ban on Russia Today and Sputnik, judiciary viability needs to be particularly considered. \cite{baade_eus_2022} argued that the ban can be viably justified based on the UN's prohibition of ``propaganda for war'',\footnote{See Article 20 of the International Covenant on Civil and Political Rights.} but this argument could not be readily applied to different outlets. The European Court of Justice's upholding of the ban against a complaint by Russia Today in July 2022 indeed relied on this prohibition.\footnote{For more information, see the official \href{https://eur-lex.europa.eu/legal-content/en/TXT/?uri=CELEX:62022TJ0125}{statement} by the European Court of Justice on the ruling.} Therefore, studying this ban provides a unique natural experiment for studying the causal effects of censorship in a democratic context.

Second, we also add to the rich literature investigating the political economy of social media (see \cite{campante_political_2023} for an overview). The effects of the advent of social media on political systems are not clear yet. On one side, social media appears to facilitate the spread of populism \citep{campante_politics_2018, guriev_3g_2021}, xenophobia \citep{bursztyn_social_2019}, boosts a trend towards political polarization \citep{halberstam_homophily_2016, levy_social_2021, muller_hashtag_2023} and reduce subjective well-being \citep{allcott_welfare_2020}. On the other side, some scholars argue not only that segregation offline is stronger than that online \citep{gentzkow_ideological_2011}, but also that social media could play a role in decreasing polarization \citep{barbera_how_2014}. 

An emerging literature in this space is concerned with the effect of online content moderation  \citep{chandrasekharan_you_2017,jhaver_evaluating_2021, jimenez-duran_economics_2023}. \cite{morales_perceived_2020} studies the effect of banning bots programmed to retweet the Venezuelan president Nicolás Maduro's tweets, showing that this makes the discussion on Twitter become more critical of the president. \cite{ershov_sharing_2021} find that policies/nudges around the 2020 US election aimed at decreasing the sharing of misinformation led to a stronger decrease in the sharing of more factual news. Closest to our paper are studies by \cite{muller_effects_2022} and \cite{jimenez_duran_effect_2022}. They study the effect of banning Trump's account on reducing toxicity amongst his followers and the effects of a German regulation on removing online hate speech directed towards refugees. In line with our results, both studies show that online content moderation can curb toxicity and hate speech online. We go beyond the existing evidence on two dimensions. First, our study can exploit cross-country variation in who is affected by the ban with a more natural control group to identify effects. Second, we study the response of users -- the pre-ban \textit{suppliers} of media slant -- in filling the gap left by banned outlets.

Third, our study contributes to the understanding of media slant, which can be interpreted as an indicator of state-led narratives, often associated with state propaganda. This expands the literature on the effects of propaganda \citep{enikolopov_media_2011, yanagizawa-drott_propaganda_2014, adena_radio_2015}, which traditionally examines the impact of increasing propaganda exposure. Our research setting allows us to explore the consequences of reducing exposure to media slant, providing a new dimension to the policy debate on media regulation. Moreover, we introduce a novel, data-driven approach to measuring media slant, leveraging advances in text-as-data methodology. This new measure, inspired by the works of \cite{gentzkow_ideological_2011} and \cite{gennaro_emotion_2023}, offers a more nuanced understanding of media slant and its effects.

The remainder of the paper is structured as follows. Section \ref{sec:data} discusses the data used in our analysis while section \ref{sec:measuring} provides an overview of our measure of media slant and the approach used to obtain the measurement. Section \ref{sec:frame:method} gives an overview of the framework and methods used for the analysis. Finally, we present our results on the effect of censorship in democracy in section \ref{sec:results:impact} and provide a concluding discussion in section \ref{sec:conclusions}.

\section{Data}
\label{sec:data}

In this paper, we use two main samples of tweets. The first one consists of more than 15,000 tweets by official Ukrainian and Russian government accounts, collectively labelled as \textit{government tweets} (GT), which will be further discussed in Section \ref{sec:measuring}. The second sample constitutes the main dataset for our analysis, to which we refer from now on as to the \textit{users' tweets} (UT). This dataset covers the general European Twitter discussion about the Russo-Ukrainian conflict between February 19$^{th}$, 2022, and March 15$^{th}$, 2022. We provide detailed information about the extraction of this dataset in the Appendix \ref{appendix:method}. 

For the \textit{users' tweets} -- due to limitations in accessing the Twitter API\footnote{Free API access for researchers was unfortunately suspended shortly after Elon Musk's takeover of Twitter in October 2022 and during this research project.}  -- we restrict our extraction to the following European countries: Austria, France, Germany, Ireland, Italy, Switzerland, and the United Kingdom. We also focus exclusively on tweets in the English language for several reasons. First, the most important branches of the Russian state-led propaganda outlets were those operating in English. Second, considering the international reach of the Russo-Ukrainian conflict, much of the discussion on Twitter was happening in English. Third, for the \textit{government tweets}, we selected officials posting extensive English content to signal their willingness to spread their narrative beyond the boundaries of their countries.

The resulting sample consists of 775,616 English-language tweets by 133,276 users. We show descriptive statistics on the tweet level in Table \ref{tab:main:descrip}, Panel A. 5.7\% of the tweets in our sample can be classified as being pro-Russian slant, while 10\% are a retweet of pro-Russian slant. The number of own tweets and retweets is fairly balanced, with the latter representing 53\% of the sample. Notably, tweets of our sample are retweeted extensively, with the median being 26 retweets per tweet. We show descriptive statistics on the user level in Table \ref{tab:main:descrip}, Panel B. On average, each user of our sample produces 2.7 tweets and 3.1 retweets covered in our sample. Of all users, 3.7\% had direct contact with the banned outlets before the ban, as in a reply or a retweet of the outlets' accounts. In the Appendix Table \ref{app:tab:main:descrip:nobots}, we provide the same summary statistics after cleaning our sample of 2,489 potential bots. In Appendix \ref{appendix:lateacc} Table \ref{app:tab:main:descrip:lateacc}, we also show the descriptive statistics after dropping 389 accounts in our sample that were created only after the ban was enforced.

\begin{table}[t]
\centering
\caption{\textit{Summary statistics}}
\label{tab:main:descrip}

\begin{minipage}{0.61\textwidth}
\footnotesize{\textbf{Panel A: Tweet level}}
\end{minipage}\\[0.2cm]
\begin{tabular}{lcccc}
    \resizebox{0.63\linewidth}{!}{{
\def\sym#1{\ifmmode^{#1}\else\(^{#1}\)\fi}
\begin{tabular}{l*{1}{ccccc}}
\toprule
                &\textbf{Mean}&\textbf{Median}&\textbf{St. Dev.}&\textbf{Min.}&\textbf{Max.}\\
\midrule
\textbf{Dependent Variables}\vspace{3pt}&         &         &         &         &         \\
Propaganda ratio&  -.00011&     .042&        1&       -4&4.8926959038\\
Russian propaganda tweet&     .057&        0&      .23&        0&        1\\
Russian propaganda retweet&       .1&        0&       .3&        0&        1\\
\\ \textbf{Tweet type}\vspace{3pt}&         &         &         &         &         \\
Retweet         &      .53&        1&       .5&        0&        1\\
No. of words    &       25&       23&       11&        1&      108\\
No. of mentions &      1.6&        1&      2.4&        0&       50\\
No. of hashtags &      .44&        0&      1.6&        0&       42\\
\midrule
No. of Observations&  775,616&         &         &         &         \\
\bottomrule
\end{tabular}
}
}
\end{tabular}\\[0.5cm] 

\begin{minipage}{0.61\textwidth}
\footnotesize{\textbf{Panel B: User level}}
\end{minipage}\\[0.2cm]
\begin{tabular}{lcccc}
    \resizebox{0.63\linewidth}{!}{{
\def\sym#1{\ifmmode^{#1}\else\(^{#1}\)\fi}
\begin{tabular}{l*{1}{ccccc}}
\toprule
                &\textbf{Mean}&\textbf{Median}&\textbf{St. Dev.}&\textbf{Min.}&\textbf{Max.}\\
\midrule
\textbf{User behavior}\vspace{3pt}&         &         &         &         &         \\
No. tweets from user&      2.7&        1&       12&        0&    1,528\\
No. retweets from user&      3.1&        1&       11&        0&      616\\
No. replies from user&      .52&        0&      2.1&        0&      202\\
No. russian propaganda tweets&      .33&        0&      1.6&        0&      300\\
No. russian propaganda retweets&       .6&        0&      2.3&        0&      138\\
Interacted with RT/Spk&     .037&        0&      .19&        0&        1\\
No. retweets of RT/Spk&     .001&        0&     .044&        0&        6\\
\\ \textbf{Region}\vspace{3pt}&         &         &         &         &         \\
European Union  &      .39&        0&      .49&        0&        1\\
\midrule
No. of Observations&  133,276&         &         &         &         \\
\bottomrule
\end{tabular}
}
}
\end{tabular}

\begin{minipage}{\textwidth} \vspace{0.4cm}   
\footnotesize{\textbf{Notes:} Panel A reports descriptive statistics for the sample of tweets used in the analysis, posted by users that we could locate in the countries of interest: Austria, France, Germany, Ireland, Italy, Switzerland and the United Kingdom. Tweets were extracted using the Historical Twitter APIv2, with the query: \textit{ukrain* OR russ* OR NATO OR OTAN}, in the time window between February 19$^{th}$ and March 15$^{th}$ 2022. Panel B reports descriptive statistics on user characteristics, for users that posted tweets used in our analysis and described in Panel A. In both panels, for all variables we report mean, median, standard deviation, minimum, and maximum values. Appendix Tables \ref{app:tab:main:descrip:nobots}, \ref{app:tab:main:descrip:lateacc} and \ref{app:tab:main:descrip:diffthres} show the same descriptive statistics excluding users that are plausible bots, excluding accounts created only after the ban and using 0, instead of 1, threshold to define the binary variables of pro-Russia tweets and retweets, respectively.}
\end{minipage}
\end{table}

\begin{figure}[t]
	\centering
	\caption{\textit{Geographical distribution of users}}\label{fig:map:full}
	\includegraphics[width=.55\textwidth, trim={0cm 0.1cm 0.1cm 3cm},clip]{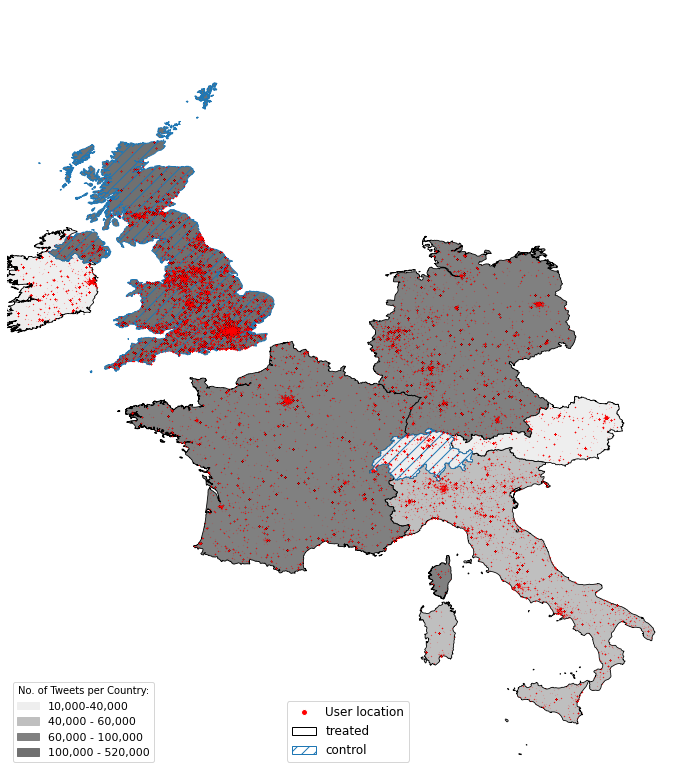}
	
	\begin{minipage}{\textwidth} \vspace{0.4cm}   \footnotesize{\textbf{Notes}: The figure shows the geographical distribution of users in our sample. In order to locate users in the countries of our analysis, Austria, France, Germany, Italy, Ireland, Switzerland and United Kingdom, we follow the geo-location pipeline proposed in \cite{gehring_analyzing_2023}.}\end{minipage}
\end{figure}

In Figure \ref{fig:map:full}, we delve into the geographical distribution of tweets in our dataset, mapping each tweet from the UT sample, according to the location of the user producing the tweet. As expected, a significant portion originates from the United Kingdom. This is attributable to the country's substantial Twitter user base and our emphasis on English-language tweets. While the UK's representation is pronounced, it is important to keep in mind the two areas that we compare in our study encompass the UK and Switzerland on the one side -- as countries where the ban was not applied -- and Austria, France, Germany, Ireland, and Italy on the other side -- as countries where the ban was applied.

\section{Measuring pro-Russian and pro-Ukrainian slant}
\label{sec:measuring}

In the context of the Russian invasion of Ukraine in 2022, we conceptualize the discourse on the war as defined by a one-dimensional continuum between two narrative poles: pro-Russia and pro-Ukraine. In our analysis, discussions about the conflict occupy positions along this spectrum, with a tweet's content being closer to one pole or the other, indicating its narrative slant. This proximity to either pole reflects the intensity of its alignment,  with content equidistant from both poles representing a neutral stance, thus supporting neither side strongly. By adopting this framework, we emphasize the importance of capturing the nuanced spectrum of discussions, arguing that it would be overly simplistic and inaccurate to categorize tweets exclusively as pro-Russia or not, hence acknowledging the complexity of public discourse and opinion formation during the conflict.

To obtain a quantifiable and tractable measure of pro-Russian and pro-Ukrainian media slant, we adopt a procedure proposed by \cite{gennaro_emotion_2023}, drawing inspiration from earlier work on media slant by \cite{gentzkow_ideological_2011}. This approach is both simple and powerful, relying on the measurement of language similarity in tweets by European users discussing the war relative to two distinct ideological poles. More specifically, we calculate the cosine similarity of a tweet's language to what we define as the 'pro-Russian pole' and compare this to its similarity to what we define as the 'pro-Ukrainian pole'. The critical decision in this method lies in determining the content that constitutes each pole. 

We construct our ideological poles using content disseminated on Twitter by key figures within the Russian and Ukrainian governments. To achieve this, we systematically gather tweets posted by accounts affiliated with these governments, the comprehensive list of which is detailed in Table \ref{tab:main:gov_accounts}. Our collection encompasses 5,993 tweets from Russian government representatives and 9,451 tweets from Ukrainian government representatives, collected over the period between January 24$^{th}$, 2022, to April 4$^{th}$. This dataset proves instrumental in establishing our measure of media slant, offering a direct insight into the narratives each government endorsed and ensuring our analytical framework's robustness.

\begin{table}[t]\centering
	\caption{\textit{Accounts of the Russian and Ukrainian governments' representatives}}
	\label{tab:main:gov_accounts}
	\def\sym#1{\ifmmode^{#1}\else\(^{#1}\)\fi}
	\resizebox{\linewidth }{!}{
		\begin{tabular}{lcccccc}
			\hline\hline
			\begin{tabular}{p{7cm}p{7cm}p{7cm}p{7cm}}
\toprule
                 Ukrainian Accounts &                                             Account Holder &                    Russian Accounts &                                       Account Holder \\
\midrule
     https://twitter.com/DI\_Ukraine &                                       Defence Intelligence &  https://twitter.com/RussianEmbassy &                                    Embassy in the UK \\
        https://twitter.com/Ukraine &                                                    Ukraine &      https://twitter.com/mfa\_russia &                          Ministry of Foreign Affairs \\
       https://twitter.com/DefenceU &                                        Ministry of Defense &      https://twitter.com/mission\_rf & Mission to the International Organizations in Vienna \\
       https://twitter.com/CinC\_AFU &                          Colonel General Oleksandr Syrskyi &         https://twitter.com/RF\_OSCE &                                  Mission to the OSCE \\
https://twitter.com/oleksiireznikov &                                        Minister of Defence &       https://twitter.com/RusEmbUSA &                                    Embassy in the US \\
    https://twitter.com/kabmin\_ua\_e &                                       Cabinet of Ministers & https://twitter.com/RussianEmbassyC &                                    Embassy in Canada \\
    https://twitter.com/MFA\_Ukraine &                                Ministry of Foreign Affairs & https://twitter.com/KremlinRussia\_E &                                Official Kremlin News \\
   https://twitter.com/DmytroKuleba &                                Minister of Foreign Affairs & https://twitter.com/EmbassyofRussia &                              Embassy in South Africa \\
   https://twitter.com/AndriyYermak &                        Head of the Office of the President &    https://twitter.com/PMSimferopol &        Ministry of Foreign Affairs' Office in Crimea \\
        https://twitter.com/NSDC\_ua & Press Service of the National Security and Defense Council &   https://twitter.com/RusMission\_EU &                                    Mission to the EU \\
       https://twitter.com/UKRinDEU &                              Embassy of Ukraine in Germany &    https://twitter.com/RusBotschaft &                                   Embassy in Germany \\
       https://twitter.com/ukrinche &                          Embassy of Ukraine in Switzerland &     https://twitter.com/RusEmbSwiss &                               Embassy in Switzerland \\
       https://twitter.com/ukrinfra &                               Embassy of Ukraine in France &    https://twitter.com/ambrusfrance &                                    Embassy in France \\
        https://twitter.com/ukrinit &                                Embassy of Ukraine in Italy &     https://twitter.com/rusembitaly &                                     Embassy in Italy \\
   https://twitter.com/UkrEmbLondon &                               Embassy of Ukraine in the UK &                                     &                                                      \\
   https://twitter.com/MelnykAndrij &                            Ukrainian Ambassador to Germany &                                     &                                                      \\
\bottomrule
\end{tabular}

		\end{tabular}
	}
 
	\begin{minipage}{\linewidth} \vspace{0.4cm}   \footnotesize \setstretch{1.0} {\textbf{Notes}: The table reports the Russian and Ukrainian government-affiliated accounts that were used as sources for the two poles used to create our slant measurement. For each of these accounts, we extracted tweets in English between January 24$^{th}$, 2022, and April 4$^{th}$, 2022. This extraction resulted in 5,993 tweets for the Russian pole and 9,451 tweets for the Ukrainian pole. Note that to increase the number of English accounts on the Russian side, we also included the embassy account for non-European English-speaking countries active on Twitter.}\end{minipage}
\end{table}

\begin{figure}[t]
	\centering
	\caption{\textit{Word frequency in the sample of government tweets}}\label{fig:keywords}
	\includegraphics[width=.65\textwidth, trim={0cm 0.1cm 0.1cm 0.1cm},clip]{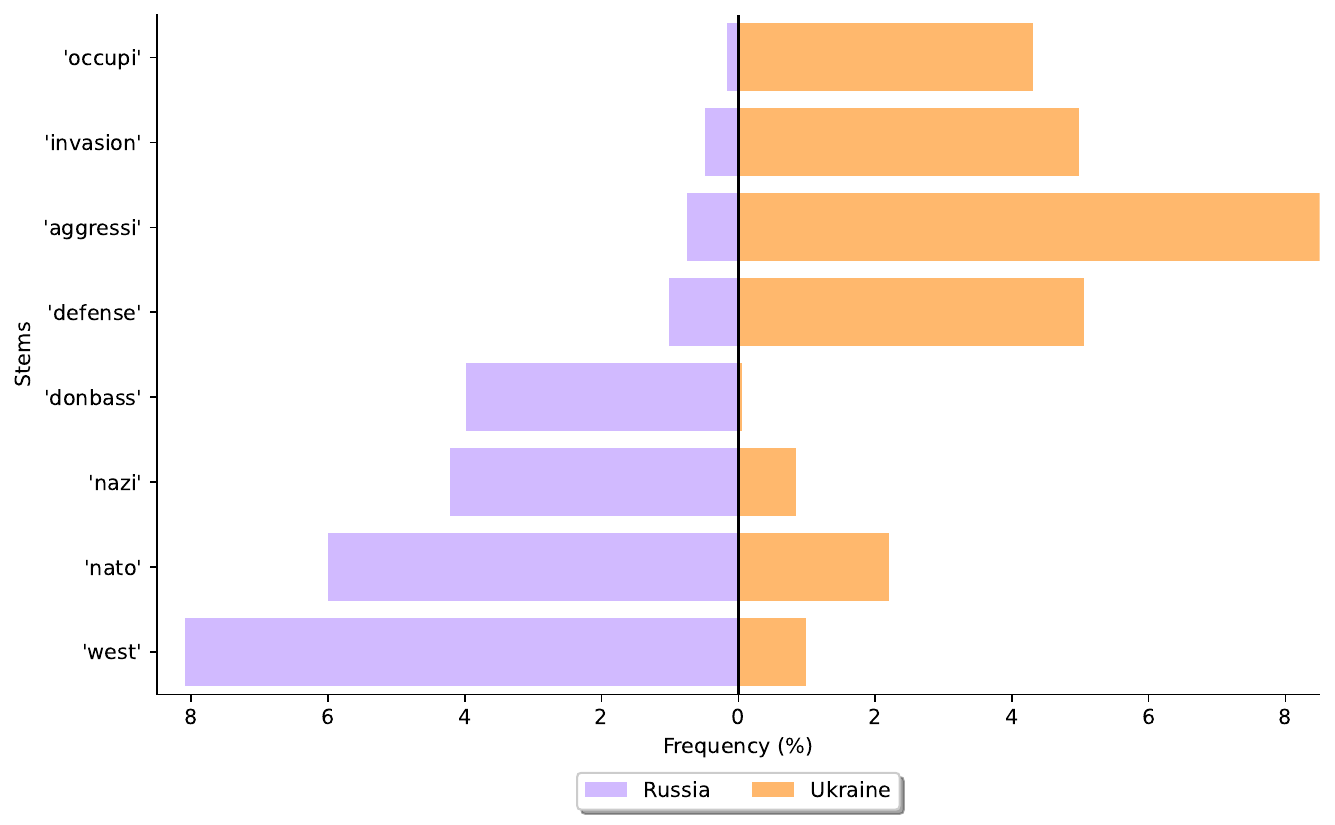}
	
	\begin{minipage}{\textwidth} \vspace{0.4cm}   \footnotesize{\textbf{Notes}: The figure shows frequencies for selected word stems in the sample of government tweets. In purple, we show the frequency in tweets coming from representatives of the Russian government, and in orange for the Ukrainian government. Frequencies represent the percentage of tweets containing the stem of each specific word of interest. Results are based on 9,451 tweets from Ukrainian government exponents and 5,993 tweets from Russian government exponents.}\end{minipage}
\end{figure}

Figure \ref{fig:keywords} provides insights into the content of these government tweets through keyword frequency analysis. We differentiate the frequencies of Russian and Ukrainian government tweets, represented in purple and orange, respectively. Keywords like 'aggression' and 'invasion' are predominantly used by Ukrainian accounts to portray the conflict as an invasion, contrasting with the Russian portrayal as a 'military operation'. Other stems like 'occupi', 'defense', 'nato', 'west', 'nazi', and 'donbass' further delineate the narratives of each side. The use of these terms underlines the slant in the content from these government accounts, making them suitable benchmarks for our measurement.

Following \cite{gennaro_emotion_2023} we take all Ukrainian tweets in GT, create a vector representation using the text embedding model sentence-t5-xl \citep{ni_sentence-t5_2021}, and average those representations to produce a single vector representing the Ukrainian government pole. We compute the Russian government pole analogously. Then, we embed all UT of our main analysis' dataset (see data section \ref{sec:data}) with sentence-t5-xl and use Equation \ref{eq:pole:ratio} to obtain a score for each input tweet. This score is a ratio measuring the language similarity between the given tweet and the Russian pole relative to the similarity between the given tweet and the Ukrainian pole. Formally, we compute this ratio as follows:

\begin{equation}
	\label{eq:pole:ratio}
	Y = \dfrac{sim(d, R) + b}{sim(d, U) + b} - 1,
\end{equation} 

\noindent where $d$ denotes a vector representing the input text, $R$ and $U$ are vectors representing the two poles, $b$ is a smoothing parameter set to 1, and $sim$ refers to the cosine similarity. We subtract 1 to center the ratio around zero. Positive values indicate tweets more similar to the Russian pole. Tweets with negative values are more similar to the Ukrainian pole. If $Y = 0$, this means that the input text is equally similar to the Ukrainian and Russian poles. We compute the poles as time-varying measures to account for possible changes over time in the official viewpoints. The comparison poles for a day t consist of all government tweets between days t-7 and t, using a decay factor of 0.5 to reduce the influence of more distant days\footnote{The decay factor of 0.5 results in the following weights: 0.5, 0.55204476, 0.60950683, 0.6729501, 0.74299714, 0.82033536, 0.90572366, 1.}. The pole ratio for each user tweet from day t is then computed based on the two corresponding daily poles. 

Finally, we standardize the resulting media slant ratio to a mean of 0 and a standard deviation of 1. Increasing our final measure by one unit implies moving one standard deviation closer to the Russian pole. Appendix Figure \ref{app:fig:examples} shows examples of tweets and their corresponding score. When in the following sections we refer to tweets and retweets of pro-Russian slant, we refer to a binary measure. This measure is obtained by assigning 1 to any tweet or retweet whose ratio score is one standard deviation above 0, hence towards the Russian pole, and 0 otherwise. We use the threshold of one standard deviation to mitigate the inclusion of potential noise around the mean value of 0. Nevertheless, we perform robustness checks using a zero threshold to further ensure the reliability of our findings.

\begin{figure}[t]
	\centering
	\caption{\textit{Time-series of our slant measure: Daily averages}}\label{fig:timeseries:slant:daily}
	\includegraphics[width=.65\textwidth, trim={0cm 0.1cm 0.1cm 0.1cm},clip]{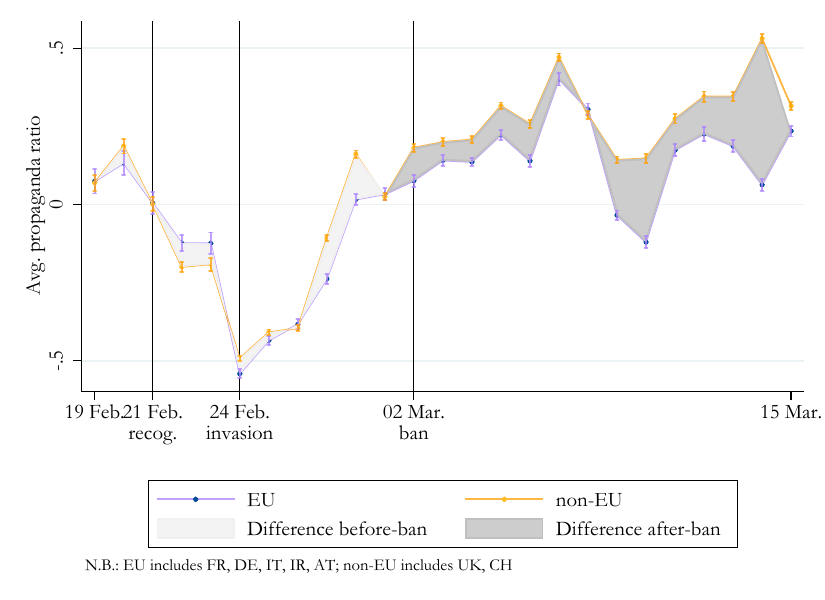}
	
	\begin{minipage}{\textwidth} \vspace{0cm}   \footnotesize{\textbf{Notes}: The figure shows the daily averages of our media slant measurement, in the time-frame of our analysis, between February 19$^{th}$, 2022 to March 15$^{th}$, 2022. The measure is normalized to have a mean of 0 and a standard deviation of 1. When positive the measure indicates content closer to the Russian pole, and when negative it indicates content closer to the Ukrainian pole. In purple, we show the daily averages in the EU countries in our study, Austria, France, Germany, Ireland, and Italy, while in orange the daily averages in non-EU countries, United Kingdom and Switzerland. In grey, the difference between the two averages. We indicate the following important dates: 21$^{st}$ Feb. for the official recognition by Putin of the Donetsk People’s Republic and the Luhansk People’s Republic, 24$^{th}$ Feb. as the beginning of the war and 2$^{nd}$ Mar. as the beginning of the ban. Appendix Figure \ref{app:fig:timeseries:slant:daily:onlytw} reproduces the same time-series but limiting the analysis to original tweets excluding retweets.}\end{minipage}
\end{figure}

In Figure \ref{fig:timeseries:slant:daily}, we present the daily average measure of media slant in both EU and non-EU countries as part of our analysis. The graph reveals several notable trends. In the days marking the onset of the invasion, both EU and non-EU regions reached a minimum in average slant, suggesting a widespread initial reaction leaning towards the Ukrainian pole. Subsequently, there is a pronounced and consistent shift towards the Russian pole, underscoring the European Commission's concern that the conflict was being waged not only on the ground but also online, with the EU particularly targeted by Russian propaganda efforts. Furthermore, the movement of average slant in EU and non-EU countries shows similar trends before the ban's implementation, after which a distinct divergence is observed. This pattern may reflect the ban's impact, a hypothesis we will explore more systematically in the subsequent sections. In Appendix Figure \ref{app:fig:timeseries:slant:daily:onlytw} we show the same using only original tweets.

\section{Conceptual framework and empirical method}
\label{sec:frame:method}

\subsection{A supply shock to the market for news}
\label{sub:sec:conceptual}

The enforcement of the ban against Russia Today and Sputnik represents a significant shock to the news market. With millions of followers and a presence across diverse online platforms, they were central nodes in spreading state-aligned Russian narratives. Their direct links to the Russian government enhanced their effectiveness, enabling them to access and disseminate information rapidly and widely. Therefore, their sudden block within the EU from a crucial platform like Twitter did not just eliminate two prolific voices, it fundamentally disrupted the network of information flow. This disruption is akin to removing a major supplier from a traditional market, with expected repercussions on both the availability of certain narratives and the overall dynamics of information dissemination.

We follow the literature on media bias \citep{mullainathan_market_2005, gentzkow_what_2010} and think of the news market about a specific subject, the Russo-Ukrainian conflict, as a Hotelling-type market structure. Consumers are distributed along a line of beliefs and have a preference to read news consistent with their beliefs. The two ends of the distribution represent the most extreme worldviews, in our case, the Russian and Ukrainian governments. News outlets cater to a particular segment of consumers along this distribution. The choice of the segment is driven by two motives: medium-internal preferences over worldviews and maximizing outreach. 

In this setup, the ban takes the form of a classic supply shock. Users who previously read and interacted with Russia Today and Sputnik lose a source of information\footnote{Importantly, the way Twitter enforced the ban did not allow for direct circumvention of the ban via a VPN on an account created before the ban. We address the potential for circumvention by creating new accounts after the ban in Appendix \ref{appendix:lateacc}.}. In a traditional news market, the affected consumers would resort to other news outlets. Additionally, outlets could adjust their reporting in an attempt to capture this vacated share of readers. However, on social media, we expect users not only to adjust their news consumption behavior but also their production of content by posting and sharing information. Notably, in the very short run, the ban leads to a mechanical decrease in the amount of pro-Russian slant content that can be shared. However, social media also allows users not affiliated with any media outlet to share or create information aligning with their worldview. Therefore, certain users may try to fill the information gap created by the ban. We test this hypothesis in Section \ref{sub:sec:mechanisms}.

To assess the ban's impact, we examine the dynamics of pro-Russian and pro-Ukrainian slant content on Twitter from three analytical angles. The first aspect is the intensive margin or the depth of intensity of the slant content. The ban might signal a broader intolerance towards extreme narratives, potentially prompting users who engage with or disseminate such content to re-calibrate their approach. This could lead to a decrease in the overall extremism of tweets, as users might moderate their language and content to avoid being flagged or banned.

The second aspect, the extensive margin, concerns the volume of slanted content production, which we quantify by counting the number of original pro-Russian slant tweets with our binary measure. With focal \textit{suppliers} of Russian-aligned narratives removed, users who previously directly or indirectly relied on these sources might experience an increased cost of searching for pro-Russian content, resulting in reduced pro-Russian content creation. In turn, this could also spur users to compensate for the loss of ready-made content by increasing their production of original tweets that align with or support the now-absent narratives. Such a response would reflect an attempt to maintain the presence of these narratives in the public discourse, despite the absence of their primary propagators. The empirical question we tackle in our analysis is which of the two forces dominates.

The third aspect focuses on the spread of slant content, particularly through retweets, and can still be conceived as part of the extensive margin. By cutting off a significant content source, the ban will likely impact the volume of retweets of pro-Russian content. Users finding fewer original posts from the banned outlets directly or from other users building on banned outlet content to share might reduce their retweeting behavior. This could result in a noticeable decline in the spread of Russian-aligned narratives on the platform, affecting the reach and penetration of these narratives among the wider audience.

By examining these three aspects -- intensity, production, and spread -- we aim to comprehensively understand the multifaceted effects of the ban. Each dimension provides a different lens through which to view the ban's impact, collectively offering a holistic picture of how a significant policy reshapes the digital landscape of media slant and information dissemination.

\subsection{Identification strategy}
\label{sub:sec:is}

We use a difference-in-difference design to estimate the causal effect of banning the Russian outlets by comparing tweets posted by users in the EU, who were affected by the ban, to tweets posted by users outside the EU, whose exposure to these outlets was not restricted. We estimate difference-in-difference specifications at the user-time level of the following form:

\begin{equation}\label{eq:did:user}
	Y_{i, t} = \alpha_{i}+\gamma_{t}+\beta EU_i \times Ban_t+ \Theta \pmb{X}_{i, t}+\epsilon_{i, t},
\end{equation}

\noindent  where $Y_{i, t}$ is a measure of user behavior, such as the average media slant ratio measure, the number of pro-Russian slant tweets, and retweets, by user $i$ in period $t$. These outcomes target the arguably most policy-relevant issues at the core of the EU's justification of the ban: the overall qualitative content of the discourse on the war in Ukraine and the quantity of pro-Russian narratives spread in the online information space. $EU_i$ is an indicator variable equal to 1 for users located in the European Union and 0 otherwise. $Ban_t$ is an indicator equal to 1 after March 2$^{nd}$, 2022. We also estimate an event study version of Equation \ref{eq:did:user} to investigate the timing of the effect. $\alpha_i$ and $\gamma_t$ are a full set of user and day-fixed effects absorbing average differences in tweet content across users and time. $X_{i, t}$ is a vector of additional control variables capturing tweet style information, i.e., the number of words, hashtags, and mentions.

To interpret $\beta$ in Equation \ref{eq:did:user} as the causal effect of the ban on Russia Today and Sputnik, we require the assumption that Twitter user behavior in EU and non-EU countries would have followed parallel trends in the absence of the ban. While this assumption is not directly testable, we provide two pieces of evidence to support it. First, we test for balance in tweet characteristics \textit{before} the ban between users located in the EU and non-EU countries in our sample. Appendix Figure \ref{app:fig:balance} shows the results of this exercise. We detect some differences and we control for the variables mentioned above. These variables do not affect our estimates. Second, in an event study specification, we investigate the link between user behavior and being located in the EU. This allows us to see whether the slant of users in the EU compared to users outside the EU followed similar trends before the ban. The results of this exercise show no meaningful differential trends in outcomes before the ban, making it less likely that the trends would have diverged in the absence of the ban.

Another potential concern is that the invasion itself might impact the use of pro-Russian or pro-Ukrainian slant on Twitter. Day-fixed effects absorb any common shock to EU and non-EU countries. We only include major Western European countries in our sample to make it less likely that differential exposure to the war itself or fear of spillover to the country of residence confounds our estimates. Eastern European countries bordering Russia or Ukraine might have a differential response to the war, so we excluded them from the analysis.

A final concern is that the ban was not enforced and a clean assignment of the treatment status in our setting also depends on the absence of ways to circumvent the ban. As outlined in detail in Appendix \ref{appendix:geo}, we use the location of Twitter users as indicated in their profiles to assign them to a country and thereby a treatment status. These locations are public information and can be retrieved via the Twitter API. However, Twitter also gathers non-public information on users' locations to determine the content that cannot be displayed to a user. According to \href{https://help.twitter.com/en/managing-your-account/how-to-change-country-settings}{Twitter's public documentation}, such information does not only include IP addresses, which can be easily changed via a VPN but, for example, also wireless networks or cell towers near a user. Crucially, manually changing this non-public country setting does not change the content withheld by Twitter due to local laws. Therefore, even if readers of RT and Sputnik who reside in one of the countries that enacted the ban and had their location assigned by Twitter use a VPN to reach the website of the outlets, they still cannot access, read, or interact with any account of RT and Sputnik.

\section{Results: The impact of the ban}
\label{sec:results:impact}

\subsection{Direct effect}
\label{sub:sec:consumers}

The first step of our analysis involves assessing the impact of the ban on individuals who previously engaged with Russia Today and Sputnik. These users -- that from now on we indicate as the \textit{interaction users} -- are pivotal for several reasons. First, the ban directly impacts them by removing a critical source of information and news content from their online social network. Second, while they represent a minor segment of our sample, their contribution to the overall volume of tweets is substantial. Third, this group predominantly disseminated pro-Russia content prior to the ban -- as shown in Appendix Figure \ref{app:fig:rusinteract:preban} -- placing them at the core of the EU's regulatory measure.

\begin{figure}[t]
	\centering
	\caption{\textit{Daily event-study on our slant measure: Interaction users}}\label{fig:eventstudy:intensive:interact}
	\includegraphics[width=.65\textwidth, trim={0cm 0.1cm 0.1cm 0.1cm},clip]{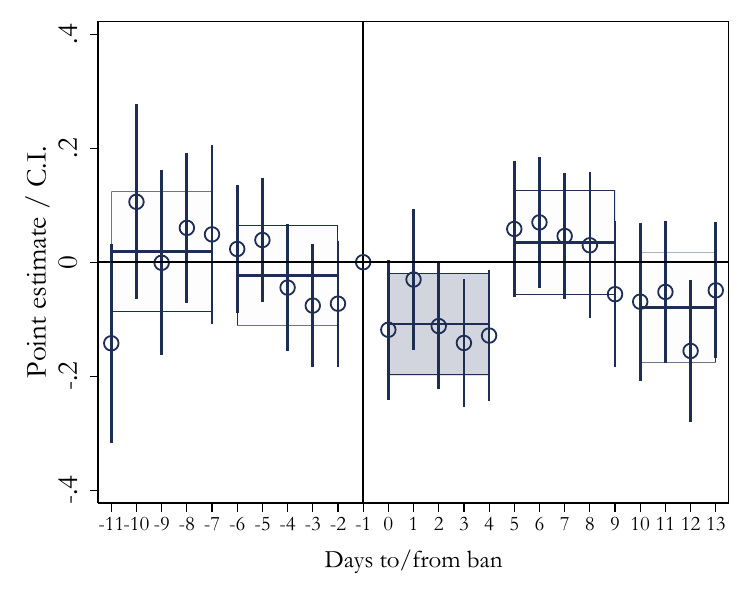}
	\begin{minipage}{\textwidth} \vspace{0.4cm}   \footnotesize{\textbf{Notes}: The figure displays coefficients and 95\% confidence intervals from estimating a daily event study version of Equation \ref{eq:did:user}. The sample consists of \textit{interaction users} -- users who interacted with the banned outlets before the ban -- and includes users located in the UK and Switzerland as control group and users located in Austria, France, Germany, Ireland, and Italy as treatment group. The dependent variable is each user's daily average of slant in tweets. We use tweets from the period between 19$^{th}$ February to 15$^{th}$ March 2022. We estimate Equation \ref{eq:did:user} including user- and day-fixed effects relative to the omitted day, 1$^{st}$ March 2022 immediately before the ban's implementation, and controlling for word count, mentions count, and hashtags count. In the aggregate specification, coefficients of interest are interactions between a dummy variable for aggregated intervals for 19$^{th}$ to 23$^{th}$ February, 24$^{th}$ to 28 February$^{th}$, 2$^{nd}$ to 6$^{th}$ March, 7$^{th}$ to 11$^{th}$ March and 12$^{th}$ to 15$^{th}$ March, relative to the omitted day, 1$^{st}$ March 2022 immediately before the implementation of the ban. Coefficient estimates on the day interactions are plotted as dots with their 95\% confidence intervals indicated with vertical lines. Coefficient estimates on the aggregate interactions are shown with horizontal lines, and their 95\% confidence intervals are indicated as boxes. We cluster standard errors at the user level. Appendix Figure \ref{fig:eventstudy:intensive:interact:allestim} displays regression results using alternative estimators proposed by \cite{callaway_difference--differences_2021} and \cite{borusyak_revisiting_2024}. Appendix Figures \ref{app:fig:eventstudy:interaction:altembeddings1} and \ref{app:fig:eventstudy:interaction:altembeddings2} reproduce the same results using four different alternative models of vectorization. Appendix Figures \ref{app:fig:eventstudy:intensive:interact:bots}, \ref{app:fig:eventstudy:intensive:interact:lateacc} show the same analysis excluding users that are plausible bots and excluding accounts created only after the ban, respectively.}\end{minipage}
\end{figure}

Figure \ref{fig:eventstudy:intensive:interact} shows results of estimating a daily event study version of Equation \ref{eq:did:user} with our media slant measure as the dependent variable. This approach captures what we call the intensive margin of the effect, focusing on shifts in content slant. The analysis solely includes tweets from February 19$^{th}$, 2022, to March 15$^{th}$, 2022, produced by the \textit{interaction users}. Each regression incorporates user- and day-fixed effects, with standard errors clustered at the user level. In our main specification, we estimate regressions using OLS. Appendix \ref{appendix:altebeestimators} provides results using alternative estimators proposed by \cite{callaway_difference--differences_2021} and by \cite{borusyak_revisiting_2024} and outlines some considerations on the use of the different estimators in our setting.

Figure \ref{fig:eventstudy:intensive:interact} provides three key insights. First, we do not observe discernible pre-trends in the period leading up to the ban. Second, immediately after the ban's implementation, we observe a negative shift indicating a movement toward the Ukrainian pole or, at the very least, a move away from the most extreme pro-Russian positions. This shift is statistically significant, with a point estimate of 0.11 standard deviations for the aggregated estimate of the first five days after the ban. Third, the effect is notably very short-lived, with differences between treated and untreated users reverting to being closer to zero and not statistically significant from day six onwards. Alternative DID estimators presented in Figure \ref{fig:eventstudy:intensive:interact:allestim} show a similar pattern and suggest somewhat stronger effects.

After examining the impact on the content's slant, our analysis next focuses on the extensive margin. We investigate whether there is a change in the total volume of pro-Russia content produced by \textit{interaction users} in the European Union compared to those located outside the EU. In this part of our study, we define a tweet or retweet as pro-Russia if our media slant measure places it one standard deviation above 0 towards the Russian pole. This threshold is chosen to minimize the risk of including tweets that might be mistakenly identified as pro-Russia due to only slight variations in the slant measure. For robustness, we also perform analyses with a threshold of 0, as detailed in Appendix \ref{appendix:diffthreshold}.

\begin{figure}[t]
	\centering
	\caption{\textit{Daily event-study on share of slanted tweets and retweets: Interaction users}}
	\label{fig:eventstudy:extensive:interact}
	\begin{subfigure}{.49\textwidth}
		\centering
		\caption{\textit{Effect on pro-Russian slanted tweets}}
		\includegraphics[width=\textwidth]{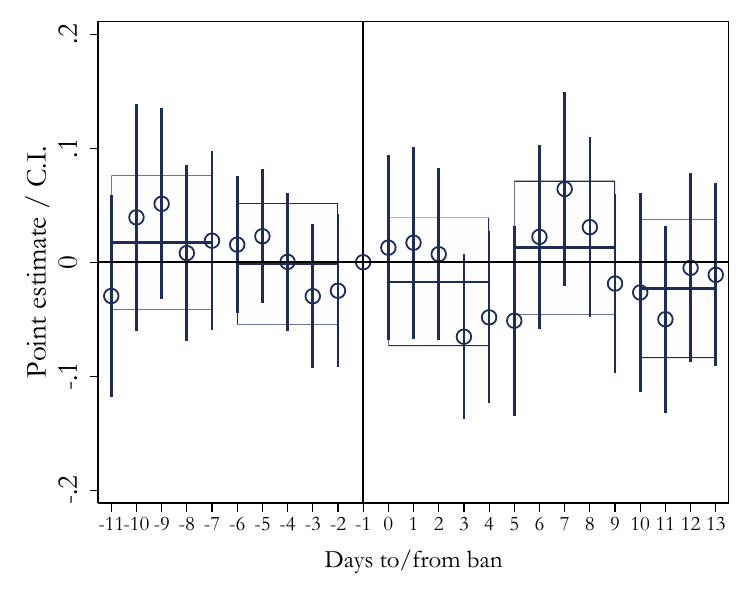}
		\label{fig:eventstudy:extensive:tw:interact}
	\end{subfigure}
	\begin{subfigure}{.49\textwidth}
		\centering
		\caption{\textit{Effect on pro-Russian slanted retweets}}
		\includegraphics[width=\textwidth]{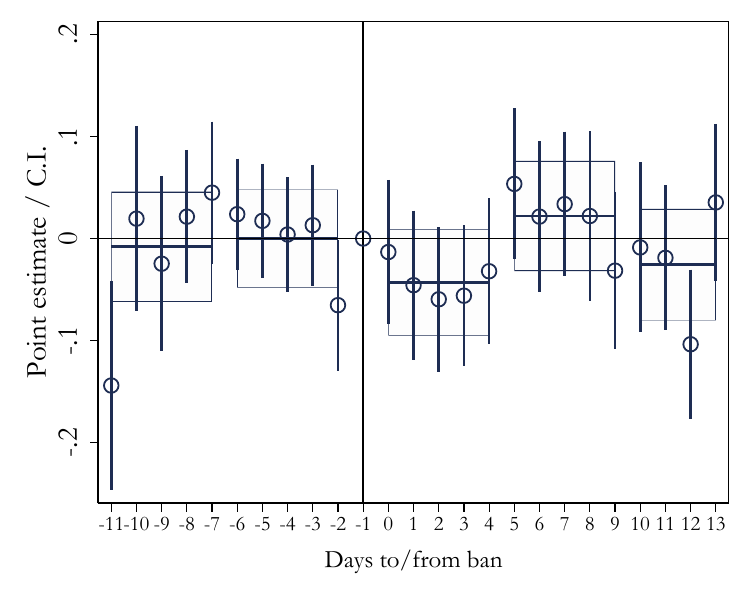}
		\label{fig:eventstudy:extensive:rtw:interact}
	\end{subfigure}
	
	\begin{minipage}{\textwidth} \vspace{-.2cm}   \footnotesize{\textbf{Notes}: The figure displays coefficients and 95\% confidence intervals from estimating a daily event study version of Equation \ref{eq:did:user}. The sample consists of \textit{interaction users} -- users who interacted with the banned outlets before the ban -- and includes users located in the UK and Switzerland as control group and users located in Austria, France, Germany, Ireland, and Italy as treatment group. We use tweets from the period between 19$^{th}$ February to 15$^{th}$ March 2022. We estimate Equation \ref{eq:did:user} including user- and day-fixed effects relative to the omitted day, 1$^{st}$ March 2022 immediately before the ban's implementation, and controlling for word count, mentions count, and hashtags count. In the aggregate specification, coefficients of interest are interactions between a dummy variable for aggregated intervals for 19$^{th}$ to 23$^{th}$ February, 24$^{th}$ to 28 February$^{th}$, 2$^{nd}$ to 6$^{th}$ March, 7$^{th}$ to 11$^{th}$ March and 12$^{th}$ to 15$^{th}$ March, relative to the omitted day, 1$^{st}$ March 2022 immediately before the implementation of the ban. Coefficient estimates on the day interactions are plotted as dots with their 95\% confidence intervals indicated with vertical lines. Coefficient estimates on the aggregate interactions are shown with horizontal lines, and their 95\% confidence intervals are indicated as boxes. We cluster standard errors at the user level. Figure \ref{fig:eventstudy:extensive:tw:interact} display estimates using user's daily share of pro-Russian slanted tweets -- defined as having a media slant measure above 1 -- as dependent variable. Figure \ref{fig:eventstudy:extensive:rtw:interact} display estimates using user's daily share of pro-Russian slanted retweets -- defined as having a media slant measure above 1 -- as dependent variable. Appendix Figure \ref{fig:eventstudy:extensive:interact:allestim} displays regression results using alternative estimators proposed by \cite{callaway_difference--differences_2021} and \cite{borusyak_revisiting_2024}. Appendix Figures \ref{app:fig:eventstudy:extensive:interact:bots}, \ref{app:fig:eventstudy:extensive:interact:lateacc} and \ref{app:fig:eventstudy:extensive:interact:diffthres} show the same analysis excluding users that are plausible bots, excluding accounts created only after the ban and using 0, instead of 1, threshold to define the binary variables of pro-Russia tweets and retweets, respectively.}\end{minipage}
\end{figure}

Figures \ref{fig:eventstudy:extensive:tw:interact} and \ref{fig:eventstudy:extensive:rtw:interact} display the results of exploring the effects of the ban on the extensive margin. Specifically, Figure \ref{fig:eventstudy:extensive:tw:interact} evaluates the ban's impact on the share of a user's total tweets classified as pro-Russia. Similarly, Figure \ref{fig:eventstudy:extensive:rtw:interact} assesses how the ban influenced the share of retweets by a user that falls into the pro-Russia category. In contrast to the insights derived from the intensive margin analysis, we find that shares of tweets and retweets identified as pro-Russia do not indicate a noticeable effect of the ban. Similarly to the previous results, no clear pre-trend is observable in the period preceding the ban. 

Next, we delve deeper into the temporary nature of the ban's impact on the \textit{interaction users}. Table \ref{tab:did:rusinteract:week} displays results of regressions using specification \ref{eq:did:user}, with additional interaction terms for each of the two weeks after the ban. Column 1 shows the effect of the ban in the two weeks after the ban itself on the average slant measure, the same dependent variable used for results in Figure \ref{fig:eventstudy:intensive:interact}. Columns 3 and 4 show respectively the effect on the share of pro-Russia tweets and pro-Russia retweets out of all tweets produced in a day by a user (and captured by our query), respectively the same dependent variables considered in Figure \ref{fig:eventstudy:extensive:tw:interact} and \ref{fig:eventstudy:extensive:rtw:interact}. Columns 4 and 5 show the impact of the total number -- instead of share -- of pro-Russia tweets and retweets produced daily by each user. 

The findings offer a consistent picture across all measures of the prevalence of Russian propaganda amongst \textit{interaction users}. We find a statistically significant reduction in the average slant and volume of Russian propaganda created and spread in the first week after the ban's implementation. As shown in Column 1, the media slant decreased by 73.8\% from its pre-ban average, a statistically significant change. Columns 2 and 3 illustrate that the point estimates suggest that the share of pro-Russia tweets and retweets declined by approximately 17.4\% and 10.2\%, respectively, compared to their pre-ban averages -- both estimates are, however, not statistically significant. While more modest, the reduction in the absolute number of pro-Russia tweets by 3.1\%, documented in Column 4, is statistically significant in the first week after the ban. Column 5 shows that our point estimates suggest a statistically not significant 1.8\% reduction in the absolute number of pro-Russia retweets. However, all effects diminish and lose statistical significance in the second week, suggesting a short-lived effect of the ban.

\begin{table}[t]\centering
	\caption{\textit{User-day level two-periods TWFE with post-ban weeks interactions: Interaction users}}
	\label{tab:did:rusinteract:week}
	\def\sym#1{\ifmmode^{#1}\else\(^{#1}\)\fi}
	\resizebox{1\linewidth }{!}{
		\begin{tabular}{lcccccc}
			\hline\hline
			                &\multicolumn{1}{c}{(1)}&\multicolumn{1}{c}{(2)}&\multicolumn{1}{c}{(3)}&\multicolumn{1}{c}{(4)}&\multicolumn{1}{c}{(5)}\\
                &Avg. media slant&\% pro-Russia tweets&\% pro-Russia retweets&Tot. Pro-Russia tweets&Tot. pro-Russia retweets\\
\hline
EU $\times$ 1st week after-ban&   -0.050&   -0.020&   -0.017&   -0.041&   -0.034\\
                &  [0.023]&  [0.014]&  [0.013]&  [0.020]&  [0.026]\\
EU $\times$ 2nd week after-ban&   -0.034&   -0.002&   -0.010&    0.004&   -0.018\\
                &  [0.025]&  [0.016]&  [0.014]&  [0.017]&  [0.024]\\
User FEs        &      yes&      yes&      yes&      yes&      yes\\
Day FEs         &      yes&      yes&      yes&      yes&      yes\\
\hline
Observations    &    29704&    16508&    19614&    29704&    29704\\
\(R^{2}\)       &    0.343&    0.236&    0.247&    0.215&    0.375\\
Pre-period mean of DV&   -0.068&    0.113&    0.162&    1.324&    1.861\\
1st week \% of mean&   -73.83&   -17.39&   -10.23&    -3.07&    -1.83\\
\hline\hline    &         &         &         &         &         \\

		\end{tabular}
	}
	\begin{minipage}{\textwidth} \vspace{0cm}   \footnotesize{\textbf{Notes}: The table displays coefficients from estimating Equation \ref{eq:did:user} examining the ban's differential impact in the two weeks following the ban, on users who interacted with Russia Today and Sputnik before the ban. Column 1 shows the effects on our measure of media slant, the intensive margin. Columns 2 and 3 show effects on our measure of the extensive margin, respectively, the share of pro-Russia tweets and retweets. Columns 4 and 5 show the effect on the total number of pro-Russia tweets and retweets produced by the author in the time period. The sample consists of \textit{interaction users} -- users who interacted with the banned outlets before the ban -- and includes users located in the UK and Switzerland as control group and users located in Austria, France, Germany, Ireland, and Italy as treatment group. We use tweets from the period between 19$^{th}$ February to 15$^{th}$ March 2022. We estimate Equation \ref{eq:did:user} including user- and day-fixed effects, and controlling for word count, mentions count, and hashtags count. Standard errors clustered at the user level are reported in brackets. Appendix Tables \ref{app:tab:did:rusinteract:week:bots}, \ref{app:tab:did:rusinteract:week:lateacc} and \ref{app:tab:did:rusinteract:week:diffthres} show the same analysis excluding users that are plausible bots, excluding accounts created only after the ban and using 0, instead of 1, the threshold to define the binary variables of pro-Russia tweets and retweets, respectively.}\end{minipage}
\end{table}

In sum, we document a statistically significant change in the average slant of content within the EU shifting away from the Russian pole, indicating an effect at the intensive margin in the time frame of our sample. This overall reduction is primarily driven by an immediate response in the first week after the ban and is reduced quickly in the second week after the ban. At the extensive margin, we find only small effects on the volume of clearly pro-Russian content that are not statistically significant and, if anything, strongest in the first week after the ban. This discrepancy suggests that the ban's overall impact on users who interacted with the banned outlets is subtle, with a short lived effect on the average slant of content but no meaningful and statistically significant impact on the overall level of clearly pro-Russian propaganda generated and spread by \textit{interaction users}.

In the next step, we explore the heterogeneity of the ban's effects on users with different levels of slant in their pro-Russian content before the ban to provide a clearer understanding of how content adjustments span across the spectrum of user types.

\subsection{Heterogeneous effect among \textit{interaction users}}
\label{sub:sec:hetero:interact}

\begin{figure}[t]
	\centering
	\caption{\textit{Heterogeneous effects of the ban by pre-ban level of pro-Russian slant: Interaction users}}\label{fig:coefplot:interact:byext}
	\includegraphics[width=.45\textwidth, trim={0cm 0.1cm 0.1cm 0.1cm},clip]{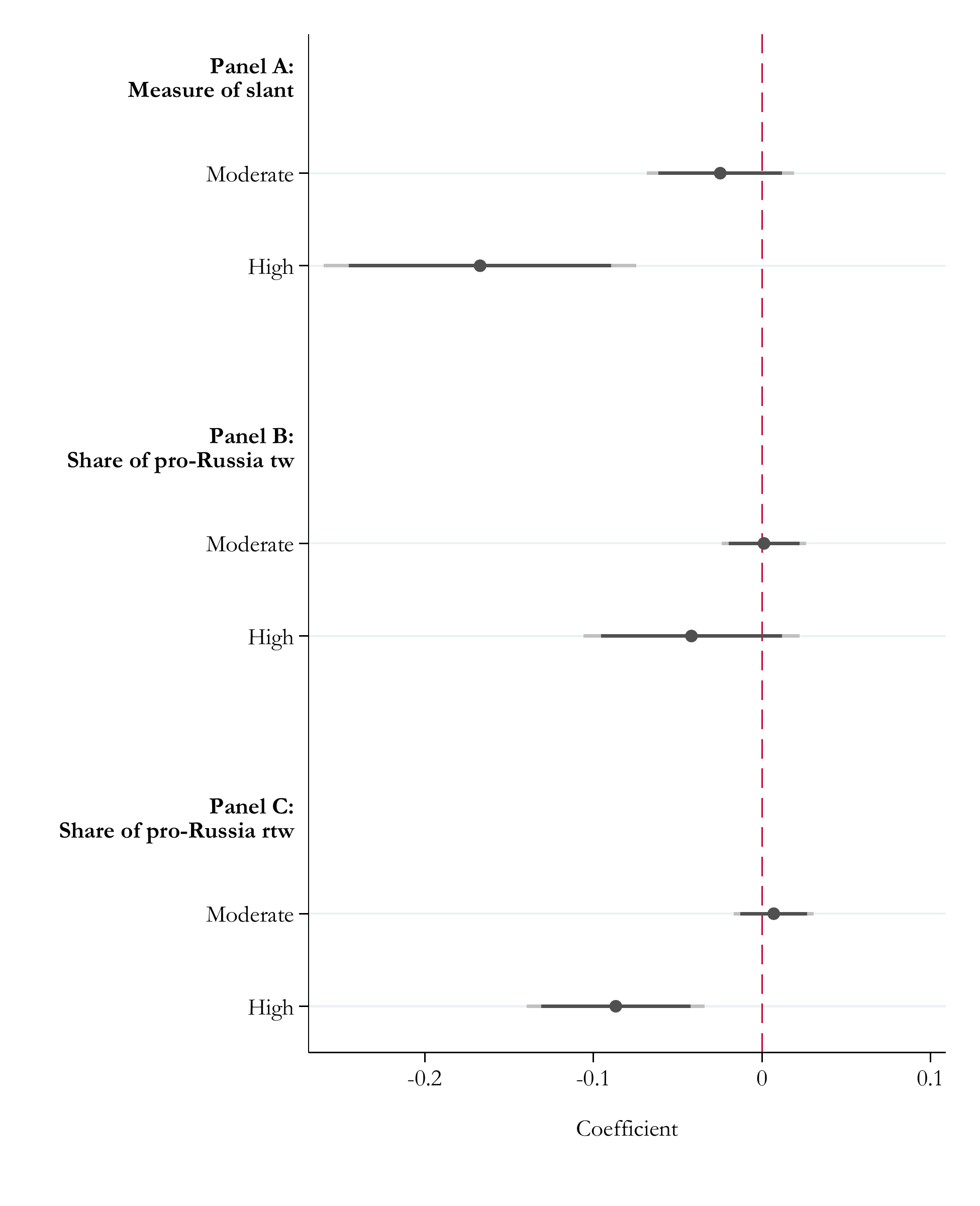}
	\begin{minipage}{\textwidth} \vspace{0cm}   \footnotesize{\textbf{Notes}: The figure presents coefficients from estimating Equation \ref{eq:did:user} assessing the ban's heterogeneous impact on users that used to interact with the outlets before the ban. Users are divided into two groups: moderate when the average slant in their tweets before the ban is in the bottom 75\% of the distribution, and high when the average slant in their tweets before the ban is in the top 25\%. The coefficients are shown with 90\% and 95\% confidence intervals (95\% in light grey). Panel A shows the results of our measure of media slant, the intensive margin, obtained by taking daily averages of media slant for each user. Panel B and C show results on the daily proportion of tweets/retweets classified as pro-Russia, out of all tweets/retweets produced by the user and captured by our query. The sample consists of \textit{interaction users} -- users who interacted with the banned outlets before the ban -- and includes users located in the UK and Switzerland as control group and users located in Austria, France, Germany, Ireland, and Italy as treatment group. We use tweets from the period between 19$^{th}$ February to 15$^{th}$ March 2022. We estimate Equation \ref{eq:did:user} including user- and day-fixed effects, clustering standard errors at the user level, and controlling for word count, mentions count, and hashtags count. Appendix Figures \ref{app:fig:coefplot:interact:byext:bots}, \ref{app:fig:coefplot:interact:byext:lateacc} and \ref{app:fig:coefplot:interact:byext:diffthres} show the same analysis excluding users that are plausible bots, excluding accounts created only after the ban and using 0, instead of 1, threshold to define the binary variables of pro-Russia tweets and retweets, respectively.}\end{minipage}
\end{figure}

Figure \ref{fig:coefplot:interact:byext} illustrates the ban's heterogeneous effects, based on the average level of the pro-Russian slant of users interacting with the outlets before the ban. We categorize \textit{interaction users} into two subgroups for analysis using the specification outlined in Equation \ref{eq:did:user}: a high pro-Russian slant group, identified as those in the top 25$^{th}$ percentile of the slant distribution, and a moderate group, encompassing the remaining 75\% of the distribution. Panel A addresses the intensive margin, reflecting changes in the average media slant. Panels B and C examine the extensive margin, focusing on the volume of pro-Russia tweets and retweets produced by these groups. The figure shows the coefficients with 90\% and 95\% confidence intervals, the latter in light grey, using standard errors clustered at the user level. 

We find that users with a moderate level of pro-Russia slant before the ban display no significant shifts across the measured dimensions. Importantly, these are always relative changes to the corresponding control group -- moderate users who used to interact with the outlets -- in the non-EU countries in our analysis. In contrast, users who, before the ban, displayed the highest levels of average pro-Russia slant demonstrate a marked response to the ban. In particular, these users show a decrease in both the intensive margin captured by the average media slant post-ban and the extensive margin captured by the number of pro-Russia retweets post-ban. Neither moderate group users nor the most extreme ones display a statistically significant response in their production of own pro-Russia tweets.

Taken together, these results seem to indicate that the ban had an effect on those \textit{interaction users} that used to engage with the outlets -- Russia Today and Sputnik -- before the ban. Despite being short-lived, the point estimates suggest a meaningful effect. In particular, the ban mainly had an impact on the intensive margin -- shifting the slant of the content rather than the amount of it. The shift seems to have mainly occurred among those \textit{interaction users} that, before the ban, were producing and spreading highly slanted pro-Russian content. There are several potential explanations for the observed pattern. These users may have been more actively engaged with the content from the banned outlets or heavily relied on these sources for their information. This interpretation is supported by the fact that these highly pro-Russian users also experienced a reduction in the extensive margin, notably in the percentage of pro-Russia retweets, likely due to the removal of two significant sources from their information network. An alternative explanation could be that the ban acted as a signal, making potential repercussions of disseminating extremely pro-Russian content more tangible. This interpretation would suggest that the most extreme users were the ones most influenced by the signalling effect of the ban.

\FloatBarrier

\subsection{Indirect effect}
\label{sub:sec:limeffect}

In the previous sections, we explored how the ban affects media slant and the volume of pro-Russia content among the \textit{interaction users}. To give deeper insight into the reach of the ban beyond the users directly affected, we explore the indirect effect on users who had not interacted directly with the banned outlets -- we label these as \textit{non-interaction users}. It is crucial to clarify that this absence of interaction does not imply that these users were disengaged from creating, consuming, or sharing pro-Russia content. Instead, their involvement with such content did not occur through direct engagement with the outlets now under the ban. The ban may indirectly influence them through the spread of the outlet's content in the social media network. This distinction allows us to explore the ban's broader effects on the online discourse beyond the immediate circle of the banned outlets' direct users.
\FloatBarrier
\begin{table}[t]\centering
	\caption{\textit{User-day level two-periods TWFE: Interaction and non-interaction users}}
	\label{tab:did:comparison}
	\def\sym#1{\ifmmode^{#1}\else\(^{#1}\)\fi}
    \begin{minipage}{\textwidth}
       \footnotesize{\textbf{Panel A: Interaction users}}
     \end{minipage}\\[0.2cm]
     \resizebox{1\linewidth }{!}{
		\begin{tabular}{lcccccc}
			\hline
			                &\multicolumn{1}{c}{(1)}&\multicolumn{1}{c}{(2)}&\multicolumn{1}{c}{(3)}&\multicolumn{1}{c}{(4)}&\multicolumn{1}{c}{(5)}\\
                &Avg. media slant&\% pro-Russia tweets&\% pro-Russia retweets&Tot. Pro-Russia tweets&Tot. pro-Russia retweets\\
\hline
EU $\times$ after-ban&   -0.043&   -0.012&   -0.014&   -0.021&   -0.027\\
                &  [0.020]&  [0.012]&  [0.011]&  [0.015]&  [0.021]\\
User FEs        &      yes&      yes&      yes&      yes&      yes\\
Day FEs         &      yes&      yes&      yes&      yes&      yes\\
\hline
Observations    &    29704&    16508&    19614&    29704&    29704\\
\(R^{2}\)       &    0.343&    0.236&    0.247&    0.215&    0.375\\
Pre-period mean of DV&   -0.068&    0.113&    0.162&    1.324&    1.861\\
\% of mean      &   -63.13&   -10.88&    -8.45&    -1.60&    -1.45\\
\hline\hline    &         &         &         &         &         \\

		\end{tabular}
	}
     \begin{minipage}{\textwidth}
       \footnotesize{\textbf{Panel A: Non-interaction users}}
     \end{minipage}\\[0.2cm]
	\resizebox{1\linewidth }{!}{
		\begin{tabular}{lcccccc}
			\hline
                            &\multicolumn{1}{c}{(1)}&\multicolumn{1}{c}{(2)}&\multicolumn{1}{c}{(3)}&\multicolumn{1}{c}{(4)}&\multicolumn{1}{c}{(5)}\\
                &Avg. media slant&\% pro-Russia tweets&\% pro-Russia retweets&Tot. Pro-Russia tweets&Tot. pro-Russia retweets\\
\hline
EU $\times$ after-ban&   -0.034&    0.002&   -0.038&   -0.004&   -0.011\\
                &  [0.007]&  [0.004]&  [0.004]&  [0.006]&  [0.005]\\
User FEs        &      yes&      yes&      yes&      yes&      yes\\
Day FEs         &      yes&      yes&      yes&      yes&      yes\\
\hline
Observations    &   312779&   147536&   181353&   312779&   312779\\
\(R^{2}\)       &    0.424&    0.328&    0.313&    0.299&    0.297\\
Pre-period mean of DV&   -0.199&    0.101&    0.140&    0.934&    1.110\\
\% of mean      &   -17.27&     1.75&   -26.85&    -0.44&    -1.00\\
\hline\hline    &         &         &         &         &         \\

		\end{tabular}
	}
	\begin{minipage}{\textwidth} \vspace{0cm}   \footnotesize{\textbf{Notes}: The table displays coefficients from estimating Equation \ref{eq:did:user} examining the ban's impact on users who interacted with Russia Today and Sputnik before the ban in Panel A, and on user that had no interactions with the outlets in Panel B. Column 1 shows the effects on our measure of media slant, the intensive margin. Columns 2 and 3 show effects on our measure of the extensive margin, respectively, the share of pro-Russia tweets and retweets. Columns 4 and 5 show the effect on the total number of pro-Russia tweets and retweets, respectively, produced by the author in the time period. The sample includes users located in the UK and Switzerland as control group and users located in Austria, France, Germany, Ireland, and Italy as treatment group. We use tweets from the period between 19$^{th}$ February to 15$^{th}$ March 2022. We estimate Equation \ref{eq:did:user} including user- and day-fixed effects, and controlling for word count, mentions count, and hashtags count. Standard errors clustered at the user level are reported in brackets. Appendix Tables \ref{app:tab:did:comparison:bots}, \ref{app:tab:did:comparison:lateacc} and \ref{app:tab:did:comparison:diffthres} show the same analysis excluding users that are plausible bots, excluding accounts created only after the ban and using 0, instead of 1, threshold to define the binary variables of pro-Russia tweets and retweets, respectively.}\end{minipage}
\end{table}

Table \ref{tab:did:comparison} reports results of regressions of estimating Equation \ref{eq:did:user}. We report results for the \textit{interaction users} -- in Panel A -- and for \textit{non-interaction users} -- in Panel B. In both panels, Column 1 shows the effect of the ban on our measure of content slant. Columns 3 and 4 show, respectively, the effect on the share of pro-Russia tweets and pro-Russia retweets out of all tweets produced in a day by a user. Columns 4 and 5 show the impact of the total number -- instead of share -- of pro-Russia tweets and retweets produced daily by each user. 

As already shown above, the ban had a strong impact in decreasing media slant among users who used to interact with the outlets before the ban. However, we also find an effect, albeit less pronounced, for users only indirectly affected by the ban. Compared to the 63.1\% reduction relative to the pre-ban average, the \textit{non-interaction users} only display a decrease of 17.3\% in media slant. It is important to notice that this is still a statistically significant effect despite being more moderate. In turn, there is no effect on either group of users for both the share and the absolute number of pro-Russia tweets. Interestingly, pro-Russia retweets display a more pronounced decrease in relative terms among the users who did not interact with the outlets before the ban. In fact, the decrease among these users is about 26.9\% relative to the pre-ban average and statistically significant on a 5\% level, while it is 8.5\% the pre-ban average for \textit{interaction users} and not statistically significant. 

Some key observations stand out when assessing the ban's overall effectiveness. First, while the ban initially had a pronounced effect on users who were previously engaging with the outlets, this impact was notably short-lived. Second, the ban's indirect effect on \textit{non-interacting users} was comparatively milder, especially regarding the intensity of the media slant. Third, it is crucial to interpret these findings within the limited time-frame of our study. Our data stops on March 15$^{th}$, 2022, just as the UK joined the European ban. Despite the brevity of this period, there was an immediate adjustment in the media market following the ban. This rapid adaptation suggests a dynamic response to regulatory measures, although the long-term implications are beyond the scope of this paper.

\subsection{Mechanisms}
\label{sub:sec:mechanisms}

Moving forward, our analysis aims to delve into potential mechanisms within the Twitter media market that might explain the observed limited and short-lived impact of the ban. Specifically, in this section, we explore the role of \textit{suppliers} of slanted content. We define a \textit{supplier} of slanted content as any user that produced or shared at least one tweet or retweet whose content scores above 1 in our slant measure in the period before the ban. Focusing on these users, we aim to provide suggestive evidence to uncover dynamics that may have counteracted the effect of the ban.

The first mechanism that may counteract the effect of the ban is the emergence of new content \textit{suppliers} in the EU following the ban. This aspect is crucial for understanding the broader implications of the ban, which acts as a supply shock within Twitter's media ecosystem. By removing two major providers of state-backed pro-Russia content, a gap is created in the network. This gap may be filled by new suppliers, who could be motivated by the reduced competition for influence within the realm of slanted content or might aim to counter the ban's objectives by spreading pro-Russian slant themselves. Therefore, we first focus on exploring the entry of new \textit{suppliers} into the market.

\begin{figure}[t]
	\centering
	\caption{\textit{Share of users supplying slanted content}}
	\label{fig:suppliers:share}
	\begin{subfigure}{.47\textwidth}
		\centering
		\caption{\textit{Regular users}}
		\includegraphics[width=\textwidth]{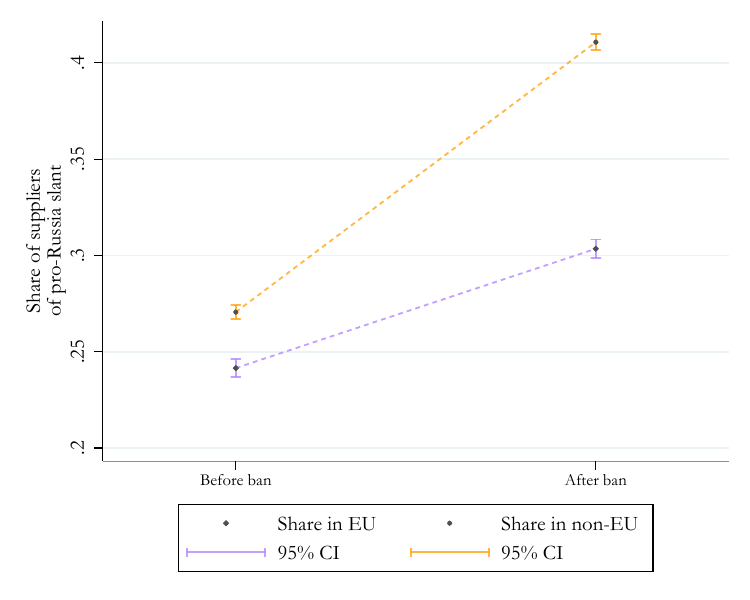}
		\label{fig:suppliers:share:users}
	\end{subfigure}
	\begin{subfigure}{.47\textwidth}
		\centering
		\caption{\textit{Potential bots}}
		\includegraphics[width=\textwidth]{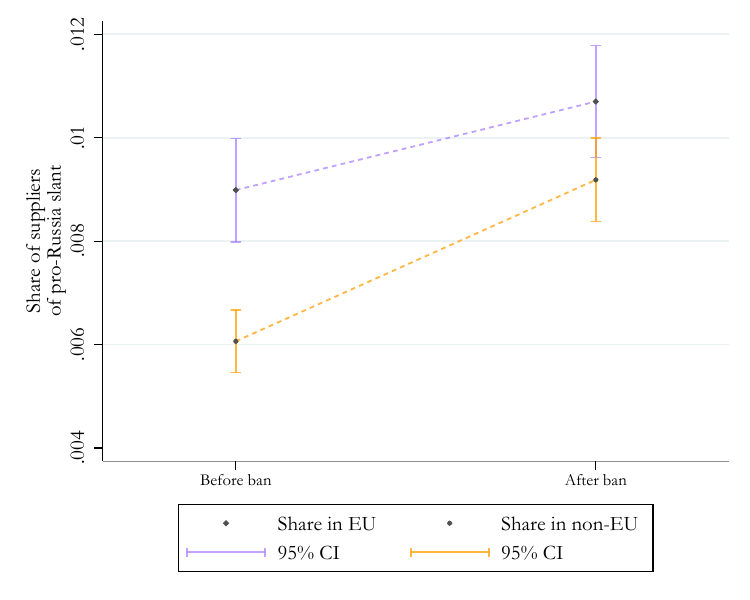}
		\label{fig:suppliers:share:bots}
	\end{subfigure}
	
	\begin{minipage}{\textwidth} \vspace{0cm}   \footnotesize{\textbf{Notes}: The figure illustrates the proportion of users classified as \textit{suppliers}, defined as those who have posted at least one tweet or retweet with a slant measure exceeding 1, before and after the ban. The shares within our sample are depicted for users in EU countries (in purple) and non-EU countries (in orange). Specifically, Figure \ref{fig:suppliers:share:users} presents the proportion of all users meeting this criterion, while Figure \ref{fig:suppliers:share:bots} focuses on those identified as \textit{suppliers} and considered plausible bots. The data encompasses the time-frame from 19$^{th}$ February to 15$^{th}$ March 2022}\end{minipage}
\end{figure}

Figure \ref{fig:suppliers:share} shows the share of users identified as \textit{suppliers} in both EU and non-EU regions within our sample before and after the ban's enforcement. Figure \ref{fig:suppliers:share:bots} narrows this focus to \textit{suppliers} presumed to be bots, based on criteria outlined by \cite{tabassum_how_2023}. In this analysis, a ``bot'' is characterized by a user profile that ranks in the upper 25\% of the distribution of activity levels while being in the lower 25\% of the distribution of the ``reputation'' metric, defined as the ratio of a user's followers to the sum of their followers and followees.

Despite being merely descriptive, there are three main insights from this analysis. First, we observe a notable increase in the share of \textit{suppliers} in both EU and non-EU countries following the ban. This rise likely aligns with the escalation of the conflict, which was intensifying both on the ground and within the digital sphere during its initial stages. Second, despite the share of \textit{suppliers} growing in both areas, the increase is more pronounced in the non-EU countries. Third, only a relatively small share of users are potential bots-suppliers, and despite starting on a higher level in the EU before the ban, the post-ban increase is more substantial in non-EU countries. These observations imply that while the number of \textit{suppliers} within the EU experienced an increase following the ban, this was less substantial compared to that observed in non-EU regions, thus suggesting other mechanisms might be at play.

The second potential mechanism we examine involves an increase in activity among users who were \textit{suppliers} of slanted content prior to the ban. This consideration is based on a simple premise: a subgroup of users exists who, even before the imposition of the ban, were disseminating pro-Russia slanted content. These individuals were likely in competition with the two banned outlets or complementing their efforts. Following the outlets' removal from the network, it is plausible that these users might have tried to fill the resulting void. Our analysis thus focuses on assessing the ban's impact on these users, also exploring potential heterogeneous effects relative to the level of activity they had before the ban.

\begin{figure}[t]
	\centering
	\caption{\textit{Heterogeneous effects of the ban by pre-ban activity: Supplier of pro-Russia slant}}\label{fig:coefplot:suppliers:byact}
	\includegraphics[width=.45\textwidth, trim={0cm 0.1cm 0.1cm 0.1cm},clip]{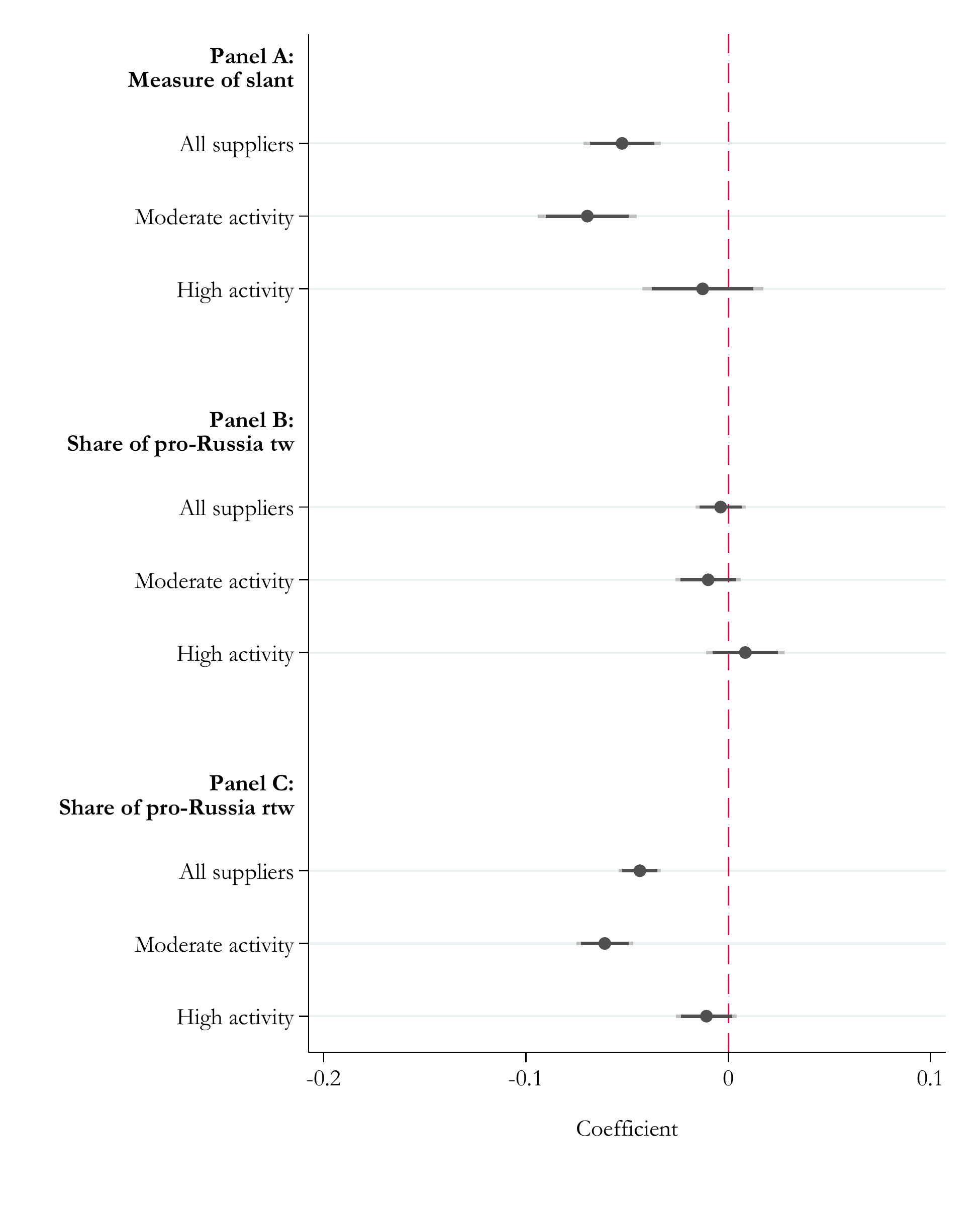}
	\begin{minipage}{\textwidth} \vspace{0.4cm}   \footnotesize{\textbf{Notes}: The figure displays coefficients from estimating Equation \ref{eq:did:user} assessing the ban’s heterogeneous impact on users that can be classified as \textit{suppliers}, defined as those who have posted at least one tweet or retweet with a slant exceeding 1 before the ban. We show results for the full sample of pre-ban \textit{suppliers} and for two subgroups of this group. The first sub-group is formed by \textit{suppliers} moderately active before the ban, namely, those whose activity is below the 75$^{th}$ percentile of the pre-ban activity distribution. The second sub-group is formed by \textit{suppliers} highly active before the ban, namely, those whose activity is above the 75$^{th}$ percentile of the pre-ban activity distribution. The coefficients are shown with 90\% and 95\% confidence intervals (95\% in light grey). Panel A shows the results of our measure of media slant, the intensive margin, obtained by taking daily averages of media slant for each user that used to interact with Russia Today and Sputnik before the ban. Panel B and C show results on the daily proportion of tweets/retweets that can be classified as pro-Russia, out of all tweets/retweets produced by the user and captured by our query. The sample includes users located in the UK and Switzerland as control group and users located in Austria, France, Germany, Ireland, and Italy as treatment group. We use tweets from the period between 19$^{th}$ February to 15$^{th}$ March 2022. We estimate Equation \ref{eq:did:user} including user- and day-fixed effect, clustering standard errors at the user level, and controlling for word count, mentions count, and hashtags count.}\end{minipage}
\end{figure}

Focusing exclusively on users who were \textit{suppliers} of slanted content before the implementation of the ban in the EU or non-EU area, we estimate Equation \ref{eq:did:user}. In Figure \ref{fig:coefplot:suppliers:byact}, Panel A illustrates the ban's effect on our slant measure, Panel B on the share of pro-Russia tweets generated daily by each user, and Panel C on the share of pro-Russia retweets. Within each panel, we present findings for the entire sample of pre-ban \textit{suppliers}, as well as for two distinct subgroups: moderately active suppliers, constituting those in the lower 75\% of the activity level distribution pre-ban, and highly active suppliers, represented by those in the upper 25\% of the distribution.

We find that the ban, on average, does not significantly alter the share of pro-Russia tweets among European \textit{suppliers} compared to their non-European counterparts. However, it notably affects the average slant of their tweets and their retweet patterns. Analyzing the two subgroups -- moderately active and highly active \textit{suppliers} -- reveals the nuanced way the ban influenced these users. The observed impact of the ban primarily comes from \textit{suppliers} who exhibited lower activity levels before the ban. In contrast, it seems to have negligible effects on those more actively engaged in disseminating pro-Russia content. This suggests that moderately active \textit{suppliers}, who may have been willing to spread pro-Russia content at a low cost, might have encountered obstacles in their supply network due to the ban, thereby increasing the cost required to find and produce such content. Conversely, the most active \textit{suppliers} before the ban appear to be individuals prepared to sustain their activity levels, even when confronted with higher costs.

Finally, we ask whether there is a subset of these highly active \textit{suppliers} that not only were not affected by the ban but also reacted to the ban by increasing the levels of activity to counteract it actively. Hence, we focus on the absolute top of this group of users, the 0.5\% most active \textit{suppliers} in our sample. This includes only very few suppliers, roughly 50 users, distributed almost equally between EU and non-EU countries. Despite the very low number of users, it is important to mention that these are very active suppliers, with an average activity of 72 posts before the ban, as shown in Appendix Figure \ref{app:fig:avg:top05}.

Table \ref{tab:did:top:suppliers} presents results focusing on the top 0.5\% of \textit{suppliers} in both EU and non-EU countries, following the estimations used in Table \ref{tab:did:comparison}. Despite the limitations due to the small sample size, the pattern of response among EU \textit{suppliers} appears to actively counter the ban's intended effects. Across all metrics, the point estimates are consistently positive. Despite the small number of observations, the share of pro-Russia tweets produced by EU \textit{suppliers} compared to their non-EU counterparts post-ban presents a positive and statistically significant coefficient. These findings, while compelling, need to be interpreted with caution due to the limited scope of the dataset. However, it is also critical to acknowledge that our data collection process captures only a subset of overall activity. Therefore, these \textit{suppliers} could exemplify a broader segment actively compensating for the ban's impact, thereby contributing to the resilience of pro-Russia narratives within the Twitter space.

\begin{table}[t]\centering
	\caption{\textit{User-day level two-periods TWFE: Top 0.5\% suppliers}}
	\label{tab:did:top:suppliers}
	\def\sym#1{\ifmmode^{#1}\else\(^{#1}\)\fi}
	\resizebox{1\linewidth }{!}{
		\begin{tabular}{lcccccc}
			\hline\hline
			                &\multicolumn{1}{c}{(1)}&\multicolumn{1}{c}{(2)}&\multicolumn{1}{c}{(3)}&\multicolumn{1}{c}{(4)}&\multicolumn{1}{c}{(5)}\\
                &Avg. media slant&\% pro-Russia tweets&\% pro-Russia retweets&Tot. Pro-Russia tweets&Tot. pro-Russia retweets\\
\hline
EU $\times$ after-ban&    0.094&    0.146&    0.092&    0.256&    0.259\\
                &  [0.179]&  [0.078]&  [0.075]&  [1.205]&  [0.508]\\
User FEs        &      yes&      yes&      yes&      yes&      yes\\
Day FEs         &      yes&      yes&      yes&      yes&      yes\\
\hline
Observations    &      640&      561&      541&      640&      640\\
\(R^{2}\)       &    0.473&    0.293&    0.261&    0.460&    0.407\\
Pre-period mean of DV&    0.398&    0.295&    0.294&   10.860&   10.899\\
\% of mean      &    23.63&    49.53&    31.24&     2.36&     2.38\\
\hline\hline    &         &         &         &         &         \\

		\end{tabular}
	}
	\begin{minipage}{\textwidth} \vspace{0cm}   \footnotesize{\textbf{Notes}: The table displays coefficients from estimating Equation \ref{eq:did:user} to assess the ban's impact on the top 0.5\% of \textit{suppliers}. These \textit{suppliers} are defined as users who posted at least one tweet or retweet with a slant over 1 prior to the ban and ranked in the top 0.5\% for activity among all \textit{suppliers} in the pre-ban period. Column 1 shows the effects on our measure of media slant, the intensive margin. Columns 2 and 3 show effects on our measure of the extensive margin, respectively, share of tweets and pro-Russia retweets. Columns 4 and 5 show the effect on the total number of pro-Russia tweets and retweets, respectively, produced by the author in the time period. The sample includes users in the UK and Switzerland as control group and users in Austria, France, Germany, Ireland, and Italy as treatment group. We use tweets from the period between 19$^{th}$ February to 15$^{th}$ March 2022. We estimate Equation \ref{eq:did:user} including user- and day-fixed effects, and controlling for word count, mentions count, and hashtags count. Standard errors are clustered at the user level are reported in brackets. }\end{minipage}
\end{table}

\section{Conclusions}
\label{sec:conclusions}

Whether and how policymakers should regulate media remains a contentious issue within democratic societies. On one hand, the dissemination of propaganda, fake news, and biased narratives by national and foreign actors poses a substantial threat to the foundation of democratic systems. On the other hand, a core principle of democratic governance is the promotion and protection of freedom of speech. Critics of media regulation are particularly concerned about the potential for censorship and the violation of free speech rights, as well as questioning the effectiveness of such policies within democratic frameworks. This paper offers an empirical perspective on evaluating the impacts of media censorship within democratic contexts.

To investigate the effect of censorship in democracy, our study examines the consequences of an unprecedented decision taken by the European Union in March 2022, during the early stages of the Russo-Ukrainian conflict initiated by Russia's invasion of Ukraine. It quickly became evident that the conflict extended beyond physical confrontations, encompassing a digital dimension characterized by a surge of propaganda, misinformation, and biased narratives about the conflict flooding the European internet space. In response to this influx, European institutions enacted a ban on two of the principal channels of state-led propaganda, Russia Today and Sputnik, along with their affiliates. Leveraging the natural experiment emerging from the differential timing and geographical spread of the ban's implementation, our analysis investigates the extent to which the prevalence and intensity of pro-Russia media slant were reduced in some of the countries enforcing the ban -- Austria, France, Germany, Italy, and Ireland -- compared to Switzerland and the United Kingdom, which did not adopt similar measures at the same time.

We find that the ban is associated with a reduction of the average pro-Russian slant of users who had previously engaged with the banned outlets, with a notable decrease of roughly 63.1\% relative to the pre-ban mean in average slant. In turn, we find no meaningful effect on the extensive margin of slant measured as the share of tweets and retweets classified as pro-Russia slanted content. Moreover, our study suggests a short-lived effect of the ban. Further, we document indirect effects on users who did not directly interact with the outlets but find them to be limited in magnitude. This suggests that while the ban initially moderated pro-Russian slant within the directly affected subset of users, its broader and lasting effect on the overall discussion was limited.

Our study further explores potential mechanisms that might have offset the ban's intended effect. One plausible explanation for the ban's limited impact could be the emergence of new \textit{suppliers} of slanted content into the media market after the ban. Although our analysis shows an increase in the share of users actively disseminating slanted content post-ban -- likely in reaction to the escalating conflict -- this increase was significantly more pronounced in countries not enforcing the ban, as opposed to those within the EU. Next, we explore an alternative  explanation: users who were already disseminating pro-Russian content before the ban might have increased their activity to fill the gap left by Russia Today and Sputnik. Our findings indicate that this is a more probable route through which the ban's effects were counteracted. Specifically, we provide suggestive evidence that the most active \textit{suppliers} in the EU prior to the ban increase their activity relative to their counterparts in non affected countries.

To be clear, the findings of this study should not be taken as an endorsement of banning media outlets nor as conclusive evidence of the effectiveness of these measures. Instead, we interpret our results as suggestive evidence that regulating the media market using censorship in a democratic context can have an effect on online discourse. However, its effectiveness might be limited due to adjustments by consumers and producers of slanted content. As our study is limited to a short time window around the ban, the results should be interpreted with caution when evaluating the long-term effects or any overall welfare effects, considering the potential cost of using censorship as a policy tool.

\newpage
\printbibliography[heading=bibintoc]

\newpage
\setcounter{section}{0}
\setcounter{page}{1}
\renewcommand \thesection {\Alph{section}}
\renewcommand{\theHsection}{chX.\the\value{section}}

\begin{center}
	{\LARGE Censorship in Democracy} 
\end{center}

\vspace{.3cm}
\begin{center}
	{\large Marcel Caesmann*, Janis Goldzycher°, Matteo Grigoletto**, Lorenz Gschwent°°}
\end{center}

\vspace{2cm}

\begin{center}
	{\LARGE Appendix}
\end{center}

\mbox{}
\vfill

\noindent *University of Zurich, e-mail: \textit{marcel.caesmann@econ.uzh.ch}
\newline °University of Zurich, e-mail: \textit{goldzycher@cl.uzh.ch}
\newline **University of Bern, Wyss Academy for Nature at the University of Bern, e-mail: \textit{matteo.grigoletto@unibe.ch}
\newline °°University of Duisburg-Essen, RTG Regional Disparities and Economic Policy, e-mail: \textit{lorenz.gschwent@uni-due.de}

\appendix
\newpage 
\renewcommand \thepart{}  
\renewcommand \partname{} 
\appendix 
\addcontentsline{toc}{section}{Appendix} 
\part{Appendix}  
\parttoc 

\renewcommand*{\theHsection}{chY.\the\value{section}}

\renewcommand{\thefigure}{\Alph{section}.\arabic{figure}}
\renewcommand{\thetable}{\Alph{section}.\arabic{table}}


\newpage
\section{Materials and Methods} 
\label{appendix:method}
\setcounter{figure}{0} 
\setcounter{table}{0} 

In this Appendix, we offer additional information on the materials and methods used in our analysis. Specifically, we detail the process of data retrieval in Section \ref{appendix:method:twitter} and describe our method for user localization in Section \ref{appendix:method:geoloc}.

\subsection{Twitter API}
\label{appendix:method:twitter}
This section provides insights into the process of extraction and processing of data from Twitter. The download of tweets is done via the Twitter APIv2 that allows researchers to extract any tweets posted and not deleted in the platform since 2006, with a monthly cap of ten million tweets\footnote{For more information see the \href{https://developer.twitter.com/en/docs/twitter-api/tweets/search/introduction}{Twitter Developer Platform documentation of the Search Tweets endpoint}.}. All the tweets we downloaded were posted between January 24$^{th}$, 2022, and April 4$^{th}$, 2022. Using the extracted data we create two datasets on data collected from Twitter: (1) a dataset of Ukrainian and Russian government-associated accounts and (2) a sample of tweets posted by users involved in the discussion about the unfolding conflict and invasion of Ukraine. 

The first dataset consists of tweets posted by accounts affiliated with the Russian and Ukrainian governments. We provide a full list of the accounts in Table \ref{tab:main:gov_accounts} in Section \ref{sec:measuring}. We refer to the 5,993 Tweets from Russian government exponents and 9,451 Tweets from Ukrainian government exponents as \textit{government tweets} (GT). This sample is key in constructing our measure of propaganda. The only filtering we do for this sample is on the language. To be part of our \textit{government tweets} sample, a tweets has to come from the selected accounts, in the period of interest, and has to be in English language.

The second dataset consists of tweets posted by users involved in the discussion about the unfolding conflict and the later invasion of Ukraine. When extracting this data, a clear trade-off emerges. On one side, we want to ensure that we capture a representative sample of the conversation about the conflict. Hence, it is necessary to use a query that is not too restrictive. On the other side, we need to impose some restrictions to avoid false positives -- tweets not primarily concerned with the conflict. Hence, our keyword query to solve this issue focuses on the main entities involved in the conflict: Russia, Ukraine, and NATO; this led to the following query: \textit{russ* OR ukrain* OR nato OR otan}. We initially downloaded all tweets fulfilling these conditions posted between January 24$^{th}$, 2022, and April 4$^{th}$, 2022 and the following day-hours windows: 9 a.m. to 12 a.m., 3 p.m. to 6 p.m., and 8 p.m. to 11 p.m. This results in 7,865,321 extracted tweets by 1,942,979 users.

\subsection{Geolocation}
\label{appendix:method:geoloc}
To create the final dataset comprising tweets from general Twitter users, we employ a geo-location process for user identification. It is crucial to remember that Twitter data acquired via the API does not automatically include geo-tag information. This means that while our query seeks tweets related to the conflict, these tweets could originate from users all around the world. To construct a dataset specifically from users in the EU, we proceed as follows. Our initial download resulted in data from 1,942,979 users. Utilizing the geo-location method outlined in \cite{gehring_analyzing_2023}, we then identify users located in our target countries: Austria, France, Germany, Ireland, Italy, Switzerland, and the UK, narrowing the group to 133,276 users. For these users, we subsequently download all English language tweets matching our query: \textit{russ* OR ukrain* OR nato OR otan}. The emphasis on English tweets aligns with the language of the propaganda poles in our study. This process yields a dataset of 775,616 tweets, to which we refer to as the \textit{user's tweets} (UT). For detailed information on the geo-location methodology, please see \cite{gehring_analyzing_2023}.

\section{Additional descriptive output} 
\label{appendix:geo}
\setcounter{figure}{0} 
\setcounter{table}{0} 

\begin{figure}[H]
\centering
\caption{\textit{Example tweets}}
\label{app:fig:examples}
\begin{subfigure}{.32\textwidth}
    \centering
    \caption{\textit{Pro-Ukrainian tweet}}
    \label{app:fig:examples:ukraine}
    \includegraphics[width=\textwidth]{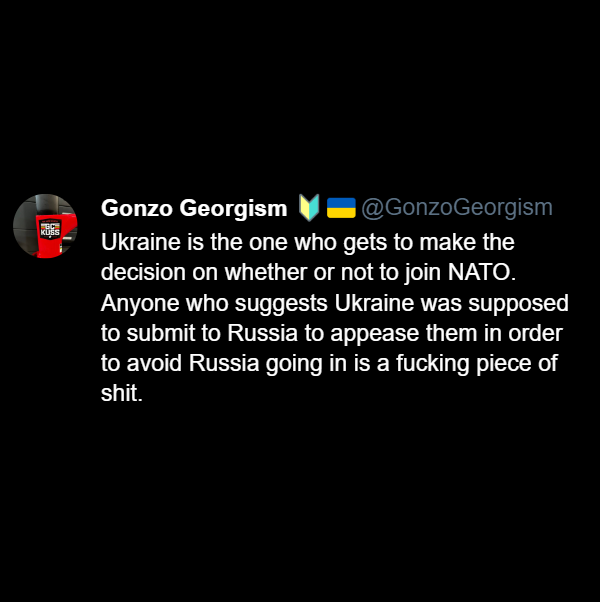}
    \subcaption*{Index value: -1.95}
\end{subfigure}
\begin{subfigure}{.32\textwidth}
    \centering
    \caption{\textit{Neutral tweet}}
    \label{app:fig:examples:neutral}
    \includegraphics[width=\textwidth]{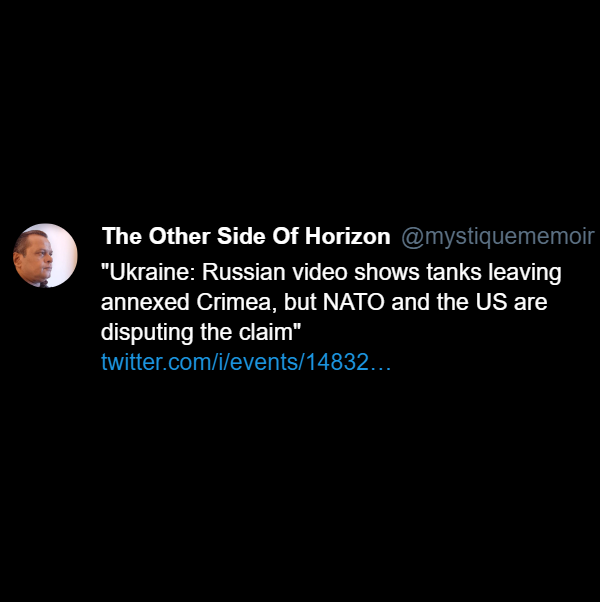}
    \subcaption*{Index value: -0.01}
\end{subfigure}
\begin{subfigure}{.32\textwidth}
    \centering
    \caption{\textit{Pro-Russian tweet}}
    \label{app:fig:examples:russia}
    \includegraphics[width=\textwidth]{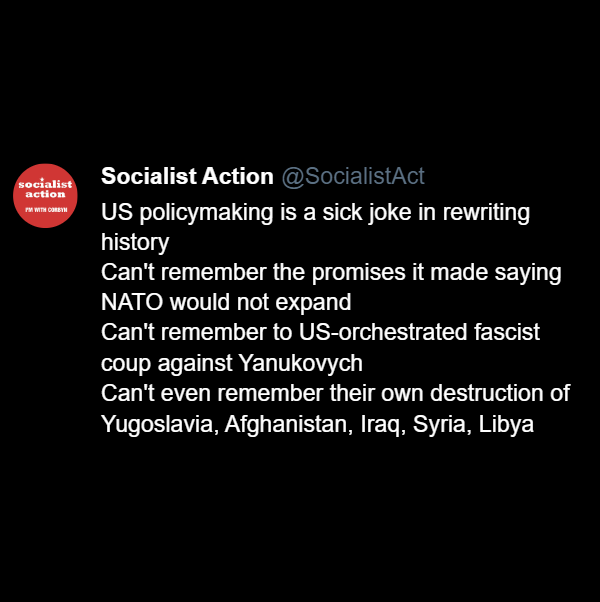}
    \subcaption*{Index value: 2.34}
\end{subfigure}
\begin{minipage}{\textwidth} \vspace{0.25cm}   \footnotesize{\textbf{Notes}: The Figure shows the examples of tweets and their corresponding slant measure. Back to Paper Section \ref{sec:measuring}.}\end{minipage}
\end{figure}

In this Appendix, we present additional descriptive data concerning the variables used in our analysis. Figure \ref{app:fig:examples} shows examples of tweets and their corresponding slant score.

\begin{figure}[t]
	\centering
	\caption{\textit{Time-series of average slant ratio in original tweets}}\label{app:fig:timeseries:slant:daily:onlytw}
	\includegraphics[width=.65\textwidth, trim={0cm 0.1cm 0.1cm 0.1cm},clip]{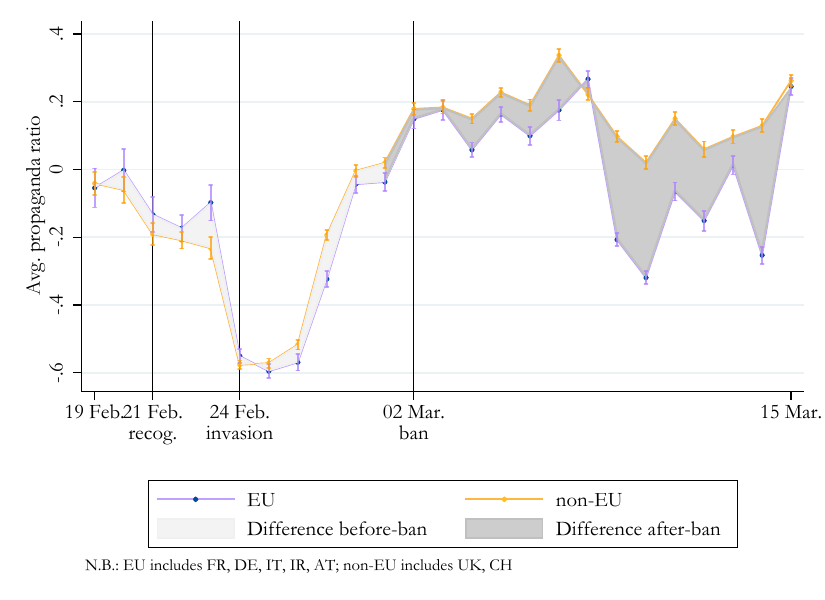}
	
	\begin{minipage}{\textwidth} \vspace{0.4cm}   \footnotesize{\textbf{Notes}: The figure shows the daily averages of our media slant measurement only in original tweets and excluding retweets, in the time-frame of our analysis, between 19$^{th}$ February, 2022, to March 15$^{th}$, 2022. The measure is normalized to have a mean of 0 and a standard deviation of 1. When positive the measure indicates content closer to the Russian pole, and when negative it indicates content closer to the Ukrainian pole. In purple, we show the daily averages in the EU countries in our study, Austria, France, Germany, Ireland, and Italy, while in orange the daily averages in non-EU countries, United Kingdom and Switzerland. In grey, the difference between the two averages. We indicate in the graph the most relevant dates in our time period. Paper Figure \ref{fig:timeseries:slant:daily} reproduces the same time-series but limiting the analysis to original tweets. }\end{minipage}
\end{figure}

Figure \ref{app:fig:timeseries:slant:daily:onlytw} depicts the time-series of the daily average levels of our slant measure for both EU and non-EU countries within our sample, focusing solely on original tweets. Conversely, Paper Figure \ref{fig:timeseries:slant:daily} accounts for all tweet types in our dataset. A comparison of these figures reveals that the notable increase in the slant measure immediately preceding the ban, as observed in the latter figure, predominantly results from an increase in retweets. Such an increase is absent when only original tweets are considered.

\begin{figure}[H]
\centering
\caption{\textit{Pro-Russia activity of users before the ban}}
\label{app:fig:rusinteract:preban}
\begin{subfigure}{.49\textwidth}
    \centering
    \caption{\textit{Pro-Russia tweets}}
    \label{app:fig:bar:rusinteract:tweets}
    \includegraphics[width=\textwidth]{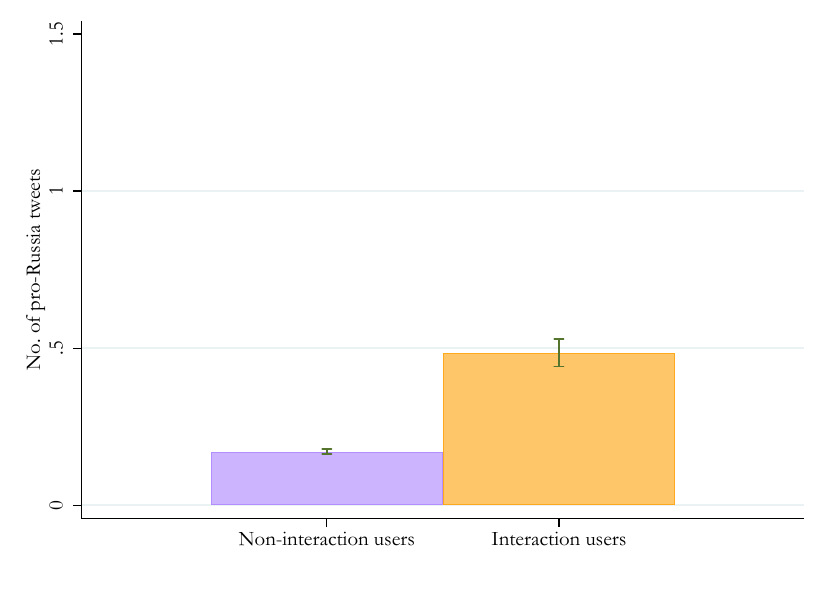}
\end{subfigure}
\begin{subfigure}{.49\textwidth}
    \centering
    \caption{\textit{Pro-Russia retweets}}
    \label{app:fig:bar:rusinteract:retweets}
    \includegraphics[width=\textwidth]{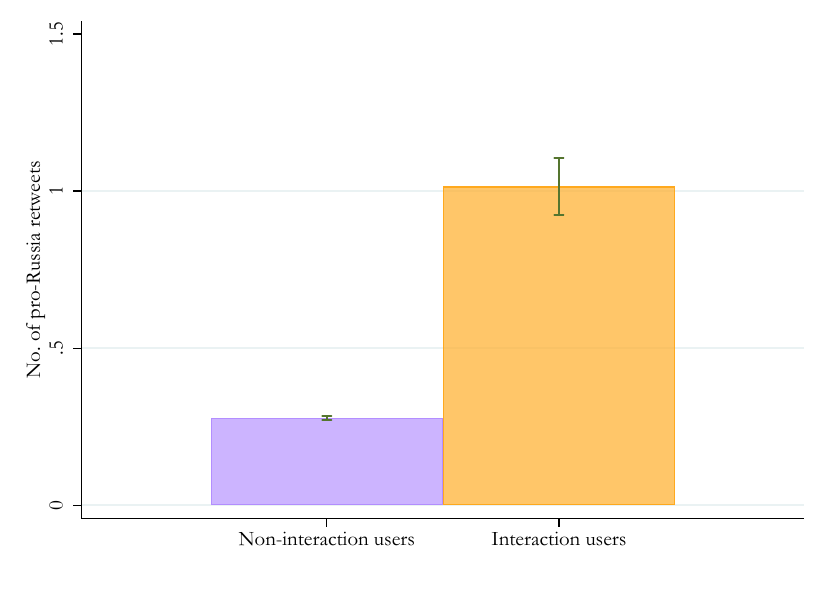}
\end{subfigure}

\begin{minipage}{\textwidth} \vspace{0.25cm}   \footnotesize{\textbf{Notes}: The Figure shows the average No. of pro-Russia tweets and the average No. of pro-Russia retweets -- respectively sub-Figure \ref{app:fig:bar:rusinteract:tweets} and sub-Figure \ref{app:fig:bar:rusinteract:retweets} -- of the \textit{non-interaction users} vs. \textit{interaction users} before the ban. Back to Paper Section \ref{sec:results:impact}.}\end{minipage}
\end{figure}

Figure \ref{app:fig:rusinteract:preban} provides insights into the pro-Russia activity of users before the ban. In particular, Figure \ref{app:fig:bar:rusinteract:tweets} shows the average number of pro-Russia slanted tweets produced before the ban, by users that had no contacts with the banned outlets, and users that used to engage with the outlets, before the ban. Figure \ref{app:fig:bar:rusinteract:retweets} shows the same for \textit{non-interaction} and \textit{interaction} users before the ban, for retweets containing pro-Russia slanted content. These figures distinctly highlight the variance in pro-Russia content production between the two groups of users. Notably, users who previously engaged with the outlets produced three to four times more pro-Russia content than those who did not.

\begin{figure}[H]
	\centering
	\caption{\textit{Average activity of the top 0.5\% suppliers}}\label{app:fig:avg:top05}
	\includegraphics[width=.55\textwidth, trim={0cm 0.1cm 0.1cm 0.1cm},clip]{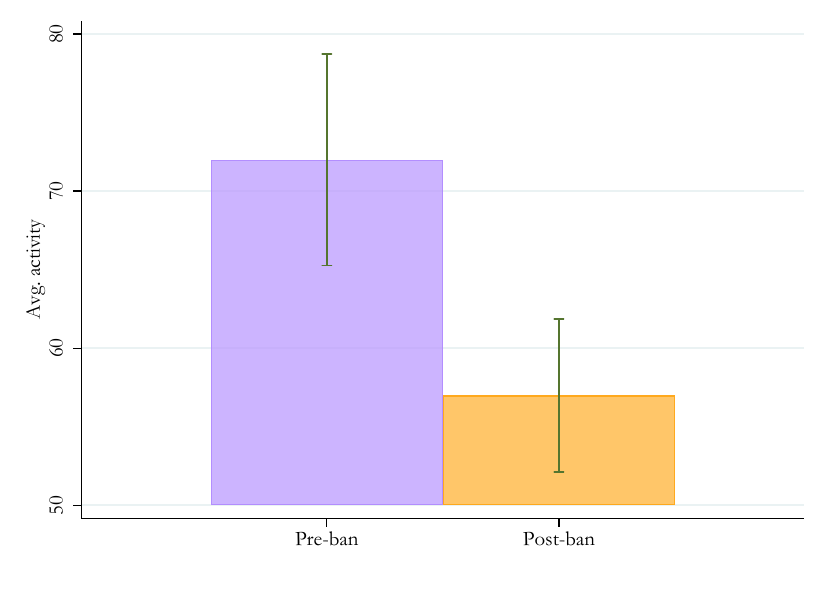}
	
	\begin{minipage}{\textwidth} \vspace{0.4cm}   \footnotesize{\textbf{Notes}: The figure shows the average activity before and after the ban of the top 0.5\% \textit{suppliers} of pro-Russia slant. To define this \textit{suppliers} we select all users that before the ban had produced or shared at least one tweet/retweet with slant above 1; among these suppliers, we take the distribution of total activity before the ban, and then select the top 0.5\% most active users, comprising around 50 users. Back to Paper Section \ref{sec:results:impact}.}\end{minipage}
\end{figure}

Figure \ref{app:fig:avg:top05} delves into the activities of the most engaged pro-Russia users, specifically the top 0.5\%, identified through a precise filtering process. Initially, we include all users who, before the ban, produced or shared any tweets or retweets exhibiting a pro-Russia slant score exceeding 1. From this pool, we examine the distribution of their activity levels prior to the ban. The top 0.5\% represents users with the highest levels of pro-Russia slanted activity. The figure presents a comparison of average activities for these users before and after the ban, revealing that they are exceptionally active. Despite a noticeable decline in their activity following the ban, these users still produce, on average, almost 60 tweets per user, underscoring their significant engagement even in the face of restrictions.

\FloatBarrier

\begin{figure}[H]
	\centering
	\caption{\textit{Balance in tweets characteristic before the ban: EU vs. non-EU countries}}\label{app:fig:balance}
	\includegraphics[width=.55\textwidth, trim={0cm 0.1cm 0.1cm 0.1cm},clip]{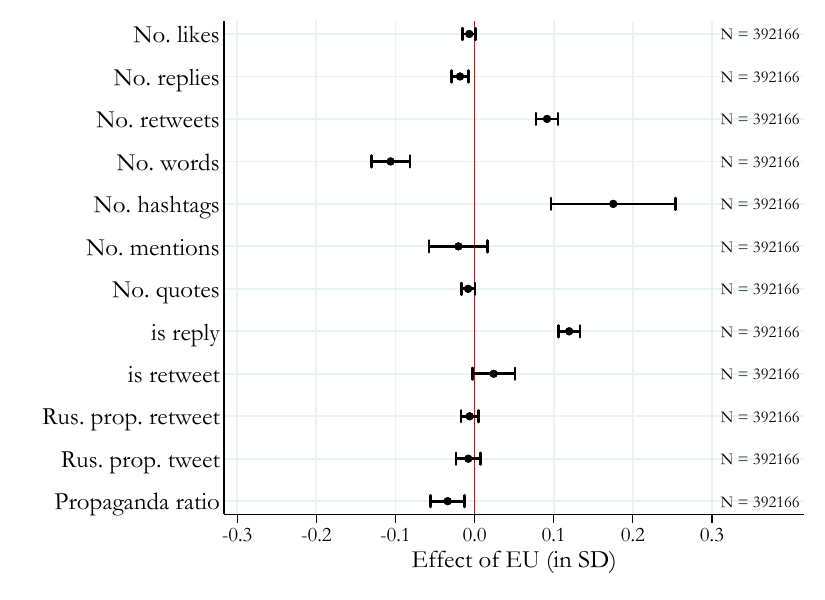}
	
	\begin{minipage}{\textwidth} \vspace{0.4cm}   \footnotesize{\textbf{Notes}: The figure shows tests of balance on a number of pre-ban tweet features, between EU and non-EU countries in our sample. Back to Section \ref{sec:frame:method}.}\end{minipage}
\end{figure}
\FloatBarrier

\section{Robustness check: Alternative estimators} 
\label{appendix:altebeestimators}
\setcounter{figure}{0} 
\setcounter{table}{0} 

In this section, we delve into the rationale behind the use of some alternative estimators for the analysis. It is crucial to acknowledge that our research context presents significant challenges. Firstly, our dataset comprises a series of repeated cross-sections, not a standard panel. Secondly, while users may feature across multiple days, this results in an unbalance panel due to the extraction method. Thirdly, the users within our sample, despite originating from diverse locations, may be interconnected, adding another layer of complexity to our analysis.

As indicated in Equation \ref{eq:did:user}, our analysis relies on a conditional parallel trends assumption, specifically conditioning on time-varying covariates. \cite{caetano_difference--differences_2023} discuss the challenges of applying traditional TWFE regressions in such contexts. Notably, TWFE regressions with time-varying covariates may encounter issues with negative weighting, similar to those identified in studies of staggered treatment designs. Moreover, TWFE regressions are vulnerable when parallel trends are influenced by the levels of time-varying covariates, rather than their changes. Consequently, we opt not to incorporate control variables in our TWFE regressions showcased in the paper.

\cite{caetano_difference--differences_2023} demonstrate that "imputation estimators", akin to those proposed by \cite{borusyak_revisiting_2024}, are capable of accurately estimating the treatment effect.\footnote{They further note that doubly-robust estimators, like those suggested by \cite{chernozhukov_doubledebiased_2018, chang_doubledebiased_2020}, are also applicable in such scenarios.} However, the existing implementations of these estimators typically compare outcomes for units observed just before the treatment, at $t-1$, with those post-treatment. Given our unbalanced panel and the absence of observations for all users right before the ban, this approach excludes a significant number of users. Conversely, the implementation by \cite{callaway_difference--differences_2021} treats the data as repeated cross-sections, yet assumes covariates are time-invariant. Nevertheless, neither method perfectly suits our context.

Below, we show daily event studies displayed in the paper using two alternative estimators described above. For comparison we report on the graphs also estimates from a regression using OLS including user and day fixed effects. We are aware some recent work (see \cite{roth_interpreting_2024}) suggests to not display these coefficient together, nevertheless we do so for convenience of space and because they show a very similar pattern. Generally, the alternative estimators suggest very similar results and interpretation to what shown by the TWFE OLS estimation in the paper, despite presenting somewhat stronger coefficients.

\FloatBarrier
\begin{figure}[H]
	\centering
	\caption{\textit{Daily event-study on our slant measure: Interaction users}}\label{fig:eventstudy:intensive:interact:allestim}
	\includegraphics[width=.65\textwidth, trim={0cm 0.1cm 0.1cm 0.1cm},clip]{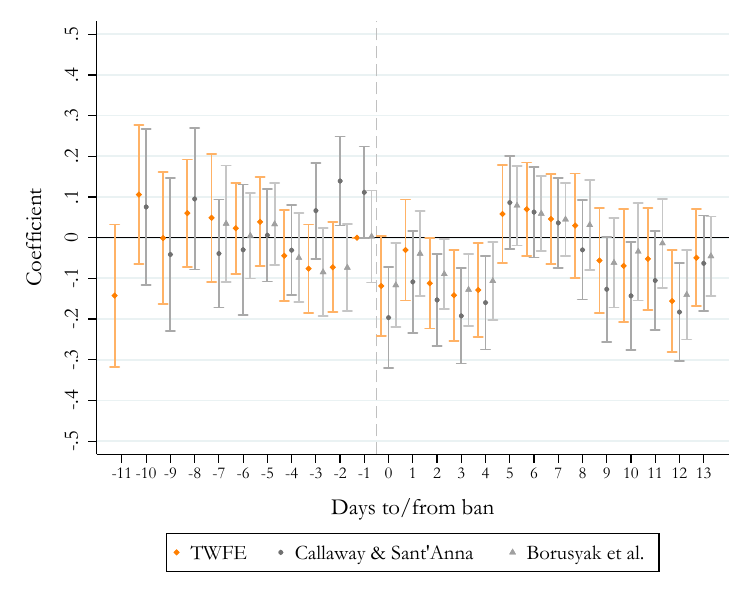}
	\begin{minipage}{\textwidth} \vspace{0.4cm}   \footnotesize{\textbf{Notes}: The figure displays coefficients and 95\% confidence intervals from a daily event study, estimating regressions to assess the effect of the ban on our media slant measure, referred to as the intensive margin. The dependent variable is obtained by taking daily averages of media slant for each user that used to interact with Russia Today and Sputnik before the ban. In orange, we display regression results using a TWFE OLS estimator, while in grey, we present outcomes from the estimators proposed by \cite{callaway_difference--differences_2021} and \cite{borusyak_revisiting_2024}, which additionally incorporate time-varying control variables: word count, mentions count, and hashtags count. All models include the UK and Switzerland as control countries and Austria, France, Germany, Ireland, and Italy as treatment countries. They include tweets from the period between 19$^{th}$ February to 15$^{th}$ March 2022, they incorporate both user- and day-fixed effects and cluster standard errors at the user level. We control for word count, mentions count, and hashtags count. Back to Figure \ref{fig:eventstudy:intensive:interact}.}\end{minipage}
\end{figure}

\begin{figure}[H]
	\centering
	\caption{\textit{Daily event-study on share of slanted tweets and retweets: Interaction users}}
	\label{fig:eventstudy:extensive:interact:allestim}
	\begin{subfigure}{.49\textwidth}
		\centering
		\caption{\textit{Effect on pro-Russian slanted tweets}}
		\includegraphics[width=\textwidth]{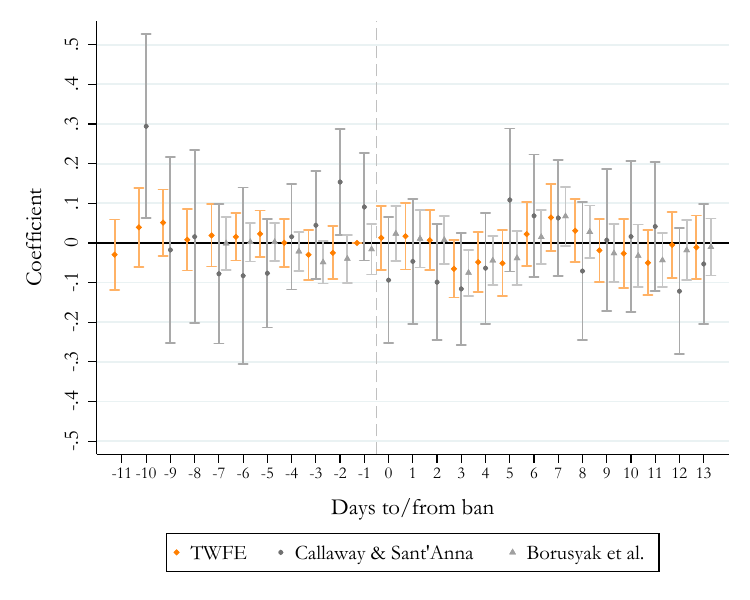}
		\label{fig:eventstudy:extensive:tw:interact:allestim}
	\end{subfigure}
	\begin{subfigure}{.49\textwidth}
		\centering
		\caption{\textit{Effect on pro-Russian slanted retweets}}
		\includegraphics[width=\textwidth]{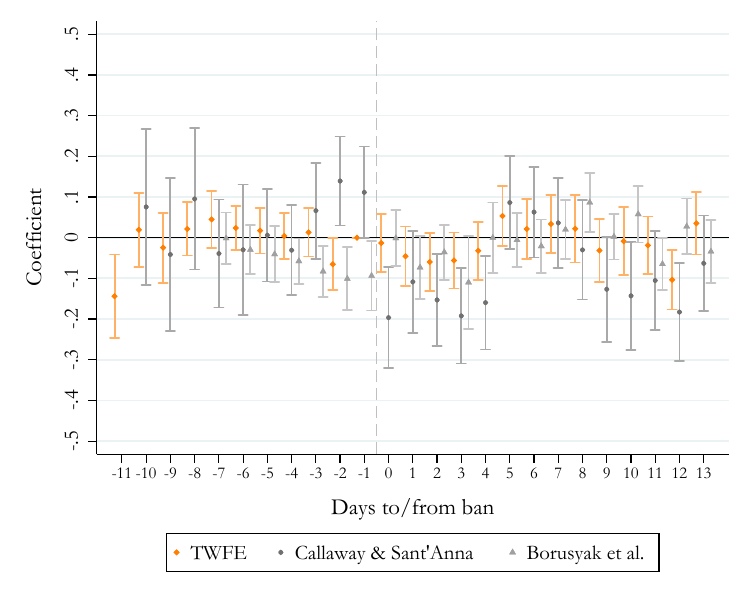}
		\label{fig:eventstudy:extensive:rtw:interact:allestim}
	\end{subfigure}
	
	\begin{minipage}{\textwidth} \vspace{0cm}   \footnotesize{\textbf{Notes}: The figure displays coefficients and 95\% confidence intervals from a daily event study, estimating regressions to assess the effect of the ban on measures capturing the extensive margin of our analysis, for the \textit{interaction users}. Figure \ref{fig:eventstudy:extensive:tw:interact} shows the impact on the share of daily tweets produced by a user and captured by our query, that can be classified as pro-Russia slant, hence having a media slant above 1. Figure \ref{fig:eventstudy:extensive:rtw:interact} shows the same for retweets. In orange, we display regression results using a standard TWFE OLS estimation, while in grey, we present outcomes from the estimators proposed by \cite{callaway_difference--differences_2021} and \cite{borusyak_revisiting_2024}, which additionally incorporate time-varying control variables: word count, mentions count, and hashtags count. All models include the UK and Switzerland as control countries and Austria, France, Germany, Ireland, and Italy as treatment countries. They include tweets from the period between 19$^{th}$ February to 15$^{th}$ March 2022, they incorporate both user- and day-fixed effects and cluster standard errors at the user level. We control for word count, mentions count, and hashtags count. Back to Figure \ref{fig:eventstudy:extensive:interact}.}\end{minipage}
\end{figure}

\section{Robustness check: Alternative text embedding models} 
\label{appendix:altebemddings}
\setcounter{figure}{0} 
\setcounter{table}{0} 

In this Appendix, we provide alternative versions of the main findings from our study, utilizing different embedding models. Specifically, we explore several variations of Paper Figure \ref{fig:eventstudy:intensive:interact}. This figure originally investigates the effect of the ban on media slant within the \textit{interaction users} group, where media slant is calculated by embedding user tweets through the sentence-t5-xl model. This involves computing the proximity ratio between tweets and the dynamically defined Russian and Ukrainian reference poles, which are updated daily based on data from the previous 14 days with a decay factor of 0.5. Here, we adjust various elements of this methodology to offer a broader perspective on our analysis.

\begin{figure}[H]
	\centering
	\caption{\textit{Daily event-study on our slant measure for interaction users: Alternative text embedding models (i.)}}
	\label{app:fig:eventstudy:interaction:altembeddings1}
	\begin{subfigure}{.49\textwidth}
		\centering
		\caption{\textit{Model: sentence-t5-xl static}}
		\includegraphics[width=\textwidth]{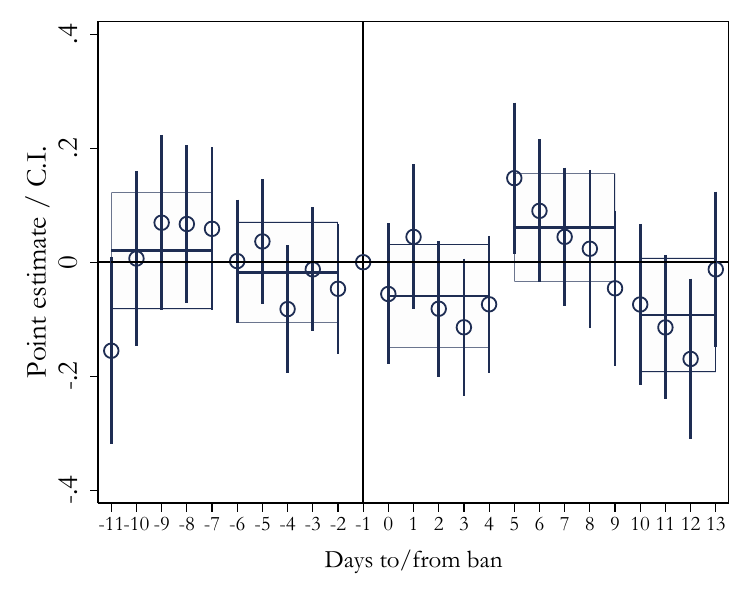}
		\label{app:fig:xlstatic}
	\end{subfigure}
	\begin{subfigure}{.49\textwidth}
		\centering
		\caption{\textit{Model: sentence-t5-xl static pre-treatment}}
		\includegraphics[width=\textwidth]{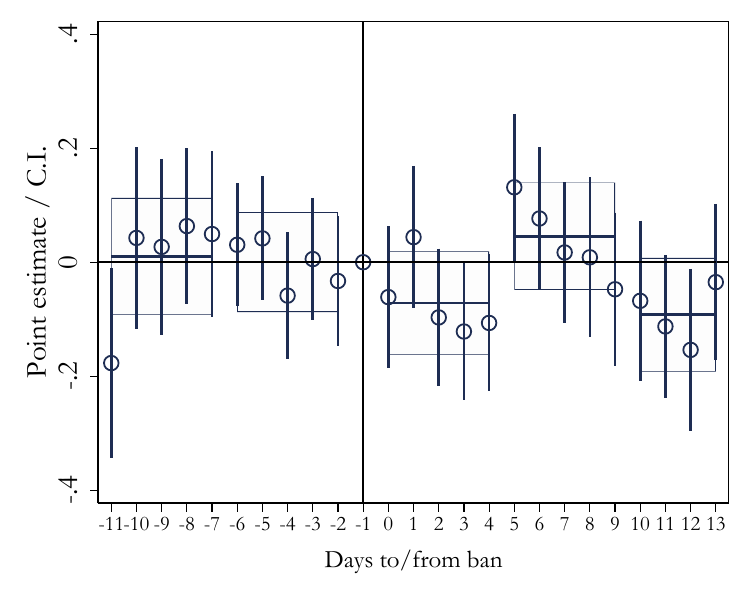}
		\label{app:fig:xlstaticpre}
	\end{subfigure}
	
	\begin{minipage}{\textwidth} \vspace{-.1cm}   \footnotesize{\textbf{Notes}: The figure displays results reproducing what is shown in the paper Figure \ref{fig:eventstudy:intensive:interact}, but using alternative models to compute our media slant measure. Sub-figure \ref{app:fig:xlstatic} shows results using a slant measure computed by only computing a single comparison pole for all Ukrainian and Russian government tweets over the full-time period. Sub-figure \ref{app:fig:xlstaticpre} is similar but only considers government tweets that were posted before the ban.}\end{minipage}
\end{figure}

\begin{figure}[t]
	\centering
	\caption{\textit{Daily event-study on our slant measure for interaction users: Alternative text embedding models (ii.)}}
	\label{app:fig:eventstudy:interaction:altembeddings2}
	\begin{subfigure}{.49\textwidth}
		\centering
		\caption{\textit{Model: simCSE}}
		\includegraphics[width=\textwidth]{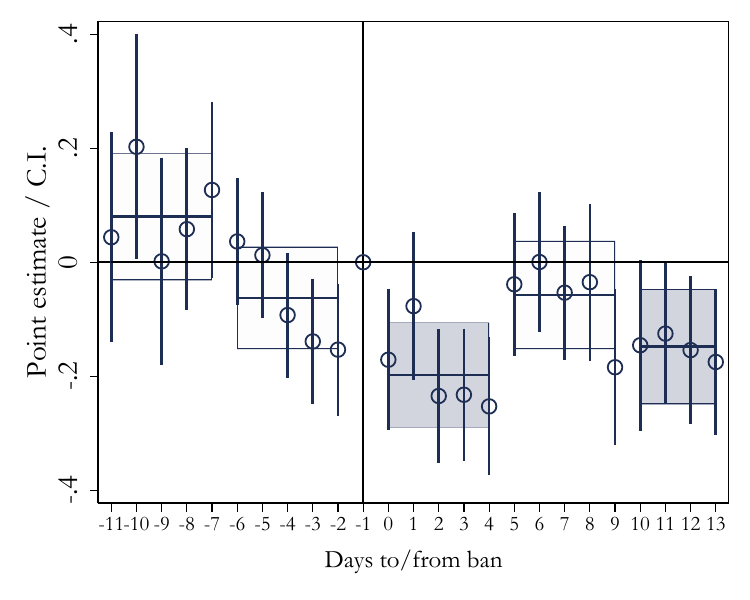}
		\label{app:fig:simcse}
	\end{subfigure}
	
	\begin{minipage}{\textwidth} \vspace{-.1cm}   \footnotesize{\textbf{Notes}: The figure displays results reproducing what is shown in the paper Figure \ref{fig:eventstudy:intensive:interact}, but using alternative models to compute our media slant measure. Sub-figure \ref{app:fig:simcse} shows results substituting sentence-t5-xl with an alternative text embedding model, SimCSE \citep{gao_simcse_2022}.}\end{minipage}
\end{figure}

In Figure \ref{app:fig:eventstudy:interaction:altembeddings1} and \ref{app:fig:eventstudy:interaction:altembeddings2}, we show the robustness of our results. We computed the measurement -- in Figure \ref{app:fig:eventstudy:interaction:altembeddings1} Panel A -- by only computing a single comparison pole for all Ukrainian and Russian government tweets over the full-time period (instead of 28 time-varying ones). The measure in Panel B of the same figure is similar to the one in Panel A but only considers government tweets that were posted before the ban. In Figure \ref{app:fig:eventstudy:interaction:altembeddings2} Panel A is the same as in the main part of the paper with an additional centered seven-day covering average smoothing. In Panel B of the same figure, we are substituting the vector representations from sentence-t5-xl with an alternative text embedding model, SimCSE, \citep{gao_simcse_2022}. After sentence-t5-xl, SimCSE has shown the next best performances on various sentence embedding benchmarks. Therefore, we evaluated it as a suitable substitute.

The results in the Figure \ref{app:fig:eventstudy:interaction:altembeddings1} show very similar patterns as in our main results, though the estimates after treatment are shifted upwards. We explain this by the static nature of these measures, which fail to capture the evolution of the government tweets as the full-scale war developed further. Nonetheless, the initial reduction in pro-Russian slant due to the ban is also captured by these measures. In Panel A of Figure \ref{app:fig:eventstudy:interaction:altembeddings2}, results are hard to interpret given the strong pre-trends. Figure \ref{app:fig:eventstudy:interaction:altembeddings2} Panel B shows that the results also qualitatively hold for the alternative sentence embedding model, but with this model, pre-trends appear to be more problematic.

\FloatBarrier

\FloatBarrier
\FloatBarrier

\section{Robustness check: Excluding plausible bots} 
\label{appendix:robbots}
\setcounter{figure}{0} 
\setcounter{table}{0} 

In this Appendix, we presents the findings from our analysis replicated from the main body of the paper, adjusting our sample with the removal of potential bots. To identify these bots, we rely on the criteria established by recent studies, which suggest that Twitter bots typically exhibit a high frequency of tweets per day \citep{tabassum_how_2023} and a low 'reputation' ratio, calculated as the number of followers divided by the sum of the number of followers and the number of accounts followed \citep{chu_detecting_2012}. Following the recommendations of \cite{gehring_analyzing_2023}, we classify potential bots as accounts ranking in the upper 25\% for daily tweet frequency and in the lower 25\% for the reputation metric. 

\begin{figure}[H]
	\centering
	\caption{\textit{Daily event-study on share of slanted tweets and retweets: Interaction users excluding plausible bots}}\label{app:fig:eventstudy:intensive:interact:bots}
	\includegraphics[width=.65\textwidth, trim={0cm 0.1cm 0.1cm 0.1cm},clip]{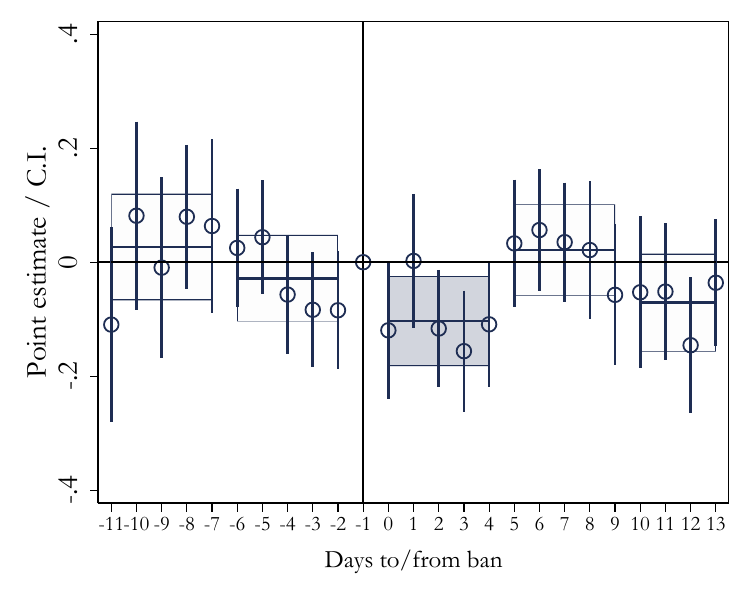}
	\begin{minipage}{\textwidth} \vspace{-.3cm}   \footnotesize{\textbf{Notes}: The figure displays coefficients and 95\% confidence intervals from estimating a daily event study version of Equation \ref{eq:did:user} excluding potential pots. The sample consists of \textit{interaction users} -- users who interacted with the banned outlets before the ban -- and includes users located in the UK and Switzerland as control group and users located in Austria, France, Germany, Ireland, and Italy as treatment group. We exclude potential bots from the sample. Dependent variable is the daily average of slant in tweets for each user. We use tweets from the period between 19$^{th}$ February to 15$^{th}$ March 2022. We estimate Equation \ref{eq:did:user} including user- and day-fixed effects relative to the omitted day, 1$^{st}$ March 2022 immediately before the implementation of the ban, and controlling for word count, mentions count, and hashtags count. In the aggregate specification, coefficients of interest are interactions between a dummy variable for aggregated intervals for 19$^{th}$ to 23$^{th}$ February, 24$^{th}$ to 28 February$^{th}$, 2$^{nd}$ to 6$^{th}$ March, 7$^{th}$ to 11$^{th}$ March and 12$^{th}$ to 15$^{th}$ March, relative to the omitted day, 1$^{st}$ March 2022 immediately before the implementation of the ban. Coefficient estimates on the day interactions are plotted as dots with their 95\% confidence intervals indicated with vertical lines. Coefficient estimates on the aggregate interactions are shown with horizontal lines, and their 95\% confidence intervals are indicated as boxes. We cluster standard errors at the user level. Paper Figure \ref{fig:eventstudy:intensive:interact} shows results for all \textit{interaction users}.}\end{minipage}
\end{figure}

\begin{figure}[H]
	\centering
	\caption{\textit{Daily event-study on share of slanted tweets and retweets: Interaction users excluding plausible bots}}
	\label{app:fig:eventstudy:extensive:interact:bots}
	\begin{subfigure}{.49\textwidth}
		\centering
		\caption{\textit{Effect on pro-Russian slanted tweets}}
		\includegraphics[width=\textwidth]{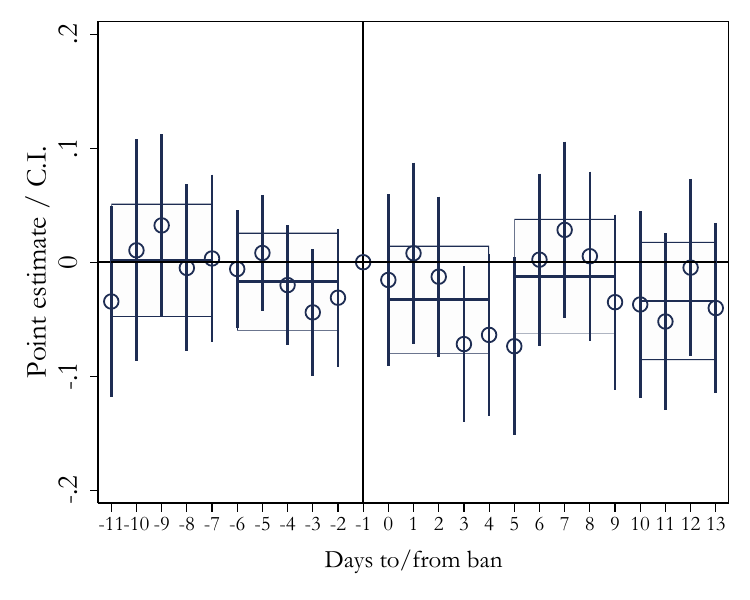}
		\label{app:fig:eventstudy:extensive:tw:interact:bots}
	\end{subfigure}
	\begin{subfigure}{.49\textwidth}
		\centering
		\caption{\textit{Effect on pro-Russian slanted retweets}}
		\includegraphics[width=\textwidth]{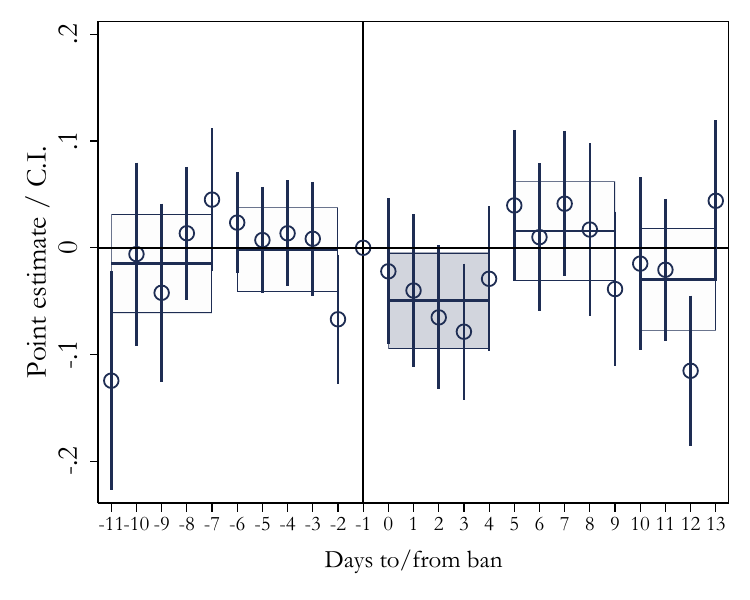}
		\label{app:fig:eventstudy:extensive:rtw:interact:bots}
	\end{subfigure}
	
	\begin{minipage}{\textwidth} \vspace{-.7cm}   \footnotesize{\textbf{Notes}: The figure displays coefficients and 95\% confidence intervals from estimating a daily event study version of Equation \ref{eq:did:user} excluding potential bots. The sample consists of \textit{interaction users} -- users who interacted with the banned outlets before the ban -- and includes users located in the UK and Switzerland as control group and users located in Austria, France, Germany, Ireland, and Italy as treatment group. We exclude potential bots from the sample. We use tweets from the period between 19$^{th}$ February to 15$^{th}$ March 2022. We estimate Equation \ref{eq:did:user} including user- and day-fixed effects relative to the omitted day, 1$^{st}$ March 2022 immediately before the implementation of the ban, and controlling for word count, mentions count, and hashtags count. In the aggregate specification, coefficients of interest are interactions between a dummy variable for aggregated intervals for 19$^{th}$ to 23$^{th}$ February, 24$^{th}$ to 28 February$^{th}$, 2$^{nd}$ to 6$^{th}$ March, 7$^{th}$ to 11$^{th}$ March and 12$^{th}$ to 15$^{th}$ March, relative to the omitted day, 1$^{st}$ March 2022 immediately before the implementation of the ban. Coefficient estimates on the day interactions are plotted as dots with their 95\% confidence intervals indicated with vertical lines. Coefficient estimates on the aggregate interactions are shown with horizontal lines, and their 95\% confidence intervals are indicated as boxes. We cluster standard errors at the user level. Figure \ref{fig:eventstudy:extensive:tw:interact} display estimates using user's daily share of pro-Russian slanted tweets -- defined as having a media slant measure above 1 -- as dependent variable. Figure \ref{fig:eventstudy:extensive:rtw:interact} display estimates using user's daily share of pro-Russian slanted retweets -- defined as having a media slant measure above 1 -- as dependent variable. Paper Figure \ref{fig:eventstudy:extensive:interact} shows results for all \textit{interaction users}.}\end{minipage}
\end{figure}

\begin{figure}[H]
	\centering
	\caption{\textit{Heterogeneous effects of the ban by pre-ban slant: Interaction users excluding plausible bots}}\label{app:fig:coefplot:interact:byext:bots}
	\includegraphics[width=.45\textwidth, trim={0cm 0.1cm 0.1cm 0.1cm},clip]{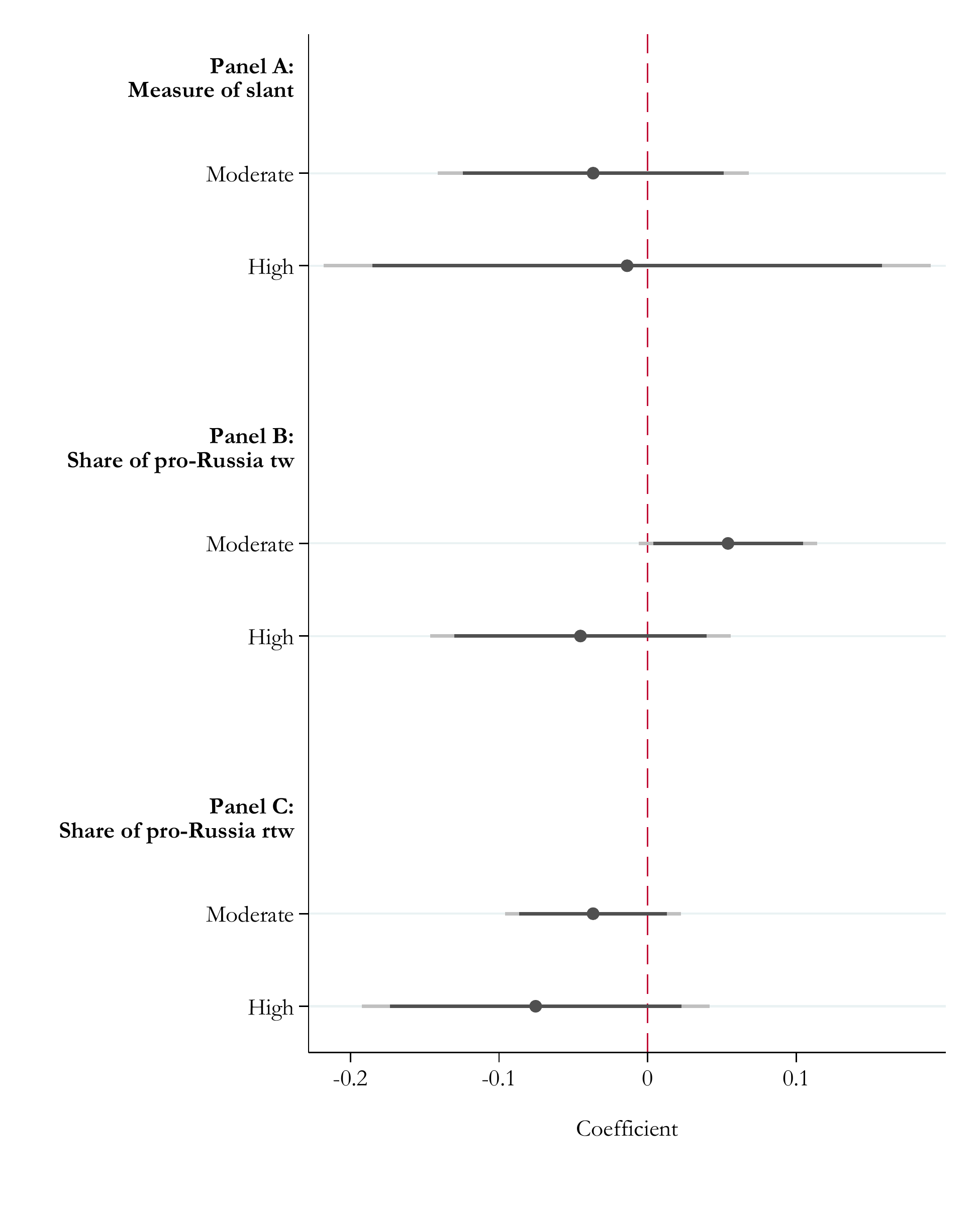}
	\begin{minipage}{\textwidth} \vspace{0.4cm}   \footnotesize{\textbf{Notes}: The figure presents coefficients from difference-in-differences regressions estimating Equation \ref{eq:did:user} and assessing the ban’s heterogeneous impact on the \textit{interaction users} that are not plausible bots. Users are divided into two groups: moderate, when the average slant in their tweets before the ban is in the bottom 75\% of the distribution, and high, when the average slant in their tweets before the ban is in the top 25\%. The coefficients are shown with 90\% and 95\% confidence intervals (95\% in light grey). Panel A shows the results of our measure of media slant, the intensive margin, obtained by taking daily averages of media slant for each user that used to interact with Russia Today and Sputnik before the ban. Panel B and C show results on the daily proportion of tweets/retweets that can be classified as pro-Russia, out of all tweets/retweets produced by the user and captured by our query. All models include the UK and Switzerland as control countries and Austria, France, Germany, Ireland, and Italy as treatment countries. They include tweets from the period between 19$^{th}$ February to 15$^{th}$ March 2022, they incorporate both user- and day-fixed effects and cluster standard errors at the user level. We control for word count, mentions count, and hashtags count. Paper Figure \ref{fig:coefplot:interact:byext} shows results for all \textit{interaction users}.}\end{minipage}
\end{figure}

The exclusion of 32,432 tweets from 2,489 users identified by these criteria results in minimal changes to the distributions of descriptive statistics reported in Table \ref{app:tab:main:descrip:nobots}. Moreover, we demonstrate that the primary conclusions drawn in the main section of the paper remain consistent even after these accounts are omitted. In particular, we reproduce the results for \textit{interaction users} excluding the potential bots. Figure \ref{app:fig:eventstudy:intensive:interact:bots} shows the intensive margin analysis, Figure \ref{app:fig:eventstudy:extensive:interact:bots} the extensive margin, Table \ref{app:tab:did:rusinteract:week:bots} the weekly interactions and Table \ref{app:tab:did:comparison:bots} the comparison with \textit{non-interaction users}. An exception is represented by the heterogeneous effects of Figure \ref{app:fig:coefplot:interact:byext:bots}. It is seems the exclusion of bots, does change the effect on the most highly slanted \textit{interaction users}, suggesting the bots population might be the most active in this sub-group.

\begin{table}[H]
\centering
\caption{\textit{Summary statistics without plausible bots}}
\label{app:tab:main:descrip:nobots}

\begin{minipage}{0.61\textwidth}
\footnotesize{\textbf{Panel A: Tweets}}
\end{minipage}\\[0.2cm]
\begin{tabular}{lcccc}
    \resizebox{0.63\linewidth}{!}{{
\def\sym#1{\ifmmode^{#1}\else\(^{#1}\)\fi}
\begin{tabular}{l*{1}{ccccc}}
\toprule
                &\textbf{Mean}&\textbf{Median}&\textbf{St. Dev.}&\textbf{Min.}&\textbf{Max.}\\
\midrule
\textbf{Dependent Variables}\vspace{3pt}&         &         &         &         &         \\
Propaganda ratio& -2.1e-10&     .041&        1&       -4&4.8945765495\\
Russian propaganda tweet&     .058&        0&      .23&        0&        1\\
Russian propaganda retweet&       .1&        0&       .3&        0&        1\\
\\ \textbf{Tweet type}\vspace{3pt}&         &         &         &         &         \\
Retweet         &      .53&        1&       .5&        0&        1\\
Reply           &     .088&        0&      .28&        0&        1\\
\\ \textbf{Tweet style}\vspace{3pt}&         &         &         &         &         \\
numer of quotes of tweet&      .12&        0&      5.8&        0&    2,369\\
No. of mentions &      1.6&        1&      2.5&        0&       50\\
No. of hashtags &      .44&        0&      1.6&        0&       42\\
No. of words    &       25&       23&       11&        1&      108\\
\midrule
No. of Observations&  762,332&         &         &         &         \\
\bottomrule
\end{tabular}
}
}
\end{tabular}\\[0.5cm] 

\begin{minipage}{0.61\textwidth}
\footnotesize{\textbf{Panel B: Users}}
\end{minipage}\\[0.2cm]
\begin{tabular}{lcccc}
    \resizebox{0.63\linewidth}{!}{{
\def\sym#1{\ifmmode^{#1}\else\(^{#1}\)\fi}
\begin{tabular}{l*{1}{ccccc}}
\toprule
                &\textbf{Mean}&\textbf{Median}&\textbf{St. Dev.}&\textbf{Min.}&\textbf{Max.}\\
\midrule
\textbf{User behavior}\vspace{3pt}&         &         &         &         &         \\
No. tweets from user&      2.7&        1&       12&        0&    1,543\\
No. retweets from user&        3&        1&       11&        0&      676\\
No. replies from user&      .51&        0&      2.1&        0&      209\\
No. russian propaganda tweets&      .34&        0&      1.7&        0&      303\\
No. russian propaganda retweets&      .59&        0&      2.3&        0&      151\\
Interacted with RT/Spk&     .036&        0&      .19&        0&        1\\
No. retweets of RT/Spk&   .00096&        0&     .041&        0&        5\\
\\ \textbf{Region}\vspace{3pt}&         &         &         &         &         \\
European Union  &      .39&        0&      .49&        0&        1\\
\midrule
No. of Observations&  132,081&         &         &         &         \\
\bottomrule
\end{tabular}
}
}
\end{tabular}

\begin{minipage}{\textwidth} \vspace{0.4cm}   
\footnotesize{\textbf{Notes:} Panel A reports descriptive statistics for the sample of tweets used in the analysis, posted by users that are not plausible bots and that we could locate in the countries of interest: Austria, France, Germany, Ireland, Italy, Switzerland and the United Kingdom. Tweets were extracted using the Historical Twitter APIv2, with the query: \textit{ukrain* OR russ* OR NATO OR OTAN}, in the time window between February 19$^{th}$ and March 15$^{th}$ 2022. Panel B reports descriptive statistics on user characteristics, for users that posted tweets used in our analysis and described in Panel A. In both panels, for all variables we report mean, median, standard deviation, minimum, and maximum values. Paper Table \ref{tab:main:descrip} shows the same for the full sample.}
\end{minipage}
\end{table}

\FloatBarrier
\newpage
\vspace*{-2cm}

\FloatBarrier

\FloatBarrier

\FloatBarrier
\newpage
\vspace*{-2.5cm}
\begin{table}[H]\centering
	\caption{\textit{User-day level two-periods TWFE with post-ban weeks interactions: Interaction users excluding plausible bots}}
	\label{app:tab:did:rusinteract:week:bots}
	\def\sym#1{\ifmmode^{#1}\else\(^{#1}\)\fi}
	\resizebox{1\linewidth }{!}{
		\begin{tabular}{lcccccc}
			\hline\hline
			                &\multicolumn{1}{c}{(1)}&\multicolumn{1}{c}{(2)}&\multicolumn{1}{c}{(3)}&\multicolumn{1}{c}{(4)}&\multicolumn{1}{c}{(5)}\\
                &Avg. media slant&\% pro-Russia tweets&\% pro-Russia retweets&Tot. Pro-Russia tweets&Tot. pro-Russia retweets\\
\hline
EU $\times$ 1st week after-ban&   -0.052&   -0.023&   -0.022&   -0.035&   -0.051\\
                &  [0.023]&  [0.014]&  [0.014]&  [0.022]&  [0.027]\\
EU $\times$ 2nd week after-ban&   -0.031&   -0.008&   -0.010&    0.012&   -0.023\\
                &  [0.025]&  [0.016]&  [0.014]&  [0.021]&  [0.024]\\
User FEs        &      yes&      yes&      yes&      yes&      yes\\
Day FEs         &      yes&      yes&      yes&      yes&      yes\\
\hline
Observations    &    29544&    16508&    19217&    29544&    29544\\
\(R^{2}\)       &    0.333&    0.230&    0.239&    0.208&    0.366\\
Pre-period mean of DV&   -0.056&    0.118&    0.165&    1.309&    1.747\\
1st week \% of mean&   -91.74&   -19.84&   -13.59&    -2.64&    -2.92\\
\hline\hline    &         &         &         &         &         \\

		\end{tabular}
	}
	\begin{minipage}{\textwidth} \vspace{0cm}   \footnotesize{\textbf{Notes}: The table displays coefficients from a two-period difference-in-differences regression analysis estimating Equation \ref{eq:did:user}, examining the ban's differential impact in the two weeks following the ban, on the \textit{interaction users} that are not plausible bots. Column 1 shows the effects on our measure of media slant, the intensive margin. Columns 2 and 3 show effects on our measure of the extensive margin, respectively share of tweets and retweets that are pro-Russia. Columns 4 and 5 show the effect on the total number of pro-Russia tweets and retweets, respectively, produced by the author in the time period. All models include the UK and Switzerland as control countries and Austria, France, Germany, Ireland, and Italy as treatment countries. They include tweets from the period between 19$^{th}$ February to 15$^{th}$ March 2022, they incorporate both user- and day-fixed effects and cluster standard errors at the user level. We control for word count, mentions count, and hashtags count. Paper Table \ref{tab:did:rusinteract:week} shows results including all \textit{interaction users}.}\end{minipage}
\end{table}

\begin{table}[H]\centering
	\caption{\textit{User-day level two-periods TWFE: Interaction and non-interaction users excluding potential bots}}
	\label{app:tab:did:comparison:bots}
	\def\sym#1{\ifmmode^{#1}\else\(^{#1}\)\fi}
    \begin{minipage}{\textwidth}
       \footnotesize{\textbf{Panel A: Consumers of content from outlets}}
     \end{minipage}\\[0.2cm]
     \resizebox{1\linewidth }{!}{
		\begin{tabular}{lcccccc}
			\hline
			                &\multicolumn{1}{c}{(1)}&\multicolumn{1}{c}{(2)}&\multicolumn{1}{c}{(3)}&\multicolumn{1}{c}{(4)}&\multicolumn{1}{c}{(5)}\\
                &Avg. media slant&\% pro-Russia tweets&\% pro-Russia retweets&Tot. Pro-Russia tweets&Tot. pro-Russia retweets\\
\hline
EU $\times$ after-ban&   -0.043&   -0.017&   -0.017&   -0.014&   -0.039\\
                &  [0.021]&  [0.012]&  [0.012]&  [0.018]&  [0.022]\\
User FEs        &      yes&      yes&      yes&      yes&      yes\\
Day FEs         &      yes&      yes&      yes&      yes&      yes\\
\hline
Observations    &    29544&    16508&    19217&    29544&    29544\\
\(R^{2}\)       &    0.333&    0.230&    0.239&    0.208&    0.365\\
Pre-period mean of DV&   -0.056&    0.118&    0.165&    1.309&    1.747\\
\% of mean      &   -75.99&   -14.22&   -10.35&    -1.08&    -2.23\\
\hline\hline    &         &         &         &         &         \\

		\end{tabular}
	}
     \begin{minipage}{\textwidth}
       \footnotesize{\textbf{Panel A: Non-consumers of content from outlets}}
     \end{minipage}\\[0.2cm]
	\resizebox{1\linewidth }{!}{
		\begin{tabular}{lcccccc}
			\hline
                            &\multicolumn{1}{c}{(1)}&\multicolumn{1}{c}{(2)}&\multicolumn{1}{c}{(3)}&\multicolumn{1}{c}{(4)}&\multicolumn{1}{c}{(5)}\\
                &Avg. media slant&\% pro-Russia tweets&\% pro-Russia retweets&Tot. Pro-Russia tweets&Tot. pro-Russia retweets\\
\hline
EU $\times$ after-ban&   -0.030&    0.000&   -0.037&   -0.000&   -0.012\\
                &  [0.008]&  [0.004]&  [0.004]&  [0.005]&  [0.005]\\
User FEs        &      yes&      yes&      yes&      yes&      yes\\
Day FEs         &      yes&      yes&      yes&      yes&      yes\\
\hline
Observations    &   312429&   148704&   178921&   312429&   312429\\
\(R^{2}\)       &    0.419&    0.325&    0.311&    0.308&    0.293\\
Pre-period mean of DV&   -0.190&    0.105&    0.142&    0.930&    1.070\\
\% of mean      &   -15.97&     0.20&   -25.96&    -0.01&    -1.16\\
\hline\hline    &         &         &         &         &         \\

		\end{tabular}
	}
	\begin{minipage}{\textwidth} \vspace{0cm}   \footnotesize{\textbf{Notes}: The table displays coefficients from a two-period difference-in-differences regression analysis estimating Equation \ref{eq:did:user}, examining the ban's impact on users who interacted with Russia Today and Sputnik before the ban in Panel A, and on user that had no interactions with the outlets in Panel B. We exclude in both panels users that are plausible bots. Column 1 shows the effects on our measure of media slant, the intensive margin. Columns 2 and 3 show effects on our measure of the extensive margin, respectively share of tweets and retweets that are pro-Russia. Columns 4 and 5 show the effect on the total number of pro-Russia tweets and retweets, respectively, produced by the author in the time period. All models include the UK and Switzerland as control countries and Austria, France, Germany, Ireland, and Italy as treatment countries. They include tweets from the period between 19$^{th}$ February to 15$^{th}$ March 2022, they incorporate both user- and day-fixed effects and cluster standard errors at the user level. We control for word count, mentions count, and hashtags count. Paper Table \ref{tab:did:comparison} shows results including also plausible bots.}\end{minipage}
\end{table}
\FloatBarrier

\newpage
\section{Robustness check: Without accounts created after the ban} 
\label{appendix:lateacc}
\setcounter{figure}{0} 
\setcounter{table}{0}

\begin{figure}[H]
	\centering
	\caption{\textit{Daily event-study on share of slanted tweets and retweets: Interaction users excluding accounts created post-ban}}\label{app:fig:eventstudy:intensive:interact:lateacc}
	\includegraphics[width=.65\textwidth, trim={0cm 0.1cm 0.1cm 0.1cm},clip]{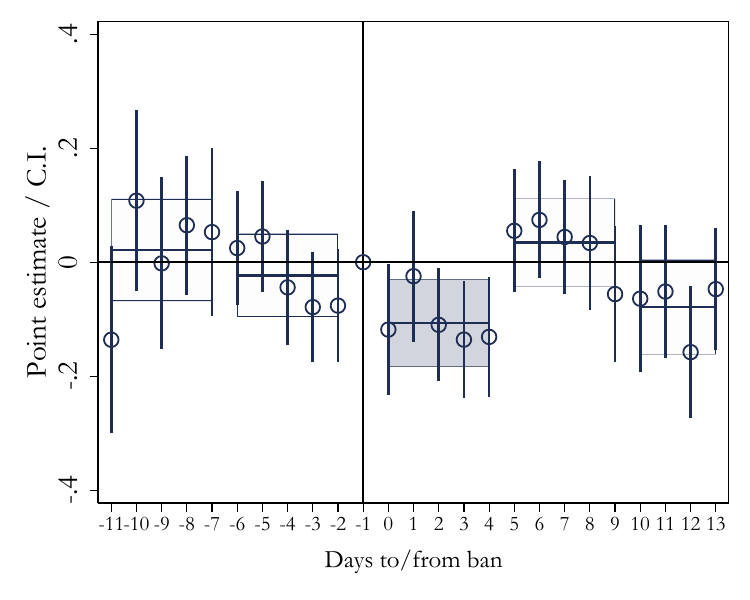}
	\begin{minipage}{\textwidth} \vspace{-.3cm}   \footnotesize{\textbf{Notes}: The figure displays coefficients and 95\% confidence intervals from estimating a daily event study version of Equation \ref{eq:did:user} excluding accounts created post-ban. The sample consists of \textit{interaction users} -- users who interacted with the banned outlets before the ban -- and includes users located in the UK and Switzerland as control group and users located in Austria, France, Germany, Ireland, and Italy as treatment group. We exclude all accounts created after the implementation of the ban. Dependent variable is the daily average of slant in tweets for each user. We use tweets from the period between 19$^{th}$ February to 15$^{th}$ March 2022. We estimate Equation \ref{eq:did:user} including user- and day-fixed effects relative to the omitted day, 1$^{st}$ March 2022 immediately before the implementation of the ban. In the aggregate specification, coefficients of interest are interactions between a dummy variable for aggregated intervals for 19$^{th}$ to 23$^{th}$ February, 24$^{th}$ to 28 February$^{th}$, 2$^{nd}$ to 6$^{th}$ March, 7$^{th}$ to 11$^{th}$ March and 12$^{th}$ to 15$^{th}$ March, relative to the omitted day, 1$^{st}$ March 2022 immediately before the implementation of the ban. Coefficient estimates on the day interactions are plotted as dots with their 95\% confidence intervals indicated with vertical lines. Coefficient estimates on the aggregate interactions are shown with horizontal lines, and their 95\% confidence intervals are indicated as boxes. We cluster standard errors at the user level. We control for word count, mentions count, and hashtags count. Paper Figure \ref{fig:eventstudy:intensive:interact} shows results for all \textit{interaction users}.}\end{minipage}
\end{figure}

\begin{figure}[H]
	\centering
	\caption{\textit{Daily event-study on share of slanted tweets and retweets: Interaction users excluding accounts created post-ban}}
	\label{app:fig:eventstudy:extensive:interact:lateacc}
	\begin{subfigure}{.49\textwidth}
		\centering
		\caption{\textit{Effect on pro-Russian slanted tweets}}
		\includegraphics[width=\textwidth]{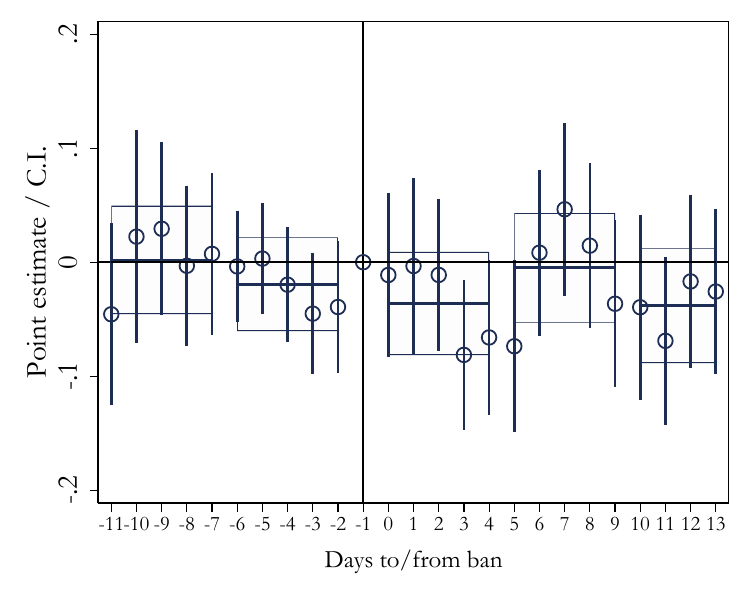}
		\label{app:fig:eventstudy:extensive:tw:interact:bots}
	\end{subfigure}
	\begin{subfigure}{.49\textwidth}
		\centering
		\caption{\textit{Effect on pro-Russian slanted retweets}}
		\includegraphics[width=\textwidth]{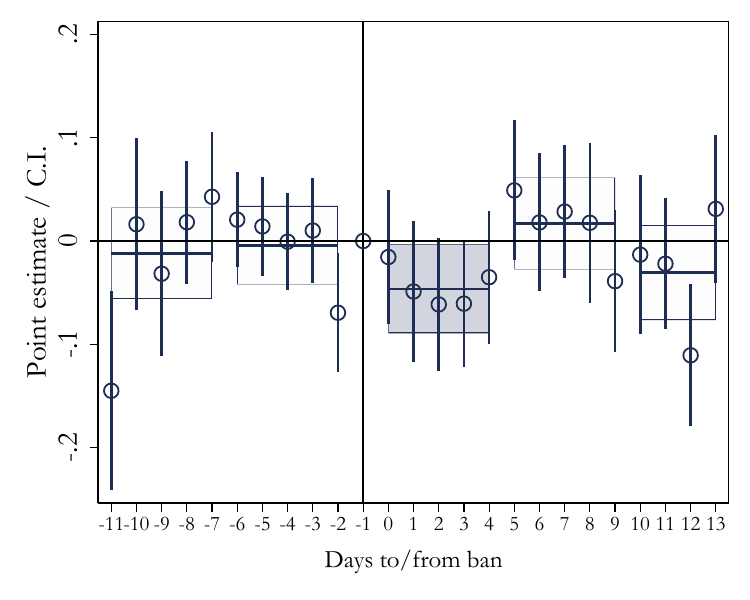}
		\label{app:fig:eventstudy:extensive:rtw:interact:bots}
	\end{subfigure}
	
	\begin{minipage}{\textwidth} \vspace{-.7cm}   \footnotesize{\textbf{Notes}: The figure displays coefficients and 95\% confidence intervals from estimating a daily event study version of Equation \ref{eq:did:user} excluding accounts created post-ban. The sample consists of \textit{interaction users} -- users who interacted with the banned outlets before the ban -- and includes users located in the UK and Switzerland as control group and users located in Austria, France, Germany, Ireland, and Italy as treatment group. We exclude all accounts created after the implementation of the ban. We control for word count, mentions count, and hashtags count. We use tweets from the period between 19$^{th}$ February to 15$^{th}$ March 2022. We estimate Equation \ref{eq:did:user} including user- and day-fixed effects relative to the omitted day, 1$^{st}$ March 2022 immediately before the implementation of the ban. In the aggregate specification, coefficients of interest are interactions between a dummy variable for aggregated intervals for 19$^{th}$ to 23$^{th}$ February, 24$^{th}$ to 28 February$^{th}$, 2$^{nd}$ to 6$^{th}$ March, 7$^{th}$ to 11$^{th}$ March and 12$^{th}$ to 15$^{th}$ March, relative to the omitted day, 1$^{st}$ March 2022 immediately before the implementation of the ban. Coefficient estimates on the day interactions are plotted as dots with their 95\% confidence intervals indicated with vertical lines. Coefficient estimates on the aggregate interactions are shown with horizontal lines, and their 95\% confidence intervals are indicated as boxes. We cluster standard errors at the user level. Figure \ref{fig:eventstudy:extensive:tw:interact} display estimates using user's daily share of pro-Russian slanted tweets -- defined as having a media slant measure above 1 -- as dependent variable. Figure \ref{fig:eventstudy:extensive:rtw:interact} display estimates using user's daily share of pro-Russian slanted retweets -- defined as having a media slant measure above 1 -- as dependent variable. Paper Figure \ref{fig:eventstudy:extensive:interact} shows results for all \textit{interaction users}.}\end{minipage}
\end{figure}

\begin{figure}[H]
	\centering
	\caption{\textit{Heterogeneous effects of the ban by pre-ban level of pro-Russian slant: Interaction users excluding accounts created after the ban}}\label{app:fig:coefplot:interact:byext:lateacc}
	\includegraphics[width=.45\textwidth, trim={0cm 0.1cm 0.1cm 0.1cm},clip]{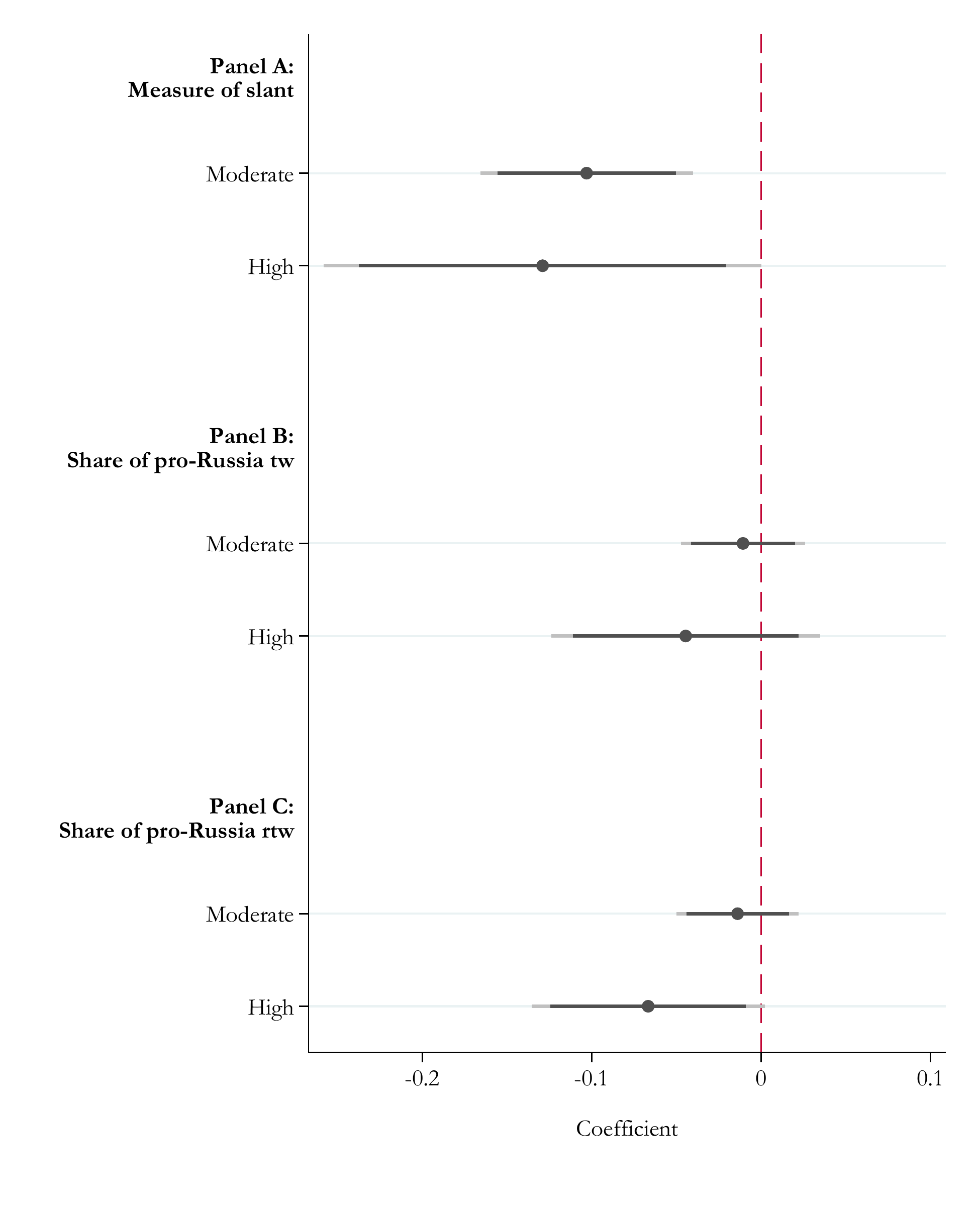}
	\begin{minipage}{\textwidth} \vspace{0.4cm}   \footnotesize{\textbf{Notes}: The figure presents coefficients from difference-in-differences regressions estimating Equation \ref{eq:did:user} and assessing the ban’s heterogeneous impact on the \textit{interaction users} that created their account any time before the ban. Users are divided into two groups: moderate, when the average slant in their tweets before the ban is in the bottom 75\% of the distribution, and high, when the average slant in their tweets before the ban is in the top 25\%. The coefficients are shown with 90\% and 95\% confidence intervals (95\% in light grey). Panel A shows the results of our measure of media slant, the intensive margin, obtained by taking daily averages of media slant for each user that used to interact with Russia Today and Sputnik before the ban. Panel B and C show results on the daily proportion of tweets/retweets that can be classified as pro-Russia, out of all tweets/retweets produced by the user and captured by our query. All models include the UK and Switzerland as control countries and Austria, France, Germany, Ireland, and Italy as treatment countries. They include tweets from the period between 19$^{th}$ February to 15$^{th}$ March 2022, they incorporate both user- and day-fixed effects and cluster standard errors at the user level. We control for word count, mentions count, and hashtags count. Paper Figure \ref{fig:coefplot:interact:byext} shows results for all \textit{interaction users}.}\end{minipage}
\end{figure}

In this Appendix, we presents the findings from our analysis replicated from the main body of the paper, adjusting our sample with the removal of accounts that were created after the ban from the sample. We do this to shed light on whether the creation of new accounts was an important mechanism of reaction to the ban. The exclusion of 2043 tweets from the 389 users that created their account after the ban, does not modify significantly the descriptive statistics and the results we presented in the main paper. Table \ref{app:tab:main:descrip:lateacc} shows the descriptive statistics. Figure \ref{app:fig:eventstudy:intensive:interact:lateacc} shows the intensive margin analysis, Figure \ref{app:fig:eventstudy:extensive:interact:lateacc} the extensive margin, Table \ref{app:tab:did:rusinteract:week:lateacc} the weekly interactions and Table \ref{app:tab:did:comparison:lateacc} the comparison with \textit{non-interaction users}. An exception is represented by heterogeneous effects in Figure \ref{app:fig:coefplot:interact:byext:lateacc}. Despite the main Paper conclusions hold for the share of retweets, there seem to be no differentiale ffect for moderate and high slant for what concerns the intensive margin.

\vspace{-.4cm}
\begin{table}[H]
\centering
\caption{\textit{Summary statistics without accounts created after the ban}}
\label{app:tab:main:descrip:lateacc}

\begin{minipage}{0.61\textwidth}
\footnotesize{\textbf{Panel A: Tweets}}
\end{minipage}\\[0.2cm]
\begin{tabular}{lcccc}
    \resizebox{0.63\linewidth}{!}{{
\def\sym#1{\ifmmode^{#1}\else\(^{#1}\)\fi}
\begin{tabular}{l*{1}{ccccc}}
\toprule
                &\textbf{Mean}&\textbf{Median}&\textbf{St. Dev.}&\textbf{Min.}&\textbf{Max.}\\
\midrule
\textbf{Dependent Variables}\vspace{3pt}&         &         &         &         &         \\
Propaganda ratio& -4.8e-10&     .041&        1&       -4&4.8913640976\\
Russian propaganda tweet&     .058&        0&      .23&        0&        1\\
Russian propaganda retweet&       .1&        0&       .3&        0&        1\\
\\ \textbf{Tweet type}\vspace{3pt}&         &         &         &         &         \\
Retweet         &      .53&        1&       .5&        0&        1\\
Reply           &      .09&        0&      .29&        0&        1\\
\\ \textbf{Tweet style}\vspace{3pt}&         &         &         &         &         \\
numer of quotes of tweet&      .11&        0&      5.7&        0&    2,369\\
No. of mentions &      1.6&        1&      2.4&        0&       50\\
No. of hashtags &      .45&        0&      1.6&        0&       42\\
No. of words    &       25&       23&       11&        1&      108\\
\midrule
No. of Observations&  792,721&         &         &         &         \\
\bottomrule
\end{tabular}
}
}
\end{tabular}\\[0.5cm] 

\begin{minipage}{0.61\textwidth}
\footnotesize{\textbf{Panel B: Users}}
\end{minipage}\\[0.2cm]
\begin{tabular}{lcccc}
    \resizebox{0.63\linewidth}{!}{{
\def\sym#1{\ifmmode^{#1}\else\(^{#1}\)\fi}
\begin{tabular}{l*{1}{ccccc}}
\toprule
                &\textbf{Mean}&\textbf{Median}&\textbf{St. Dev.}&\textbf{Min.}&\textbf{Max.}\\
\midrule
\textbf{User behavior}\vspace{3pt}&         &         &         &         &         \\
No. tweets from user&      2.8&        1&       12&        0&    1,543\\
No. retweets from user&      3.1&        1&       11&        0&      676\\
No. replies from user&      .53&        0&      2.2&        0&      209\\
No. russian propaganda tweets&      .34&        0&      1.7&        0&      298\\
No. russian propaganda retweets&       .6&        0&      2.3&        0&      151\\
Interacted with RT/Spk&     .037&        0&      .19&        0&        1\\
No. retweets of RT/Spk&     .001&        0&     .044&        0&        6\\
\\ \textbf{Region}\vspace{3pt}&         &         &         &         &         \\
European Union  &      .39&        0&      .49&        0&        1\\
\midrule
No. of Observations&  134,181&         &         &         &         \\
\bottomrule
\end{tabular}
}
}
\end{tabular}

\begin{minipage}{\textwidth} \vspace{0.4cm}   
\footnotesize{\textbf{Notes:} The table in Panel A reports descriptive statistics for the sample of tweets used in the analysis. The table in Panel B reports descriptive statistics on user characteristics, for users that posted tweets used in our analysis and described in Panel A. In both panels, for all variables we report mean, median, standard deviation, minimum, and maximum values. Paper Table \ref{tab:main:descrip} shows the same for the full sample.}
\end{minipage}
\end{table}

\FloatBarrier
\newpage
\vspace*{-2cm}

\FloatBarrier

\FloatBarrier

\FloatBarrier
\newpage
\vspace*{-2.5cm}
\begin{table}[H]\centering
	\caption{\textit{User-day level two-periods TWFE with post-ban weeks interactions: Interaction users excluding accounts created after the ban}}
	\label{app:tab:did:rusinteract:week:lateacc}
	\def\sym#1{\ifmmode^{#1}\else\(^{#1}\)\fi}
	\resizebox{1\linewidth }{!}{
		\begin{tabular}{lcccccc}
			\hline\hline
			                &\multicolumn{1}{c}{(1)}&\multicolumn{1}{c}{(2)}&\multicolumn{1}{c}{(3)}&\multicolumn{1}{c}{(4)}&\multicolumn{1}{c}{(5)}\\
                &Avg. media slant&\% pro-Russia tweets&\% pro-Russia retweets&Tot. Pro-Russia tweets&Tot. pro-Russia retweets\\
\hline
EU $\times$ 1st week after-ban&   -0.050&   -0.024&   -0.017&   -0.037&   -0.042\\
                &  [0.022]&  [0.014]&  [0.013]&  [0.021]&  [0.027]\\
EU $\times$ 2nd week after-ban&   -0.034&   -0.004&   -0.012&    0.013&   -0.029\\
                &  [0.025]&  [0.015]&  [0.014]&  [0.020]&  [0.024]\\
User FEs        &      yes&      yes&      yes&      yes&      yes\\
Day FEs         &      yes&      yes&      yes&      yes&      yes\\
\hline
Observations    &    31076&    17291&    20436&    31076&    31076\\
\(R^{2}\)       &    0.332&    0.229&    0.239&    0.207&    0.364\\
Pre-period mean of DV&   -0.058&    0.118&    0.164&    1.307&    1.799\\
1st week \% of mean&   -86.00&   -20.54&   -10.44&    -2.80&    -2.32\\
\hline\hline    &         &         &         &         &         \\

		\end{tabular}
	}
	\begin{minipage}{\textwidth} \vspace{0cm}   \footnotesize{\textbf{Notes}: The table displays coefficients from a two-period difference-in-differences regression analysis estimating Equation \ref{eq:did:user}, examining the ban's differential impact in the two weeks following the ban, on the \textit{interaction users} that created their account any time before the ban. Column 1 shows the effects on our measure of media slant, the intensive margin. Columns 2 and 3 show effects on our measure of the extensive margin, respectively share of tweets and retweets that are pro-Russia. Columns 4 and 5 show the effect on the total number of pro-Russia tweets and retweets, respectively, produced by the author in the time period. All models include the UK and Switzerland as control countries and Austria, France, Germany, Ireland, and Italy as treatment countries. They include tweets from the period between 19$^{th}$ February to 15$^{th}$ March 2022, they incorporate both user- and day-fixed effects and cluster standard errors at the user level. We control for word count, mentions count, and hashtags count. Paper Table \ref{tab:did:rusinteract:week} shows results including all \textit{interaction users}.}\end{minipage}
\end{table}
\FloatBarrier

\FloatBarrier
\begin{table}[H]\centering
	\caption{\textit{User-day level two-periods TWFE: Interaction and non-interaction users excluding accounts created after the ban}}
	\label{app:tab:did:comparison:lateacc}
	\def\sym#1{\ifmmode^{#1}\else\(^{#1}\)\fi}
    \begin{minipage}{\textwidth}
       \footnotesize{\textbf{Panel A: Consumers of content from outlets}}
     \end{minipage}\\[0.2cm]
     \resizebox{1\linewidth }{!}{
		\begin{tabular}{lcccccc}
			\hline
			                &\multicolumn{1}{c}{(1)}&\multicolumn{1}{c}{(2)}&\multicolumn{1}{c}{(3)}&\multicolumn{1}{c}{(4)}&\multicolumn{1}{c}{(5)}\\
                &Avg. media slant&\% pro-Russia tweets&\% pro-Russia retweets&Tot. Pro-Russia tweets&Tot. pro-Russia retweets\\
\hline
EU $\times$ after-ban&   -0.043&   -0.016&   -0.015&   -0.015&   -0.036\\
                &  [0.020]&  [0.012]&  [0.011]&  [0.017]&  [0.022]\\
User FEs        &      yes&      yes&      yes&      yes&      yes\\
Day FEs         &      yes&      yes&      yes&      yes&      yes\\
\hline
Observations    &    31076&    17291&    20436&    31076&    31076\\
\(R^{2}\)       &    0.332&    0.229&    0.239&    0.207&    0.364\\
Pre-period mean of DV&   -0.058&    0.118&    0.164&    1.307&    1.799\\
\% of mean      &   -74.10&   -13.34&    -9.17&    -1.15&    -2.01\\
\hline\hline    &         &         &         &         &         \\

		\end{tabular}
	}
     \begin{minipage}{\textwidth}
       \footnotesize{\textbf{Panel A: Non-consumers of content from outlets}}
     \end{minipage}\\[0.2cm]
	\resizebox{1\linewidth }{!}{
		\begin{tabular}{lcccccc}
			\hline
                            &\multicolumn{1}{c}{(1)}&\multicolumn{1}{c}{(2)}&\multicolumn{1}{c}{(3)}&\multicolumn{1}{c}{(4)}&\multicolumn{1}{c}{(5)}\\
                &Avg. media slant&\% pro-Russia tweets&\% pro-Russia retweets&Tot. Pro-Russia tweets&Tot. pro-Russia retweets\\
\hline
EU $\times$ after-ban&   -0.030&    0.001&   -0.036&   -0.004&   -0.011\\
                &  [0.007]&  [0.004]&  [0.004]&  [0.006]&  [0.005]\\
User FEs        &      yes&      yes&      yes&      yes&      yes\\
Day FEs         &      yes&      yes&      yes&      yes&      yes\\
\hline
Observations    &   321772&   152444&   185939&   321772&   321772\\
\(R^{2}\)       &    0.417&    0.324&    0.308&    0.293&    0.291\\
Pre-period mean of DV&   -0.189&    0.105&    0.142&    0.932&    1.089\\
\% of mean      &   -15.81&     0.70&   -25.43&    -0.47&    -1.05\\
\hline\hline    &         &         &         &         &         \\

		\end{tabular}
	}
	\begin{minipage}{\textwidth} \vspace{0cm}   \footnotesize{\textbf{Notes}: The table displays coefficients from a two-period difference-in-differences regression analysis estimating Equation \ref{eq:did:user}, examining the ban's impact on users who interacted with Russia Today and Sputnik before the ban in Panel A, and on user that had no interactions with the outlets in Panel B. We exclude in both panels users that created their account after the ban. Column 1 shows the effects on our measure of media slant, the intensive margin. Columns 2 and 3 show effects on our measure of the extensive margin, respectively share of tweets and retweets that are pro-Russia. Columns 4 and 5 show the effect on the total number of pro-Russia tweets and retweets, respectively, produced by the author in the time period. All models include the UK and Switzerland as control countries and Austria, France, Germany, Ireland, and Italy as treatment countries. They include tweets from the period between 19$^{th}$ February to 15$^{th}$ March 2022, they incorporate both user- and day-fixed effects and cluster standard errors at the user level. We control for word count, mentions count, and hashtags count. Paper Table \ref{tab:did:comparison} shows results including also plausible bots.}\end{minipage}
\end{table}
\FloatBarrier

\section{Robustness check: Different pro-Russia threshold} 
\label{appendix:diffthreshold}
\setcounter{figure}{0} 
\setcounter{table}{0} 

In this Appendix, we presents the findings from our analysis replicated from the main body of the paper, changing the threshold to define our binary variables of pro-Russia tweet and retweet. As a reminder, in the main analysis of the Paper, we use a threshold of 1 when we define our binary variables. This is to ensure we do not capture noise around the threshold of 0. Nevertheless, to prove additional robustness of our results we report in this appendix all results employing the binary variables, using a threshold of 0.

\begin{table}[H]
\centering
\caption{\textit{Summary statistics using alternative threshold}}
\label{app:tab:main:descrip:diffthres}

\begin{minipage}{0.61\textwidth}
\footnotesize{\textbf{Panel A: Tweets}}
\end{minipage}\\[0.2cm]
\begin{tabular}{lcccc}
    \resizebox{0.63\linewidth}{!}{{
\def\sym#1{\ifmmode^{#1}\else\(^{#1}\)\fi}
\begin{tabular}{l*{1}{ccccc}}
\toprule
                &\textbf{Mean}&\textbf{Median}&\textbf{St. Dev.}&\textbf{Min.}&\textbf{Max.}\\
\midrule
\textbf{Dependent Variables}\vspace{3pt}&         &         &         &         &         \\
Propaganda ratio& -6.2e-10&     .041&        1&       -4&4.8926959038\\
Russian propaganda tweet&      .22&        0&      .42&        0&        1\\
Russian propaganda retweet&      .29&        0&      .45&        0&        1\\
\\ \textbf{Tweet type}\vspace{3pt}&         &         &         &         &         \\
Retweet         &      .53&        1&       .5&        0&        1\\
Reply           &      .09&        0&      .29&        0&        1\\
\\ \textbf{Tweet style}\vspace{3pt}&         &         &         &         &         \\
numer of quotes of tweet&      .11&        0&      5.7&        0&    2,369\\
No. of mentions &      1.6&        1&      2.4&        0&       50\\
No. of hashtags &      .45&        0&      1.6&        0&       42\\
No. of words    &       25&       23&       11&        1&      108\\
\midrule
No. of Observations&  794,764&         &         &         &         \\
\bottomrule
\end{tabular}
}
}
\end{tabular}\\[0.5cm] 

\begin{minipage}{0.61\textwidth}
\footnotesize{\textbf{Panel B: Users}}
\end{minipage}\\[0.2cm]
\begin{tabular}{lcccc}
    \resizebox{0.63\linewidth}{!}{{
\def\sym#1{\ifmmode^{#1}\else\(^{#1}\)\fi}
\begin{tabular}{l*{1}{ccccc}}
\toprule
                &\textbf{Mean}&\textbf{Median}&\textbf{St. Dev.}&\textbf{Min.}&\textbf{Max.}\\
\midrule
\textbf{User behavior}\vspace{3pt}&         &         &         &         &         \\
No. tweets from user&      2.8&        1&       12&        0&    1,543\\
No. retweets from user&      3.1&        1&       11&        0&      676\\
No. replies from user&      .53&        0&      2.2&        0&      209\\
No. russian propaganda tweets&      1.3&        0&      5.5&        0&      712\\
No. russian propaganda retweets&      1.7&        0&      6.3&        0&      403\\
Interacted with RT/Spk&     .037&        0&      .19&        0&        1\\
No. retweets of RT/Spk&     .001&        0&     .044&        0&        6\\
\\ \textbf{Region}\vspace{3pt}&         &         &         &         &         \\
European Union  &      .39&        0&      .49&        0&        1\\
\midrule
No. of Observations&  134,570&         &         &         &         \\
\bottomrule
\end{tabular}
}
}
\end{tabular}

\begin{minipage}{\textwidth} \vspace{0.4cm}   
\footnotesize{\textbf{Notes:} The table in Panel A reports descriptive statistics for the sample of tweets used in the analysis, posted by users that we could locate in the countries of interest: the United Kingdom and Switzerland, Austria, France, Germany, Ireland, and Italy. For binary variables we use the threshold 0 instead than 1 of our slant measure, for tweets and retweets to be defined as pro-Russia. Tweets were extracted using the Historical Twitter APIv2, with the query: \textit{ukrain* OR russ* OR NATO OR OTAN}, in the time window between February 19$^{th}$ and March 15$^{th}$ 2022. The table in Panel B reports descriptive statistics on user characteristics, for users that posted tweets used in our analysis and described in Panel A. In both panels, for all variables we report mean, median, standard deviation, minimum, and maximum values. Paper Table \ref{tab:main:descrip} shows the same for the full sample.}
\end{minipage}
\end{table}

\FloatBarrier
\begin{figure}[H]
	\centering
	\caption{\textit{Daily event-study on share of slanted tweets and retweets: Interaction users using alternative threshold}}
	\label{app:fig:eventstudy:extensive:interact:diffthres}
	\begin{subfigure}{.49\textwidth}
		\centering
		\caption{\textit{Effect on pro-Russian slanted tweets}}
		\includegraphics[width=\textwidth]{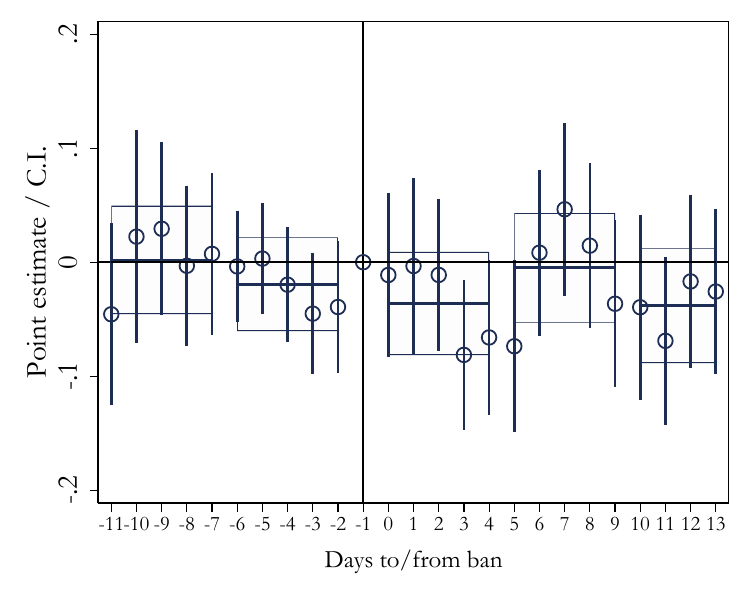}
		\label{app:fig:eventstudy:extensive:tw:interact:diffthres}
	\end{subfigure}
	\begin{subfigure}{.49\textwidth}
		\centering
		\caption{\textit{Effect on pro-Russian slanted retweets}}
		\includegraphics[width=\textwidth]{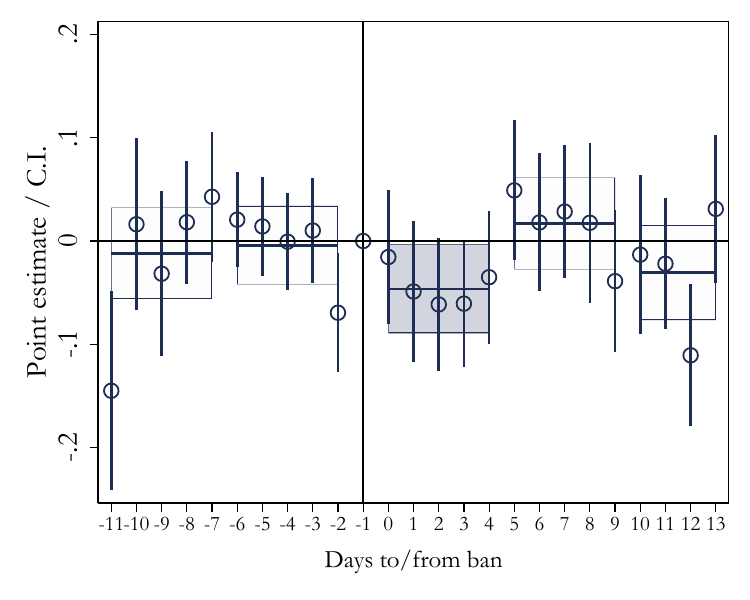}
		\label{app:fig:eventstudy:extensive:rtw:interact:diffthres}
	\end{subfigure}
	
	\begin{minipage}{\textwidth} \vspace{0cm}   \footnotesize{\textbf{Notes}: The figure displays coefficients and 95\% confidence intervals from estimating a daily event study version of Equation \ref{eq:did:user} excluding accounts created post-ban. The sample consists of \textit{interaction users} -- users who interacted with the banned outlets before the ban -- and includes users located in the UK and Switzerland as control group and users located in Austria, France, Germany, Ireland, and Italy as treatment group. We exclude all accounts created after the implementation of the ban. We use tweets from the period between 19$^{th}$ February to 15$^{th}$ March 2022. We estimate Equation \ref{eq:did:user} including user- and day-fixed effects relative to the omitted day, 1$^{st}$ March 2022 immediately before the implementation of the ban. In the aggregate specification, coefficients of interest are interactions between a dummy variable for aggregated intervals for 19$^{th}$ to 23$^{th}$ February, 24$^{th}$ to 28 February$^{th}$, 2$^{nd}$ to 6$^{th}$ March, 7$^{th}$ to 11$^{th}$ March and 12$^{th}$ to 15$^{th}$ March, relative to the omitted day, 1$^{st}$ March 2022 immediately before the implementation of the ban. Coefficient estimates on the day interactions are plotted as dots with their 95\% confidence intervals indicated with vertical lines. Coefficient estimates on the aggregate interactions are shown with horizontal lines, and their 95\% confidence intervals are indicated as boxes. We cluster standard errors at the user level. We control for word count, mentions count, and hashtags count. Figure \ref{fig:eventstudy:extensive:tw:interact} display estimates using user's daily share of pro-Russian slanted tweets -- defined as having a media slant measure above 0 -- as dependent variable. Figure \ref{fig:eventstudy:extensive:rtw:interact} display estimates using user's daily share of pro-Russian slanted retweets -- defined as having a media slant measure above 0 -- as dependent variable. Paper Figure \ref{fig:eventstudy:extensive:interact} shows results for all \textit{interaction users} using a threshold of 1 to define pro-Russian slanted tweets and retweets.}\end{minipage}
\end{figure}
\FloatBarrier

\newpage 
\vspace*{-2.5cm}
\begin{figure}[H]
	\centering
	\caption{\textit{Heterogeneous effects of the ban by pre-ban level of pro-Russian slant: Interaction users using alternative threshold}}\label{app:fig:coefplot:interact:byext:diffthres}
	\includegraphics[width=.45\textwidth, trim={0cm 0.1cm 0.1cm 0.1cm},clip]{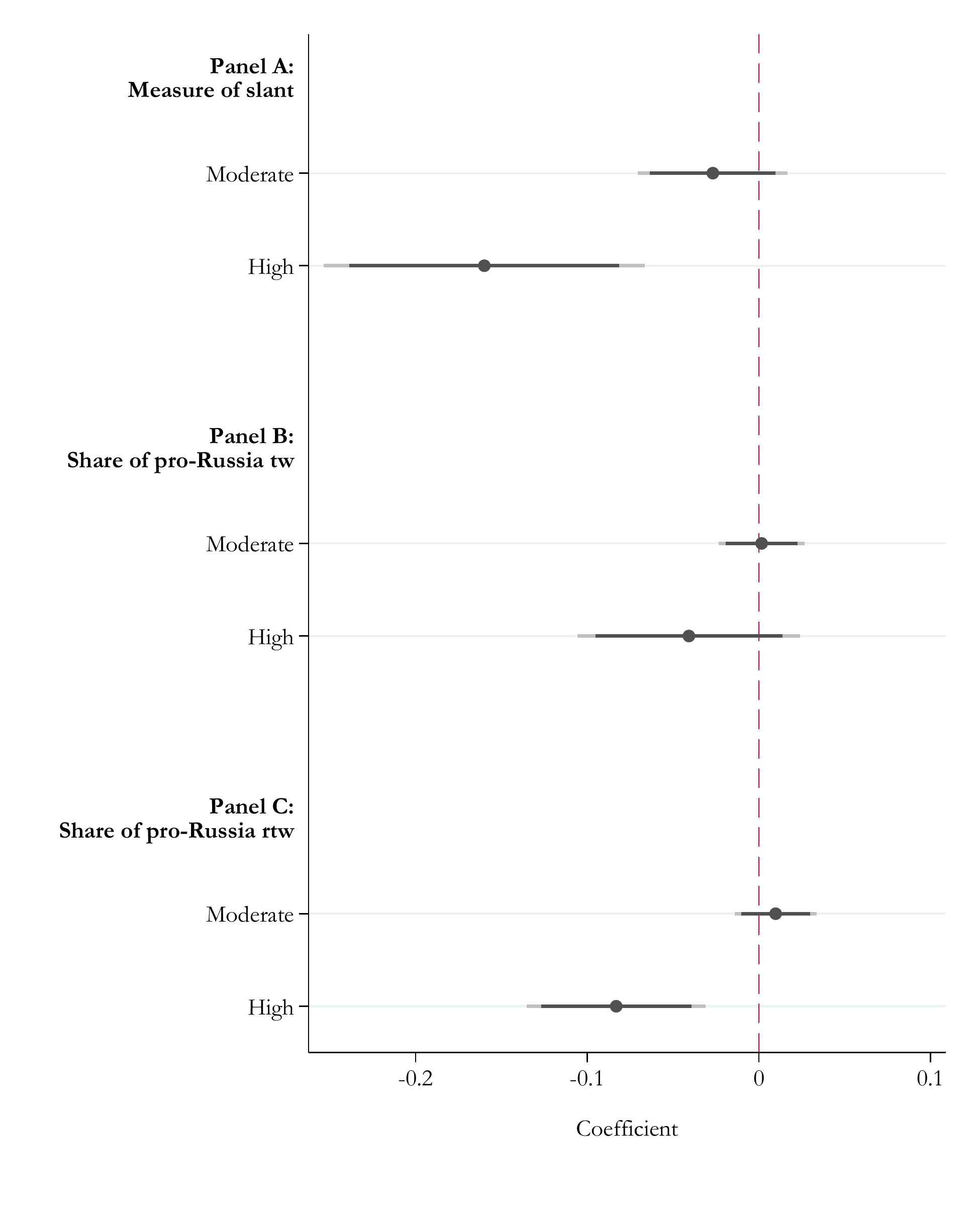}
	\begin{minipage}{\textwidth} \vspace{0.4cm}   \footnotesize{\textbf{Notes}: The figure presents coefficients from DiD regressions estimation Equation and assessing the ban’s heterogeneous impact on the \textit{interaction users}. For binary variables we use the threshold 0 instead than 1 of our slant measure, for tweets and retweets to be defined as pro-Russia. Users are divided into two groups: moderate, when the average slant in their tweets before the ban is in the bottom 75\% of the distribution, and high, for top 25\%. The coefficients are shown with 90\% and 95\% confidence intervals (95\% in light grey). Panel A shows the results for the intensive margin. Panel B and C show results on the daily proportion of tweets/retweets that can be classified as pro-Russia. All models include the UK and Switzerland as control countries and Austria, France, Germany, Ireland, and Italy as treatment countries. They include tweets from the period between 19$^{th}$ February to 15$^{th}$ March 2022, they incorporate both user- and day-fixed effects and cluster standard errors at the user level. We control for word count, mentions count, and hashtags count. Paper Figure \ref{fig:coefplot:interact:byext} shows results for all \textit{interaction users}.}\end{minipage}
\end{figure}

\vspace{-.5cm}
\begin{table}[H]\centering
	\caption{\textit{User-day level two-periods TWFE with post-ban weeks interactions: Interaction users using alternative threshold}}
	\label{app:tab:did:rusinteract:week:diffthres}
	\def\sym#1{\ifmmode^{#1}\else\(^{#1}\)\fi}
	\resizebox{1\linewidth }{!}{
		\begin{tabular}{lcccccc}
			\hline\hline
			                &\multicolumn{1}{c}{(1)}&\multicolumn{1}{c}{(2)}&\multicolumn{1}{c}{(3)}&\multicolumn{1}{c}{(4)}&\multicolumn{1}{c}{(5)}\\
                &Avg. media slant&\% pro-Russia tweets&\% pro-Russia retweets&Tot. Pro-Russia tweets&Tot. pro-Russia retweets\\
\hline
EU $\times$ 1st week after-ban&   -0.050&   -0.036&   -0.003&   -0.091&    0.014\\
                &  [0.022]&  [0.017]&  [0.015]&  [0.043]&  [0.046]\\
EU $\times$ 2nd week after-ban&   -0.034&   -0.005&    0.029&    0.044&    0.060\\
                &  [0.025]&  [0.020]&  [0.017]&  [0.051]&  [0.054]\\
User FEs        &      yes&      yes&      yes&      yes&      yes\\
Day FEs         &      yes&      yes&      yes&      yes&      yes\\
\hline
Observations    &    31082&    17297&    20436&    31082&    31082\\
\(R^{2}\)       &    0.332&    0.331&    0.254&    0.337&    0.506\\
Pre-period mean of DV&   -0.058&    0.459&    0.518&    1.307&    1.799\\
1st week \% of mean&   -87.08&    -7.76&    -0.55&    -6.97&     0.76\\
\hline\hline    &         &         &         &         &         \\

		\end{tabular}
	}
	\begin{minipage}{\textwidth} \vspace{0cm}   \footnotesize{\textbf{Notes}: The table displays coefficients from a two-period difference-in-differences regression analysis estimating Equation \ref{eq:did:user}, examining the ban's differential impact in the two weeks following the ban, on the \textit{interaction users}. For binary variables we use the threshold 0 instead than 1 of our slant measure, for tweets and retweets to be defined as pro-Russia. Column 1 shows the effects on our measure of media slant, the intensive margin. Columns 2 and 3 show effects on our measure of the extensive margin, respectively share of tweets and retweets that are pro-Russia. Columns 4 and 5 show the effect on the total number of pro-Russia tweets and retweets, respectively, produced by the author in the time period. All models include the UK and Switzerland as control countries and Austria, France, Germany, Ireland, and Italy as treatment countries. They include tweets from the period between 19$^{th}$ February to 15$^{th}$ March 2022, they incorporate both user- and day-fixed effects and cluster standard errors at the user level. We control for word count, mentions count, and hashtags count. Paper Table \ref{tab:did:rusinteract:week} shows results including all \textit{interaction users}.}\end{minipage}
\end{table}

\begin{table}[H]\centering
	\caption{\textit{User-day level two-periods TWFE: Interaction and non-interaction users using alternative threshold}}
	\label{app:tab:did:comparison:diffthres}
	\def\sym#1{\ifmmode^{#1}\else\(^{#1}\)\fi}
    \begin{minipage}{\textwidth}
       \footnotesize{\textbf{Panel A: Consumers of content from outlets}}
     \end{minipage}\\[0.2cm]
     \resizebox{1\linewidth }{!}{
		\begin{tabular}{lcccccc}
			\hline
			                &\multicolumn{1}{c}{(1)}&\multicolumn{1}{c}{(2)}&\multicolumn{1}{c}{(3)}&\multicolumn{1}{c}{(4)}&\multicolumn{1}{c}{(5)}\\
                &Avg. media slant&\% pro-Russia tweets&\% pro-Russia retweets&Tot. Pro-Russia tweets&Tot. pro-Russia retweets\\
\hline
EU $\times$ after-ban&   -0.043&   -0.023&    0.011&   -0.032&    0.034\\
                &  [0.020]&  [0.016]&  [0.013]&  [0.041]&  [0.043]\\
User FEs        &      yes&      yes&      yes&      yes&      yes\\
Day FEs         &      yes&      yes&      yes&      yes&      yes\\
\hline
Observations    &    31082&    17297&    20436&    31082&    31082\\
\(R^{2}\)       &    0.332&    0.331&    0.254&    0.336&    0.506\\
Pre-period mean of DV&   -0.058&    0.459&    0.518&    1.307&    1.799\\
\% of mean      &   -74.93&    -4.93&     2.15&    -2.47&     1.89\\
\hline\hline    &         &         &         &         &         \\

		\end{tabular}
	}
     \begin{minipage}{\textwidth}
       \footnotesize{\textbf{Panel A: Non-consumers of content from outlets}}
     \end{minipage}\\[0.2cm]
	\resizebox{1\linewidth }{!}{
		\begin{tabular}{lcccccc}
			\hline
                            &\multicolumn{1}{c}{(1)}&\multicolumn{1}{c}{(2)}&\multicolumn{1}{c}{(3)}&\multicolumn{1}{c}{(4)}&\multicolumn{1}{c}{(5)}\\
                &Avg. media slant&\% pro-Russia tweets&\% pro-Russia retweets&Tot. Pro-Russia tweets&Tot. pro-Russia retweets\\
\hline
EU $\times$ after-ban&   -0.030&    0.003&   -0.006&   -0.010&    0.055\\
                &  [0.007]&  [0.006]&  [0.005]&  [0.013]&  [0.009]\\
User FEs        &      yes&      yes&      yes&      yes&      yes\\
Day FEs         &      yes&      yes&      yes&      yes&      yes\\
\hline
Observations    &   322338&   152855&   186148&   322338&   322338\\
\(R^{2}\)       &    0.417&    0.409&    0.325&    0.293&    0.409\\
Pre-period mean of DV&   -0.189&    0.404&    0.477&    0.932&    1.089\\
\% of mean      &   -15.87&     0.83&    -1.36&    -1.12&     5.06\\
\hline\hline    &         &         &         &         &         \\

		\end{tabular}
	}
	\begin{minipage}{\textwidth} \vspace{0cm}   \footnotesize{\textbf{Notes}: The table displays coefficients from a two-period difference-in-differences regression analysis employing TWFE OLS estimator, examining the ban's impact on users who interacted with Russia Today and Sputnik before the ban in Panel A, and on user that had no interactions with the outlets in Panel B. For binary variables we use the threshold 0 instead than 1 of our slant measure, for tweets and retweets to be defined as pro-Russia. Column 1 shows the effects on our measure of media slant, the intensive margin. Columns 2 and 3 show effects on our measure of the extensive margin, respectively share of tweets and retweets that are pro-Russia. Columns 4 and 5 show the effect on the total number of pro-Russia tweets and retweets, respectively, produced by the author in the time period. All models include the UK and Switzerland as control countries and Austria, France, Germany, Ireland, and Italy as treatment countries. They include tweets from the period between 19$^{th}$ February to 15$^{th}$ March 2022, they incorporate both user- and day-fixed effects and cluster standard errors at the user level. We control for word count, mentions count, and hashtags count. Paper Table \ref{tab:did:comparison} shows results including also plausible bots.}\end{minipage}
\end{table}

\end{document}